\documentstyle[12pt]{article}
 \advance\oddsidemargin by -1in
 \advance\evensidemargin
 by -1in
 \marginparsep
 0.4cm
 \advance\topmargin by
 -0.0in
 \makeindex

 \newcommand\la{\langle}
 \newcommand\ra{\rangle}
 \newcommand\beq{\begin{equation}}
 \newcommand\noi{\noindent}
 \newcommand\eeq{\end{equation}}
 \newcommand\beqn{\begin{eqnarray}}
 \newcommand\eeqn{\end{eqnarray}}

\begin{document}
\vspace*{1.5cm}
 
\vspace*{4cm}
 
\centerline{\Large \bf Nonperturbative Effects
in Gluon Radiation}

\medskip

\centerline{\Large \bf and Photoproduction of Quark Pairs}
 
\vspace{.5cm}
\begin{center}
{\large Boris~Kopeliovich$^{1,2}$, Andreas Sch\"afer$^3$
and Alexander Tarasov$^{1,2,3}$}
 
\vspace{0.3cm}
 
 $^1${\sl Max-Planck Institut f\"ur Kernphysik, Postfach
103980, 69029 Heidelberg, Germany}\\
 $^2${\sl Joint Institute for Nuclear Research, Dubna,
141980 Moscow Region, Russia}\\
$^3${\sl Institut f\"ur Theoretische Physik,
Universit\"at Regensburg,
93040 Regensburg, Germany}

\end{center}

\vspace{1cm}
 
\begin{abstract}
We introduce a nonperturbative interaction for
light-cone fluctuations containing quarks and gluons.
The $\bar qq$ interaction squeezes the transverse size
of these fluctuations in the photon and one does not
need to simulate this effect via effective quark masses.
The strength of this interaction is fixed by data.
Data on diffractive dissociation of hadrons and photons
show that the nonperturbative interaction of gluons is much
stronger. We fix the parameters for the nonperturbative
quark-gluon interaction by data for diffractive dissociation
to large masses (triple-Pomeron regime). This allows
us to predict nuclear shadowing for gluons which turns out
to be not as strong as perturbative QCD predicts. We expect a
delayed onset of gluon shadowing at $x \leq 10^{-2}$ 
shadowing of quarks. Gluon shadowing turns
out to be nearly scale invariant up to virtualities $Q^2\sim 4\,GeV^2$
due to presence of a semihard scale characterizing the strong 
nonperturbative interaction of gluons.
We use the same concept to improve our description of gluon bremsstrahlung
which is related to the distribution function for a quark-gluon fluctuation
and the interaction cross section of a $\bar qqG$ fluctuation with a nucleon.
We expect the nonperturbative interaction to suppress dramatically the gluon
radiation at small transverse momenta compared to perturbative calculations. 

\end{abstract}

 
\newpage

\noi

\section{Introduction}

\bigskip

The light-cone representation 
introduced in \cite{ks} is nowadays a popular and powerful
tool to study the dynamics of photo-induced (real and virtual)
reactions. The central concept
of this approach is the non-normalized distribution amplitude 
of $\bar qq$ fluctuations
of the photon in the mixed $(\vec\rho,\,\alpha)$ representation, where
$\vec\rho$ is the transverse $\bar qq$ separation and $\alpha$ 
is the fraction of the light-cone momentum of the photon
carried by the quark (antiquark). For transversely and longitudinally
polarized photons it reads \cite{ks,bks},
\beq
\Psi^{T,L}_{\bar qq}(\vec\rho,\alpha)=
\frac{\sqrt{\alpha_{em}}}{2\,\pi}\,
\bar\chi\,\widehat O^{T,L}\,\chi\,K_0(\epsilon\rho)
\label{1.1}
\eeq
Here $\chi$ and $\bar\chi$ are the spinors of the quark
and antiquark respectively.
$K_0(\epsilon\rho)$ is the modified Bessel function, where
\beq
\epsilon^2 = \alpha(1-\alpha)Q^2 + m_q^2\ .
\label{1.1a}
\eeq
This is a generalization of \cite{ks,bks}
to the case of virtual photons \cite{nz,lmrt}.

The operators $\widehat O^{T,L}$ have the form,
\beq
\widehat O^{T}=m_q\,\vec\sigma\cdot\vec e +
 i(1-2\alpha)\,(\vec\sigma\cdot\vec n)\,
(\vec {e}\cdot \vec\nabla_{\rho})
+ (\vec\sigma\times\vec e)\cdot\vec\nabla_{\rho}\ ,
\label{1.2}
\eeq
\beq
\widehat O^{L}= 2\,Q\,\alpha(1-\alpha)\,\vec\sigma\cdot\vec n\ ,
\label{1.3}
\eeq
where the dimension-two operator $\vec\nabla_{\rho}$ 
acts on the transverse coordinate $\vec\rho$;
$\vec n=\vec p/p$ is a unit vector parallel to 
the photon momentum; $\vec e$ is the polarization vector
of the photon.

The advantage of the light-cone approach is the factorized form
of the interaction cross section which is given by the sum 
of the cross sections for different fluctuations weighted by 
the probabilities of these Fock states \cite{zkl,nz,mueller1}. 
The flavor independent color-dipole cross section 
$\sigma_{\bar qq}$ first
introduced in \cite{zkl} as dependent only on transverse $\bar qq$ 
separation $\rho$. It 
vanishes quadratically at $\rho\to 0$
due to color screening,
\beq
\sigma_{\bar qq}(\rho,s)\biggr|_{\rho\to 0}= 
C(\rho,s)\,\rho^2\ ,
\label{2.3a}
\eeq
where $C(\rho,s)$ is a smooth function of separation and energy.
In fact, $C(\rho,s)$ also depends on relative sharing 
by the $\bar q$ and $q$ of the total 
light cone momentum. We drop this dependence 
in what follows unless it is important ({\it e.g.} for 
diffractive gluon radiation).  
It was first evaluated assuming no energy dependence 
in pQCD \cite{zkl,kz} and phenomenologically
\cite{hp} at medium large energies and $\rho$'s and turned out to be 
$C\approx 3$. There are several models for the function $C(\rho,s)$ 
({\it e.g.} in \cite{fs-c,mark,shaw}), unfortunately neither
seems to be reliable. In this paper we 
concentrate on the principal problems how to include nonperturbative 
effects, and do not try to optimize the form of the cross section.
For practical applications it can be corrected as soon as a more
reliable model for $C(\rho,s)$ is available. 
We modify one of the models mentioned above
\cite{mark} which keeps the
calculations simple to make it more realistic and use it 
throughout this paper.

The distribution amplitudes (\ref{1.1}) control the mean transverse
$\bar qq$ separation in a virtual photon,
\beq
\la\rho^2\ra \sim \frac{1}{\alpha(1-\alpha)Q^2+m_q^2}
\label{1.4}
\eeq
Thus, even a highly virtual photon can create a large size 
$\bar qq$ fluctuation with large probability provided that
$\alpha$ (or $1-\alpha$) is very small, $\alpha Q^2 \sim m_q^2$.
This observation is central to the aligned jet model \cite{bks}.
At small $Q^2$ soft hadronic fluctuations become dominant at any $\alpha$.
In this case the perturbative distribution functions (\ref{1.1})
which are based on several assumptions including 
asymptotic freedom, are irrelevant.
One should expect that nonperturbative interactions
modify (squeeze) the distribution of
transverse separations of the $\bar qq$ pair.
In Section~2.1 we introduce a nonperturbative interaction
between the quark and antiquark into the Schr\"odinger type equation
for the Green function of the $\bar qq$ pair \cite{krt,rtv,z}.
The shape of the real part of this potential is adjusted to
reproduce the light-cone wave function of the $\rho$-meson.
We derive new light-cone distribution functions
for the interacting $\bar qq$ fluctuations of a photon,
which coincide with the known perturbative ones in the limit
of vanishing interaction.
The strength of the nonperturbative interaction can be
fixed by comparison with data sensitive to the
transverse size of the fluctuations. The observables we have chosen
in Section~2.2 
are the total photoabsorption cross sections on protons and nuclei
and the cross section for diffractive dissociation of a photon
into a $\bar qq$ pair.

For gluon bremsstrahlung we expect the transverse separation in 
a quark-gluon fluctuation to be of the order of the typical color
correlation length $\sim 0.3\,fm$ obtained by several QCD analyses
\cite{braun} - \cite{sh}. This corresponds to the radius of a constituent 
quark in many effective models. To the extend that the typical
$q-G$ separation is smaller than the $\bar qq$ one we expect gluon 
radiation to be suppressed. This results in particular in a suppression
of diffractive gluon radiation, {\it i.e.}
of the triple-Pomeron coupling, which is seen indeed in the data.

In Section~3.1 we assume a similar shape for 
the quark-gluon potential as for the $\bar qq$ one,
but with different parameters. A new light-cone distribution function
for a quark-gluon fluctuation of a quark is derived, which
correctly reproduces the known limit of perturbative QCD.

Comparison with data on diffractive excitations with large mass
fixes the strength of the nonperturbative interaction of gluons.
An intuitive physical picture of diffraction, as well 
as a simple calculation of the cross sections of different
diffractive reactions is presented in Appendix~A.
A more formal treatment of the same diffractive reactions via
calculation of Feynman diagrams is described in Appendix~B.

A crude estimate of the interaction parameters 
is given in Section~3.1.1 within the additive quark model (AQM).
For this purpose the cross section of 
diffractive gluon radiation by a quark,
$q\,N\,\to\,q\,G\,N$, is calculated in Appendices A.2 and B.1,
based on general properties of diffraction (Appendix~A.1) 
and the direct calculation of Feynman diagrams.

Quite a substantial deviation from the results for the AQM is found in
Section~3.1.2 and Appendix~C where the diffractive excitation of a nucleon
via gluon radiation, $N\,N\,\to\,X\,N$ is calculated.
The high precision of the data for this reaction allows to
fix the strength of the nonperturbative interaction of gluons
rather precisely.

The cross sections of diffractive gluon radiation
by mesons and photons are calculated in Appendices~A.3 and B.2.
In Section~3.1.3 we compare the values of the triple-Pomeron couplings
(calculated in Appendix~C) for diffractive dissociation of a 
photon and different hadrons and
find a violation of
Regge factorization by about a factor of two.

Our results for the cross section of diffractive dissociation
$\gamma^*\,N\,\to\,q\,\bar q\,G\,N$ in the limit 
of vanishing nonperturbative interaction can be compared with
previous perturbative calculations \cite{nz1,bartels}.
In this limit we are in agreement with \cite{bartels}, but
disagree with \cite{nz1}.\footnote{In spite of the claim in
\cite{bartels} that their result coincides with that of \cite{nz1},
they are quite different. We are thankful to Mark W\"usthoff 
for discussion of this controversy.}
The source of error in \cite{nz1} is the application of
Eq.~(\ref{b.6}) to an exclusive channel and a renormalization
recipe based on a probabilistic treatment of diffraction.

Diffractive radiation of photons is considered in Appendices~A.4
and B.3. It is shown that no radiation occurs without
transverse momentum transfer to the quark
(in contrast to gluon radiation). Therefore, the cross section 
for diffractive production of Drell-Yan
pairs is suppressed compared to the expectation of
\cite{k} which is also based on an improper 
application of Eq.~(\ref{b.6}) to an 
exclusive channel.

Section~3.2 is devoted to 
nuclear shadowing for the gluon distribution
function at small $x$. Calculations for many hard reactions on nuclei
(DIS, high $p_T$ jets, heavy flavor production, etc.)
desperately need the gluon distribution function
for nuclei which is expected to be shadowed at small $x$.
Many approaches \cite{glr}--\cite{fs2}
to predict nuclear shadowing for gluons
can be found in the literature (see recent review \cite{pw}). 
Our approach is based on 
Gribov's theory of inelastic
shadowing \cite{gribov} and is close to that in \cite{fs1,fs2}
which utilizes the results \cite{penn1,penn2} 
for the gluonic component of the diffractive structure function
assuming factorization and using available data. Instead,
we fix the parameters of the nonperturbative interaction 
using data on diffraction of protons and real photons. Besides,
we achieved substantial progress in understanding the evolution
of diffractively produced intermediate states in nuclear matter. 

Nuclear suppression of the gluon density which looks like a 
result of gluon fusion $G\,G\to G$ in the infinite momentum
frame of the nucleus, should be interpreted as usual
nuclear shadowing for the total interaction cross section of
fluctuations containing a gluon if seen
in the rest frame of the nucleus. We perform calculations for 
longitudinally polarized photons which are known to be a
good probe for the gluon distribution function. Although the 
physics of nuclear shadowing  and diffraction are closely related, 
even a good knowledge of single diffractive
cross section and mass distributions is not sufficient
to predict nuclear shadowing completely, but only the lowest
order shadowing correction. A technique for inclusion of 
the multiple scattering corrections was developed in
\cite{krt,rtv} which includes evolution of the intermediate
states propagating through the nucleus. These corrections are
especially important for gluon shadowing which does not
saturate even at very small $x$ in contrast to shadowing of quarks.
In Section~3.2.1 we find quite a steep $x$-dependence of gluon
shadowing at $Q^2\geq 4\,GeV^2$ 
which is rather weak compared to what have been
estimated in \cite{fs1,fs2}.
Shadowing starts at smaller values of $x < 0.01$ compared to the 
shadowing of quarks.
Such a delayed onset of gluon shadowing is a result of enlarged
mass of the fluctuations containing gluons.

As soon as our approach incorporates the nonperturbative effects
we are in position to calculate shadowing for soft gluons as well.
This is done in Section~3.2.2 using two methods. In hadronic
basis one can relate the shadowing term in the total hadron-nucleus
cross section to the known diffractive dissociation cross section.
This also give the scale for the effective absorption cross section.
A better way is to apply the Green function approach which
includes the nonperturbative gluon interaction fixed by comparison with
data for diffraction. With both methods we have arrived at a similar
shadowing, but the Green function approach leads to a delayed onset
of shadowing starting at $x < 0.01$. We conclude that gluon shadowing
is nearly scale independent up to $Q^2\sim 4\,GeV^2$.

The nonperturbative interaction of the radiated gluons
especially affects their transverse momentum distribution.
One can expect a substantial suppression of radiation with
small $k_T$ related to large transverse separations in
quark-gluon fluctuations of the projectile quark.
Indeed, in Section~3.3
we have found suppression by almost two orders of magnitude 
for radiation at $k_T=0$ compared to the perturbative QCD 
predictions. The difference remains quite large up to
a few $GeV$ of momentum transfer.
Especially strong nonperturbative effects we expect for
the $k_T$-distribution of gluon bremsstrahlung 
by a quark propagating through a nucleus. Instead of a sharp
peak at $k_T=0$ predicted by pQCD \cite{kst} now we expect 
a minimum.

\section{Virtual photoproduction of quark pairs}

\subsection{Green function of an interacting quark-antiquark pair}

\begin{figure}[tbh] 
\includegraphics{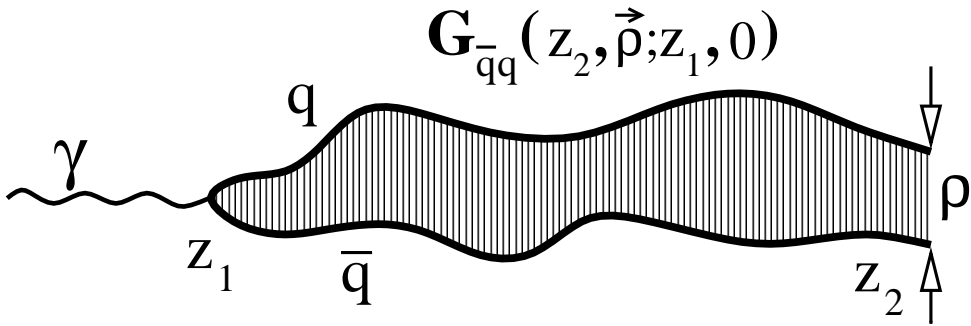}
\begin{center} 
\vspace{4cm} 
\parbox{13cm} 
{\caption[Delta]
{\sl Illustration for the Green function 
$G_{\bar qq}(z_1,\vec\rho_1=0;z_2,\vec\rho_2=\vec\rho)$
for an interacting $\bar qq$
fluctuation of a photon, as defined by Eq.~(\ref{2.1}).}
\label{fig1}} 
\end{center} 
\end{figure}

Propagation within a medium of an interacting $\bar qq$ pair which has 
been produced with initial separation $\rho=0$ from
a virtual photon 
at a point with
longitudinal coordinate $z_1$ and developed a separation $\vec\rho$
 at the point $z_2$ (see Fig.~\ref{fig1})
can be described by a light-cone Green function 
$G_{\bar qq}(z_1,\vec\rho_1=0;z_2,\vec\rho_2=\vec\rho)$.
The evolution equation for this Green
function was studied in \cite{krt}--\cite{z}\footnote{our Green function is 
related to that in \cite{krt} by\\ $G_{\bar qq}(z_1,\vec\rho_1=0
;z_2,\vec\rho_2=\vec\rho)=
exp[-i\epsilon^2(z_2-z_1)/2p\alpha(1-\alpha)]\,
W(z_1,\rho_1=0;z_2,\rho_2=\rho)$},
\beq
i\frac{d}{dz_2}\,G_{\bar qq}(z_1,\vec\rho_1;z_2,\vec\rho_2)=
\left[\frac{\epsilon^2 - \Delta_{\rho}}{2\,p\,\alpha\,(1-\alpha)}
+V_{\bar qq}(z_2,\vec \rho,\alpha)\right]
G_{\bar qq}(z_1,\vec\rho_1;z_2,\vec\rho_2)\ .
\label{2.1}
\eeq
The first term on the {\it r.h.s.} is analogous to the kinetic term in
a Schr\"odinger equation. It takes care of the phase shift
for the propagating $\bar qq$ pair. Indeed, the relevant 
phase factor is given by
${\rm exp}[i\int_{z_1}^{z_2}dz\,q_L(z)]$, with the relative 
longitudinal momentum transfer $q_L$. The latter is defined by
\beq
q_L(z)=\frac{M^2(z)+Q^2}{2p}=
\frac{\epsilon^2+k_T^2}{2p\alpha(1-\alpha)}
\label{2.2}
\eeq
Here $p$ is the photon momentum; 
$M$ is the effective mass of the $\bar qq$ pair (which
varies with $z$) and $Q^2$ is the photon virtuality. 
It depends on the transverse momentum $k_T$ 
of the quark (antiquark)
which is replaced by the Laplacian, $k_T^2 \Rightarrow - 
\Delta_{\rho}$, in the coordinate representation (\ref{2.1}).

The imaginary part of the potential $V_{\bar qq}(z_2,\vec\rho,\alpha)$
is responsible for absorption in the medium which is supposed to
be cold nuclear matter.
\beq
{\rm Im}\,V_{\bar qq}(z_2,\vec\rho,\alpha)=
-\frac{\sigma_{\bar qq}(\rho)}{2}\,\rho_A(z_2) 
\label{2.3}
\eeq
Here $\rho_A(z)$ is the nuclear density and we omit 
the dependence on the nuclear impact parameter.
$\sigma_{\bar qq}(\rho,s)$ is the 
total interaction cross section 
of a colorless $\bar qq$ pair with a nucleon \cite{zkl}
introduce in (\ref{2.3a}).
Eq.~(\ref{2.1}) with the imaginary potential (\ref{2.3}) was used in
\cite{krt} to calculate nuclear shadowing in deep-inelastic 
scattering. In other applications the quarks were treated as free, 
what is justified only in the domain of validity of perturbative QCD.

Our objective here is to include explicitly 
the nonperturbative interaction 
between the quarks in (\ref{2.1}).
We are going to rely on a nonrelativistic potential, which,
however, should be modified to be a function of the light-cone
variables $\vec\rho$ and $\alpha$. This general problem
is, however, not yet solved.
Nevertheless, we try to model the 
real part of the potential based on its general properties.
Particularly, the $\bar qq$ pair is supposed to have bound
states which are vector mesons. 

It is assumed usually that the wave function of a vector 
meson in the ground state depends on $\rho$ and $\alpha$ 
according to
\beq 
\Psi_V(\vec\rho,\alpha) = 
f(\alpha)\,{\rm exp}\left[-{1\over2}a^2(\alpha)\,
\vec\rho\;^2\right]\ .
\label{2.4}
\eeq
In order for this to be a solution of (\ref{2.1}) 
the real part of the potential should be,
\beq
{\rm Re}\,V_{\bar qq}(z_2,\vec\rho,\alpha) = 
\frac{a^4(\alpha)\,\vec\rho\;^2}
{2\,p\,\alpha(1-\alpha)}\ .
\label{2.5}
\eeq
Unfortunately, no reliable way to fix the form of $a(\alpha)$
is known. A parameterization popular in the literature is
$a(\alpha) = 2\,a\,\sqrt{\alpha(1-\alpha)}$,
which results from attempts to construct a relativistic 
approach to the problem of a $\bar qq$ bound state ( see \cite{hz}
and references therein).
In this case, however, the mean $\bar qq$ separation
$\rho\propto 1\left/\sqrt{\alpha(1-\alpha)}\right.$ increases unrestrictedly
towards the endpoints $\alpha=0,\ 1$. Such a behavior contradicts
the concept of confinement and should be corrected. The simplest way 
to do so is to add a constant term to $a(\alpha)$
(the real form of $a(\alpha)$ may be quite different, but so far
data allow only for a simple two parameter fit),
\beq
a^2(\alpha) = a^2_0 +4a_1^2\,\alpha(1-\alpha)\ .
\label{2.7}
\eeq
One can roughly evaluate $a_0$ by demanding that even at $\alpha =0,\ 1$
the transverse $\bar qq$ separation does not exceed the confinement radius,
\beq
a_0 \sim R_c^{-1}\approx \Lambda_{QCD}\ ,
\label{2.7a}
\eeq
{\it i.e.} $a_0\approx 200\,MeV$. Comparison with data (see below)
leads to a somewhat smaller value.

In what follows we study the consequences of the interaction between 
$q$ and $\bar q$ in the form (\ref{2.5}) --
(\ref{2.7}) for the quark wave function of the photon, and we discuss
several observables.

Let us denote the Green function of a $\bar qq$ pair propagating 
in vacuum (${\rm Im}\,V=0$) 
as $G_{\bar qq}(z_1,\vec\rho_1;z_2,\vec\rho_2)$.
The solution of (\ref{2.1}) has the form \cite{fg},
\beqn
G_{\bar qq}(z_1,\vec\rho_1;z_2,\vec\rho_2) &=&
\frac{a^2(\alpha)}{2\;\pi\;i\;{\rm sin}(\omega\,\Delta z)}\,
{\rm exp}\left\{\frac{i\,a^2(\alpha)}{{\rm sin}(\omega\,\Delta z)}\,
\Bigl[(\rho_1^2+\rho_2^2)\,{\rm cos}(\omega \;\Delta z) -
2\;\vec\rho_1\cdot\vec\rho_2\Bigr]\right.\nonumber\\
&-& \left.
\frac{i\;\epsilon^2\;\Delta z}{2\;p\;\alpha(1-\alpha)}\right\}\ ,
\label{2.8}
\eeqn
where $\Delta z=z_2-z_1$ and
\beq
\omega=\omega(\alpha)=
\frac{a^2(\alpha)}{p\;\alpha(1-\alpha)}\ .
\label{2.9}
\eeq
The normalization factor here is fixed by the condition
$G_{\bar qq}(z_1,\vec\rho_1;z_2,\vec\rho_2)|_{z_2=z_1}=
\delta^2(\vec\rho_1-\vec\rho_2)$.

Now we are in the position to calculate the distribution function
of a $\bar qq$ fluctuation of a photon including the 
interaction. It is given by the integral of the Green
function over the longitudinal coordinate
$z_1$ of the point at which the photon forms 
the $\bar qq$ pair (see Fig.~1),
\beq
\Psi^{T,L}_{\bar qq}(\vec\rho,\alpha)=
\frac{i\,Z_q\sqrt{\alpha_{em}}}
{4\pi\,p\,\alpha(1-\alpha)}
\int\limits_{-\infty}^{z_2}dz_1\,
\Bigl(\bar\chi\;\widehat O^{T,L}\chi\Bigr)\,
G_{\bar qq}(z_1,\vec\rho_1;z_2,\vec\rho_2)
\Bigr|_{\rho_1=0;\ \vec\rho_2=\vec\rho}
\label{2.10}
\eeq
The operators $\widehat O^{T,L}$ are defined in (\ref{1.3})--(\ref{1.4}).
Here they act on the coordinate $\vec\rho_1$.

If we write the transverse part  as
\beq
\bar\chi\;\widehat O^{T}\chi= A+\vec B\cdot\vec\nabla_{\rho_1}\ ,
\label{2.11}
\eeq
then the distribution functions read,
\beq
\Psi^{T}_{\bar qq}(\vec\rho,\alpha) =
Z_q\sqrt{\alpha_{em}}\,\left[A\,\Phi_0(\epsilon,\rho,\lambda)
+ \vec B\,\vec\Phi_1(\epsilon,\rho,\lambda)\right]\ ,
\label{2.12}
\eeq
\beq
\Psi^{L}_{\bar qq}(\vec\rho,\alpha) =
2\,Z_q\sqrt{\alpha_{em}}\,Q\,\alpha(1-\alpha)\,
\bar\chi\;\vec\sigma\cdot\vec n\;\chi\,
\Phi_0(\epsilon,\rho,\lambda)\ ,
\label{2.13}
\eeq
where
\beq
\lambda=\lambda(\alpha)=
\frac{2\,a^2(\alpha)}{\epsilon^2}\ .
\label{2.14}
\eeq

The functions $\Phi_{0,1}$ in (\ref{2.12})--(\ref{2.13})
are defined as
\beq
\Phi_0(\epsilon,\rho,\lambda) = 
\frac{1}{4\pi}\int\limits_{0}^{\infty}dt\,
\frac{\lambda}{{\rm sh}(\lambda t)}\,
{\rm exp}\left[-\frac{\lambda\epsilon^2\rho^2}{4}\,
{\rm cth}(\lambda t) - t\right]\ ,
\label{2.15}
\eeq
\beq
\vec\Phi_1(\epsilon,\rho,\lambda) = 
\frac{\epsilon^2\vec\rho}{8\pi}\int\limits_{0}^{\infty}dt\,
\left[\frac{\lambda}{{\rm sh}(\lambda t)}\right]^2\,
{\rm exp}\left[-\frac{\lambda\epsilon^2\rho^2}{4}\,
{\rm cth}(\lambda t) - t\right]\ .
\label{2.16}
\eeq

Note that the $q - \bar q$ interaction emerges in 
(\ref{2.12})--(\ref{2.13}) through the parameter $\lambda$ defined in 
(\ref{2.14}). In the limit $\lambda \to 0$ 
({\it i.e.} $Q^2\to 0$, $\alpha$ is fixed, $\alpha\not=0$ or $1$)
we get the well known
perturbative expressions (\ref{1.1}) for the distribution functions,
\beq
\Phi_0(\epsilon,\rho,\lambda)\Bigr|_{\lambda= 0} \Rightarrow
\frac{1}{2\pi}\,K_0(\epsilon\rho)\ ,
\label{2.17}
\eeq
\beq
\vec\Phi_1(\epsilon,\rho,\lambda))\Bigr|_{\lambda= 0} \Rightarrow
\frac{\epsilon\vec\rho}{2\pi\rho}\,K_1(\epsilon\rho) =
-\frac{1}{2\pi}\,\vec\nabla\,K_0(\epsilon\rho)\ .
\label{2.18}
\eeq
In contrast to these relations, in the general
case, {\it i.e.} for $\lambda \not= 0$
\beq
\vec\Phi_1(\epsilon,\rho,\lambda)\not=
-\vec\nabla\;\Phi_0(\epsilon,\rho,\lambda)\ .
\label{2.19}
\eeq

In the strong interaction limit 
$\lambda\gg m_q$ (or if both $(Q^2,\ m_q \to 0)$)
which is appropriate particularly for real 
photons and massless quarks, the functions $\Phi_{0,1}$
acquire again simple analytical forms,
\beq
\Phi_0(\epsilon,\rho,\lambda)\Bigr|_{\lambda\to\infty}
\Rightarrow \frac{1}{4\pi}\,
K_0\left[{1\over2}\,a^2(\alpha)\,\rho^2\right]\ ,
\label{2.20}
\eeq
\beq
\vec\Phi_1(\epsilon,\rho,\lambda)\Bigr|_{\lambda\to\infty}
\Rightarrow \frac{\vec\rho}{2\pi\rho^2}\,
{\rm exp}\left[-{1\over2}\,a^2(\alpha)\,\rho^2\right]\ ,
\label{2.21}
\eeq
The interaction confines even massless quarks within a finite 
range of $\rho$.

\subsection{Absorption cross section for virtual photons}

For highly virtual photons, $Q^2\gg a^2(\alpha)$, according to
(\ref{2.14}) $\lambda\to 0$ 
and the effects related to the nonperturbative 
$\bar qq$ interaction should be gone. 
Although for very asymmetric configurations,
$\alpha(1-\alpha)\ll 1$, see (\ref{1.1a}) the transverse $\bar qq$
separation increases and one may expect the nonperturbative interaction 
to be at work, it does not happen if the dipole cross section
is independent of $\rho$ at large $\rho$.

Thus, our equations 
show a smooth transition between the formalism of perturbative
QCD valid at high $Q^2$ and our model for low $Q^2$
where nonperturbative effects are important.

The absorption cross sections for transversely (T) and
longitudinally (L) polarized virtual photons ,
including the nonperturbative effects read,
\beq
\sigma_{tot}^T = 
2\;N_c\,\sum\limits_{F}Z_q^2\alpha_{em}
\int\limits_{0}^{1}d\alpha\int d^2\rho\,
\sigma_{\bar qq}(\rho,s)\,
\left\{m_q^2\,\Phi^2_0(\epsilon,\rho,\lambda) +
[\alpha^2+(1-\alpha)^2]\,
\left|\vec\Phi_1(\epsilon,\rho,\lambda)\right|^2\right\}
\label{2.22}
\eeq
\beq
\sigma_{tot}^L =
8\,Q^2\,N_c\,\sum\limits_{F}Z_q^2\alpha_{em}
\int\limits_{0}^{1}d\alpha\,\alpha\,(1-\alpha)
\int d^2\rho\,\sigma_{\bar qq}(\rho,s)\,
\Phi^2_0(\epsilon,\rho,\lambda)
\label{2.22a}
\eeq
Here $N_c$ is the number of colors, and the contributions of different 
flavors $F$ are summed up. 

According to (\ref{2.3a}) the dipole 
cross section vanishes $\sigma_{\bar qq}(\rho,s)\propto\rho^2$ at small
$\rho\ll 1\,fm$. Such a behavior approximately describes {\it e.g.}
the observed hierarchy of hadronic cross sections as functions of the 
mean hadronic radii \cite{hp}. We expect, however, that the
dipole cross section flattens off at larger separations
$\rho > 1\,fm$. Therefore, the approximation $\sigma_{\bar qq}(\rho)
\propto \rho^2$ is quite crude for the large separations 
typical for soft reactions. 
Even the simple two-gluon approximation \cite{l,n}
provides only a logarithmic growth at large $\rho$ \cite{zkl},
and confinement implies a cross section which becomes 
constant at large $\rho$. Besides, the energy dependence of the 
dipole cross
section is stronger at small $\rho$ than at large $\rho$
\cite{kp}. We use hereafter a parameterization similar to one
suggested in \cite{mark}.
\beq
\sigma_{\bar qq}(\rho,s)=\sigma_0(s)\,\left[
1-{\rm exp}\left(-\frac{\rho^2}
{\rho_0^2(s)}\right)\right]\ ,
\label{2.24}
\eeq
where $\rho_0(s)=0.88\,fm\,(s_0/s)^{0.14}$ and $s_0=1000\,GeV^2$.
In contrast to \cite{mark} all values depend on energy (as it is 
supposed to be) rather than on $x$ and
we introduce an energy dependent parameter
$\sigma_0(s)$,  
\beq
\sigma_0(s)=\sigma^{\pi p}_{tot}(s)\,
\left(1 + \frac{3\,\rho^2_0(s)}{8\,\la r^2_{ch}\ra_{\pi}}
\right)\ ,
\label{2.24a}
\eeq
otherwise one fails to reproduce hadronic
cross sections. Here 
$\la r^2_{ch}\ra_{\pi}=0.44\pm 0.01\,fm^2$ 
\cite{pion} is the mean square of the pion charge radius.
Cross section (\ref{2.24}) averaged with the pion
wave function squared automatically 
reproduces the pion-proton cross section. We use the results of 
the fit \cite{pdt} for the Pomeron part of the cross section,
\beq
\sigma^{\pi p}_{tot}(s)=
23.6\,(s/s_0)^{0.08}\, mb\ ,
\label{2.24b}
\eeq
where $s_0=1000\,GeV^2$.
We fixed the parameters comparing data with the proton structure function
calculated using Eqs.~(\ref{2.22})--(\ref{2.22a}) and 
the cross section (\ref{2.24}). Agreement is 
reasonably good up to $Q^2\sim 20\,GeV^2$ sufficient for our
purposes.

To fix from data the parameters $a_{0,1}$ 
of the potential we concentrate on
real photoabsorption which is most sensitive to nonperturbative corrections.
The photoabsorption cross section with 
free quark fluctuations in the photon diverges logarithmically at
$m_q\to 0$,
\beq
\sigma_{tot}^{T} \approx \sigma_0\,
{\rm ln}\left(\frac{1}{m_q\,\rho_0}\right)\ .
\label{2.25}
\eeq

Inclusion of interaction between the quarks in the photon
makes the photoabsorption cross section finite at 
$m_q\to 0$.
\beq
\sigma_{tot}^{T}=
\sigma_0\,\frac{\alpha_{em}N_c}{12\pi}\,
\sum\limits_F Z_q^2\, 
\biggl[\phi(x_1)-\phi(x_2)\biggr]\ ,
\label{2.26}
\eeq
where
\beq
x_1=\frac{1+a_0^2\rho_0^2}{a_1^2\rho_0^2}\ ;
\ \ \ x_2=\frac{a_0^2}{a_1^2}\ ,
\label{2.27a}
\eeq
and
\beq
\phi(x)=4\,{\rm ln}\left(\frac{x}{4}\right) -
2\,x+(4+x)\,\sqrt{1+x}\ 
{\rm ln}\left(\frac{\sqrt{1+x}+1}
{\sqrt{1+x}-1}\right)\ .
\label{2.27}
\eeq
In this case the cross section of
photoabsorption is independent of the quark mass
in the limit $m_q/a_{0,1}\ll 1$.

We adjust the values of $a_0$ and $a_1$ to the value
of the photoabsorption cross section
 $\sigma^{\gamma p}_{tot}=160\,\mu b$ at $\sqrt{s}=200\ GeV$
\cite{gamma1,gamma2}.
Eq.~(\ref{2.26}) alone does not allow to fix the two
parameters $a_0$ and $a_1$ completely, but 
it provides a relation between them. 
We found a simple way to parameterize this ambiguity. 
If we choose
\beqn
a_0^2&=&v^{
1.15}\, (112\,MeV)^2\nonumber\\
a_1^2&=&(1-v)^{1.15}\,(165\,MeV)^2\ ,
\label{2.28}
\eeqn
the total photoabsorption cross section, turns out to be
constant (within $1\%$) if $v$ varies between 0 and 1.
This covers all possible choices for $a_0$ and $a_1$.

In order to fix $v$ in (\ref{2.28}) one needs additional
 experimental information. We have tried a comparison 
with the following data:\\ 

(i) The cross section of forward diffraction dissociation
$\gamma N\to\bar qqN$ (the PPR graph in the triple-Reggeon
phenomenology \cite{kklp}),
\beq
\frac{d\sigma(\gamma N\to\bar qqN)}{dt}\biggr|_{t=0}=
\frac{1}{16\pi}\int d\alpha\, d^2\rho\,
\biggl|\Psi_{\bar qq}(\alpha,\rho)\biggr|^2
\,\sigma^2(\rho)\ .
\label{2.29}
\eeq

(ii) The total photoabsorption cross sections for nuclei
(high-energy limit),
\beq
\sigma^{\gamma A}_{tot}=2\int d^2b\int
d\alpha\, d^2\rho\,
\biggl|\Psi_{\bar qq}(\alpha,\rho)\biggr|^2
\,\left\{1-exp\left[{1\over2}
\sigma(\rho)T(b)\right]\right\}\ ,
\label{2.30}
\eeq
where 
\beq
T(b)=\int\limits_{-\infty}^{\infty}dz\,\rho_A(b,z)
\label{2.31}
\eeq
is the nuclear thickness function and the nuclear density
$\rho_A(b,z)$ depends on impact parameter $b$ and longitudinal 
coordinate $z$. This expression can be used for virtual
photons as well with a proper discrimination between transverse and
longitudinal photons.

A calculation of the observables (i) and (ii) shows, however, a 
surprising stability of the results against
variation of $v$ in (\ref{2.28}): the cross
sections change only within $\sim 1\%$ if $v$ varies between
0 and 1. Thus, we were unable 
to constrain the parameters $a_0$ and $a_1$ any further.

We have also calculated the effective interaction cross section
 of a $\bar qq$ pair with a nucleon,
\beq
\sigma^{\bar qqN}_{eff}=
\frac{\int d\alpha\,d^2\rho\,
\biggl|\Psi_{\bar qq}(\alpha,\rho)\biggr|^2
\,\sigma^2(\rho,s)}{\int d\alpha\,d^2\rho\,
\biggl|\Psi_{\bar qq}(\alpha,\rho)\biggr|^2
\,\sigma(\rho,s)}\,\equiv\, \frac{\la\sigma^2\ra}
{\la\sigma\ra}\ ,
\label{2.32}
\eeq
which is usually used to characterize 
shadowing for the interaction of the $\bar qq$ 
fluctuation of a real photon with a nucleus ({\it e.g.}
see in \cite{fs1,fs2}). 
We got at $\sigma^{\bar qqN}_{eff}= 30\,mb$ at $\sqrt{s}=200\,GeV$. 
This well corresponds to the pion-nucleon cross section (\ref{2.24b})
$\sigma^{\pi p}_{tot}= 31.7\,mb$ at this energy. This result might be
treated as success of VDM. 
On the other hand,
A calculation \cite{hkn} using VDM and 
$\sigma^{\pi p}_{tot}\approx 25\,mb$ instead
of $\sigma^{\rho p}_{tot}$ at lower energy 
for photoproduction of $\rho$-mesons off nuclei 
is in good agreement with recent HERMES measurements
\cite{hermes}. 

However, a word of caution is in order.
The nucleus to nucleon ratio of total photoabsorption cross sections  
in the approximation of frozen fluctuations (reasonably good at very 
small $x$) reads \cite{zkl,fs1,fs2},
\beq
\frac{\sigma^{\gamma^*A}_{tot}}
{A\,\sigma^{\gamma^*N}_{tot}}\, =\,
\frac{2}{\la \sigma\ra}\,\int d^2b\,
\left\la 1\,-\,{\rm exp}\left[-\frac{\sigma}{2}\,T(b)
\right]\right\ra\ ,
\label{2.33}
\eeq
Expanding the exponential up to the next term after the double
scattering one $(1/4)\,\sigma_{eff}$ one gets
$(1/24)\,\la\sigma^3\ra/\la\sigma\ra$. This is $1.5$ times larger
than $(1/24)\,\sigma_{eff}^2/\la\sigma\ra$ if to use the dipole approximation
$\sigma\propto \rho^2$ and a Gaussian distribution over $\rho$ 
for color triplet ($\bar q-q$) or color octet
($G - \bar qq$) dipoles.

\section{Gluon bremsstrahlung}

\subsection{Radiation of interacting gluons}

In processes with radiation of gluons, like
\beqn
q+N &\to& q+G+X\\
\label{3.1}
\gamma^*+N &\to& q+\bar q +G+X\ ,
\label{3.2}
\eeqn
the interaction between the radiated gluon and the parent quark
traveling in nearly the same direction 
may be important and significantly change the
radiation cross section and the transverse momentum distribution
compared to perturbative QCD calculations \cite{mueller,kst,urs}.

We describe the differential cross section of gluon 
radiation in a quark-nucleon collision
in the factorized light-cone approach \cite{kst}
\beq
\frac{d^3\sigma(q\to qG)}
{d({\rm ln}\alpha)\,d^2k_T} =
\frac{1}{(2\pi)^2}
\int d^2r_1\,d^2r_2\,
{\rm exp}\bigl[i\vec k_T(\vec r_1-\vec r_2)\bigr]\,
\Psi_{Gq}^*(\alpha,\vec r_1)\,
\Psi_{Gq}(\alpha,\vec r_2)\,
\sigma_G(\vec r_1,\vec r_2,\alpha)\ ,
\label{3.3}
\eeq
where 
\beq
\sigma_G(\vec r_1,\vec r_2,\alpha)={1\over 2}
\Bigl\{\sigma_{G\bar qq}
(\vec r_1,\vec r_1-\alpha r_2) +
\sigma_{G\bar qq}(\vec r_2,\vec r_2-\alpha r_1)
-\sigma_{\bar qq}[\alpha (\vec r_1-\vec r_2)] -
\sigma_{GG}(\vec r_1-\vec r_2)
\Bigr\}\ .
\label{3.4}
\eeq
Hereafter we assume all cross sections to depend on energy,
but do not show it explicitly for the sake of brevity 
(unless it is important).

The cross section of a colorless $G\bar qq$ system with a nucleon
$\sigma_{G\bar qq}(\vec r_1,\vec r_2)$ 
is expressed in terms of the usual $\bar qq$ dipole cross sections,
\beq
\sigma_{G\bar qq}(\vec r_1,\vec r_2)=
\frac{9}{8}\Bigl\{\sigma_{\bar qq}(r_1) +
\sigma_{\bar qq}(r_2)\Bigl\}-\frac{1}{8}\,
\sigma_{\bar qq}(\vec r_1-\vec r_2)
\label{3.5}
\eeq
$\vec r_1$ and $\vec r_2$ are the transverse separations
gluon -- quark and gluon -- antiquark respectively.
In (\ref{3.4}) $\sigma_{GG}(r)={9\over 4}\,\sigma_{\bar qq}(r)$ is the
total cross section of a colorless $GG$ dipole with a nucleon.

The cross sections of reactions (\ref{3.1})-(\ref{3.2}) integrated
over $k_T$ have simple form,
\beq
\frac{d\sigma(q\to qG)}
{d({\rm ln}\alpha)} =
\frac{1}{(2\pi)^2}
\int d^2r\,
\Bigl|\Psi_{Gq}(\alpha,\vec r)\Bigr|^2\,
\sigma_{G\bar qq}\left[\vec r,(1-\alpha)\vec r\right]\ ,
\label{3.5a}
\eeq
\beqn
&&\left.\frac{d\sigma(\gamma^*\to q\bar qG)}
{d({\rm ln}\alpha_G)}\right|_{\alpha_G\ll 1} =
\int\limits_0^1 d\alpha_q \int d^2R\,\,
\Bigl|\Psi^{\gamma^*}_{\bar qq}
(R,\alpha_q)\Bigr|^2
\nonumber\\ & \times & \,
\int d^2r
\biggl\{\Bigl|\Psi_{\bar qG}(\vec R 
+\vec r,\alpha_G)\Bigr|^2
\sigma_{GG}^N(\vec R +\vec r) + 
\Bigl|\Psi_{qG}(\vec r,\alpha_G)\Bigr|^2
\sigma_{GG}^N(r) 
\nonumber\\ & - & \,
{\rm Re}\,\Psi_{qG}^*(\vec r,\alpha_G)\,
\Psi_{\bar qG}(\vec R +\vec r,\alpha_G)\,
\Bigl[\sigma_{\bar qq}^N(\vec R +\vec r)+\sigma_{\bar qq}^N(r)-
\sigma_{GG}^N(R)\Bigr]
\biggr\}
\label{3.5b}
\eeqn
Here $\alpha_G$ is the fraction of the quark momentum carried by
the gluon; $\vec R$ and $\vec r$ are the quark-antiquark and 
quark-gluon transverse separations respectively.
The three terms in the curly brackets in (\ref{3.5b}) correspond
to the radiation of the gluon by the quark, by the antiquark 
and to their interference respectively.

The key ingredient of (\ref{3.3}), (\ref{3.5a}) and (\ref{3.5b}) 
is the distribution function
$\Psi_{Gq}(\alpha,\vec r)$ of the quark-gluon fluctuation, where
$\alpha$ is the fraction of the light-cone momentum of
the parent quark carried by the gluon, and $\vec r$ is the 
transverse quark-gluon separation. This function has a form
\cite{kst,hir,bhq} similar to (\ref{1.1}),
\beq
\Psi_{Gq}^T(\alpha,\vec r)\Bigr|_{free}={1\over\pi}\,
\sqrt{\frac{\alpha_{s}}{3}}\,
\chi_f\,\widehat\Gamma^{T,L}\,\chi_i\,K_0(\tau r_T)\ ,
\label{3.6}
\eeq
where the operator $\widehat\Gamma^{T}$ is defined in \cite{kst},
\beq
\widehat\Gamma^{T} = i\,m_q\alpha^2\,
\vec {e^*}\cdot (\vec n\times\vec\sigma)\,
 + \alpha\,\vec {e^*}\cdot (\vec\sigma\times\vec\nabla)
-i(2-\alpha)\,\vec {e^*}\cdot \vec\nabla\ ,
\label{3.6a}
\eeq
We treat the gluons as massless since we incorporate the 
nonperturbative interaction explicitly and do not need
to introduce any effective mass.

The factor $\tau$ differs from $\epsilon$ as defined in (\ref{1.1a}),
\beq
\tau^2=\alpha^2m_q^2
\label{3.7}
\eeq

In the general case the distribution function including the interaction 
between the quark and gluon can be found via the Green function
$G_{qG}(z_1,\vec\rho_1;z_2,\vec\rho_2)$ 
for the propagation of a quark-gluon pair, in analogy to (\ref{2.10}),
\beq
\Psi_{qG}(\vec\rho,\alpha)=
\frac{i\,\sqrt{\alpha_{s}/3}}
{2\pi\,p\,\alpha(1-\alpha)}\,
\int\limits_{-\infty}^{z_2}dz_1\,
\Bigl(\bar\chi\;\widehat\Gamma^{T}\chi\Bigr)\,
G_{qG}(z_1,\vec\rho_1;z_2,\vec\rho_2)
\Bigr|_{\rho_1=0;\ \vec\rho_2=\vec\rho}
\label{3.9}
\eeq

Let us add a few comments as to why this direct analogy holds. Eq.~(\ref{3.4})
might give the impression that we would have to implement the
interaction between all three partons: the gluon, the
quark and the antiquark. Checking the way in which this equation was derived,
one realizes, however, that this is not the case. 
We studied gluon bremsstrahlung from a single quark
and then expressed the radiation amplitude as a difference between the
inelastic amplitudes for a $qG$ system and an individual $\bar q$.
This is how $\sigma_{G\bar qq}$ has to be interpreted and
this is why one should only take the $q-G$ nonperturbative interaction 
into account.

The evolution equation for the Green function of an interacting
$qG$ pair originating from the parent quark at the point
with longitudinal coordinate $z_1$ with initial transverse separation
$\rho_1=0$ looks similar to (\ref{2.1}) with the replacement
$\epsilon\Rightarrow\tau$ and $V_{\bar qq}(z_2,\vec\rho,\alpha)
\Rightarrow V_{qG}(z_2,\vec\rho,\alpha)$. We parameterize the quark-gluon
potential in the same way as in (\ref{2.5}) for quark-antiquark,
\beq
{\rm Re}\,V_{qG}(z_2,\vec\rho,\alpha) =
\frac{b^4(\alpha)\,\vec\rho\;^2}
{2\,p\,\alpha(1-\alpha)}\ ,
\label{3.10}
\eeq
where $b^2(\alpha)=b_0^2+4\,b^2_1\,\alpha\,(1-\alpha)$.

The solution of the evolution equation for the quark-gluon Green function
in absence of absorption (${\rm Im}\,V_{qG}=0$) looks the same as (\ref{2.8})
with replacement $a(\alpha)\Rightarrow b(\alpha)$.

The following transformations go along with (\ref{2.10}) -- (\ref{2.21}).
The vertex function in (\ref{3.9}) is represented as,
\beq
\bar\chi\;\widehat\Gamma^{T}\chi= D+\vec E\cdot\vec\nabla_{\rho_1}\ ,
\label{3.11}
\eeq
then the result of integration in (\ref{3.9}) is,
\beq
\Psi_{qG}(\vec\rho,\alpha) =
2\,\sqrt{\left(\frac{\alpha_{s}}{3}\right)}\,
\left[D\,\Phi_0(\tau,\rho,\lambda)
+ \vec E\cdot\vec\Phi_1(\tau,\rho,\lambda)\right]\ ,
\label{3.12}
\eeq
The functions $\Phi_0(\tau,\rho,\lambda)$ and 
$\vec\Phi_1(\tau,\rho,\lambda)$ are defined in 
(\ref{2.15}) -- (\ref{2.16}).
However, $\lambda$ is now defined by
\beq
\lambda=
\frac{2\,b^2(\alpha)}{\tau^2}\ .
\label{3.14}
\eeq

One might argue that the quark-gluon potential we need (and which we 
shortly shall constrain by comparison with experimental data)
could simply be obtained by adding two quark-quark potentials
with an appropriate color factor. Such a procedure
could, however, lead to a completely wrong results as we want to illustrate
by the following example.

Motivated by perturbative QCD one might expect that
the gluon-gluon and quark-quark potentials differ
simply by a factor $9/4$ (the ratio of the Casimir factors).
However, this relation is affected by non-trivial properties of the
QCD vacuum which makes the interaction of gluons much stronger
 \cite{nsvz,sh}. The octet string tension $\kappa_8$ is related
to $\alpha'_P$, the slope of the Pomeron trajectory in the
same way as the color triplet string tension relates to the slope of
the meson Regge trajectories \cite{cnn}, 
\beq
\kappa_8=\frac{1}{2\pi\alpha'_P}
\approx 4\,GeV/fm .
\label{3.15}
\eeq
Here $\alpha'_P=0.25\,GeV^{-2}$. Thus, the value of $\kappa_8$ is
in fact four times larger than the well known 
$\bar qq$ string tension, $\kappa_3=1\,GeV/fm$, and not only by a 
factor $9/4$. 

Another piece of information about the strength of the gluon interaction 
which supports this observation
comes from data on diffractive dissociation. The triple-Pomeron
coupling turns out to be rather small \cite{kklp}. If interpreted
as a product of the Pomeron flux times the Pomeron-proton total cross section,
the latter turns out to be an order of magnitude smaller than
the proton-proton one. Naively one would again assume that 
the Pomeron as a colorless
gluonic dipole should interact $9/4$ times stronger that
an analogous $\bar qq$ dipole (such a consideration led some
authors to the conclusion that gluons are shadowed at small $x$
in nuclei stronger than sea quarks). 
The only way to explain this discrepancy
is to assume that the gluon-gluon dipole is much smaller. 
This, in turn, demands a stronger gluon-gluon interaction.
Thus, diffraction is sensitive to nonperturbative interaction of gluons.
We shall use this observation in in the next 
section to fix the corresponding parameters
$b_0$ and $b_1$.

\subsubsection{Diffractive bremsstrahlung of gluons. 
The triple-Pomeron coupling in the additive quark model}

\begin{figure}[tbh]
\includegraphics{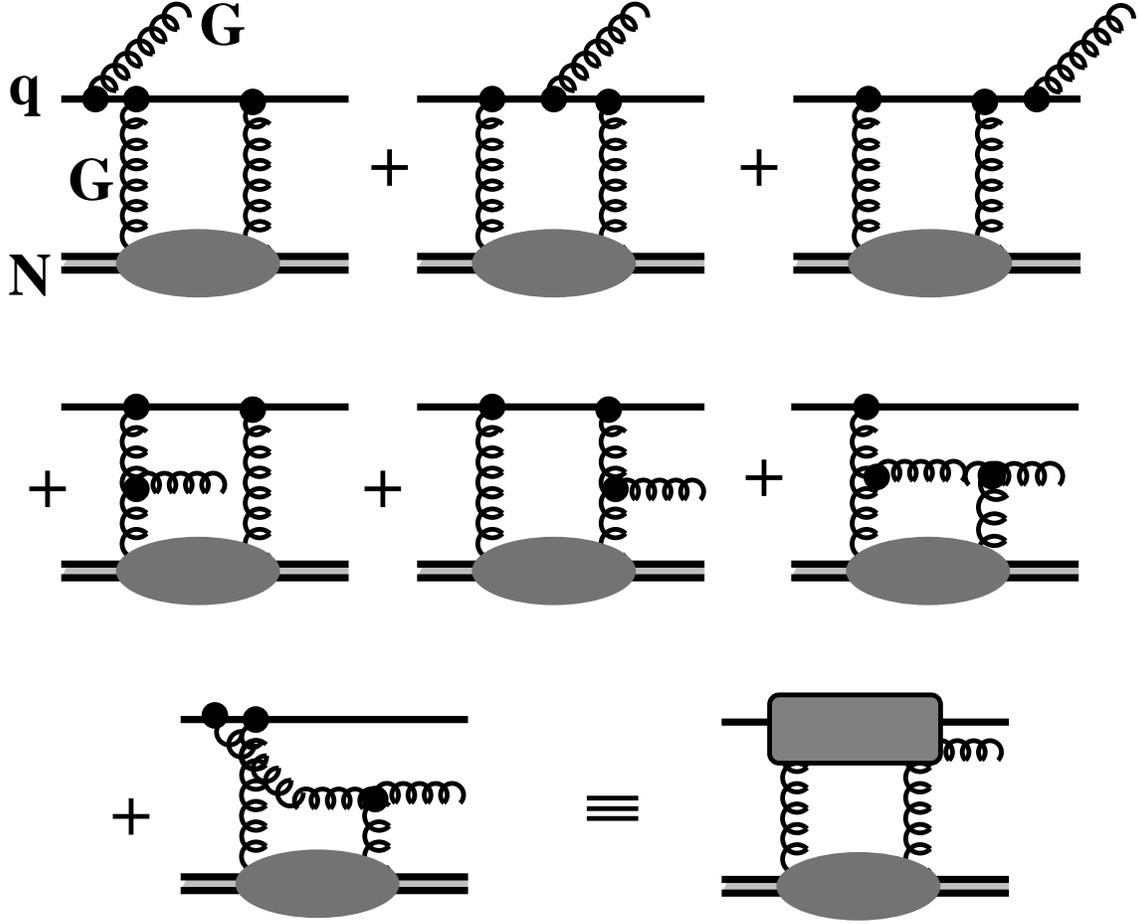}
\begin{center}
\vspace{12.5cm}
\parbox{13cm}
{\caption[Delta]
{\sl Feynman diagrams for diffractive radiation of a gluon 
in a quark-nucleon interaction, $qN\to qGN$.}
\label{fig2}}
\end{center}
\end{figure}

Let us start with diffractive dissociation of a quark,
$qN\to qGN$. We assume the diffractive amplitude to be proportional
to the gluon density $G(x,Q^2)=x\,g(x,Q^2)$ \cite{fs,barone} 
(it should be a non-diagonal distribution if the energy
is not very large) as is
shown in Fig.~\ref{fig2}.
Since the amplitude is predominantly imaginary at high energies
one can use the generalized unitarity relation known
as Cutkosky rule \cite{cut},
\beq
2\,{\rm Im}\,A_{ab} = \sum\limits_c A_{ac}A^{\dagger}_{cb}\ ,
\label{3.16}
\eeq
where $A_{ab}$ is the amplitude of the process $a\to b$ and $a$, $b$ 
denote all the particles in initial and final states respectively.
In the case under discussion $a=\{q,N\},\ b=\{q,G,N\}$, and $c$
denotes either $c_1=\{q,N_8^*\}$ or $c_2=\{q,G,N_8^*\}$, where
$N_8^*$ is a color-octet excitation of the nucleon resulting from 
gluon radiation(absorption) by a nucleon.

In what follows, we concentrate on forward diffraction, {\it i.e.} the 
transverse momentum transfer $\vec q_T=0$.
In this case the diffractive amplitude reads,
\beq
F(\alpha,k_T,q_T=0) =
\frac{i}{4\pi}\int \frac{d^2\rho}{2\pi}\,e^{i\vec k_T\vec\rho}\,
\Psi_{qG}(\alpha,\vec\rho)\,\widetilde\sigma(\vec\rho)\ ,
\label{3.17}
\eeq
where $k_T$ is the transverse momentum of the radiated gluon
 and 
\beq
\widetilde\sigma(\vec\rho)=
{9\over8}\,\sigma_{\bar qq}(\vec\rho)\ .
\label{3.18}
\eeq
Eq.~(\ref{3.17}) is derived in Appendix~A.2 in 
a simple and intuitive way
based on the general properties of a diffractive process
discussed in Appendix~A.1. A more formal derivation
based on a direct calculation of Feynman diagrams and
the Cutkosky rule (\ref{3.16}) 
is presented in Appendix~B.1.

The relation (\ref{3.17}) is valid for any value of $\alpha$. 
In contrast to the inclusive cross section for gluon 
bremsstrahlung the diffractive cross section depends on $\alpha$  
only via the distribution function.

The amplitude (\ref{3.17}) is normalized according to,
\beq
\frac{d\sigma(qN\to qGN)}{d(ln\alpha)\,d^2k_T\,d^2q_T} = 
\biggl|F(\alpha,\vec k_T,\vec q_T)\biggr|^2\ .
\label{3.19}
\eeq

The distribution for the effective mass squared, $M^2=k_T^2/\alpha(1-\alpha)
\approx k_T^2/\alpha$, at $q_T=0$ has the form,
\beq
\frac{d\sigma(qN\to qGN)}{dM^2\,dq_T^2}
\biggr|_{q_T=0} =2\pi^2\int\limits_0^{M^2}dk_T^2\,
\biggl|F(\alpha,\vec k_T,\vec q_T)\biggr|^2\ ,
\label{3.20}
\eeq
which transforms in the limit $M^2\to\infty$ into
\beq
\frac{M^2\,d\sigma(qN\to qGN)}{dM^2\,dq_T^2}
\biggr|_{q_T=0} = \frac{1}{16\pi}\,
\lim\limits_{\alpha\to 0}
\int d^2\rho\,\biggl|\Psi_{qG}(\alpha,\vec\rho)\,
\widetilde\sigma(\rho,\bar s)\biggr|^2\ ,
\label{3.21}
\eeq
where $\bar s=M_0^2/\xi$, where $M_0^2=1\,GeV^2$ and $\xi\equiv x_P
\equiv 1-x_F \approx M^2/s$. 

Since dissociation into large mass states is dominated by 
the triple-Pomeron (3P) graph the value on the {\it l.h.s.} of
(\ref{3.21}) is the effective $3P$ 
coupling $G_{3P}(qN\to XN)$ (see definition in
\cite{kklp}) at $q_T=0$. It can be evaluated using
$G_{3P}(pp\to Xp)\approx 1.5\,mb/GeV^2$ as it follows from
measurement by the CDF Collaboration \cite{cdf} (according to
\cite{dino} \footnote{We thank Doug Jansen and Thomas Nunnemann
who helped to clarify this point.} we divided the value of 
$G_{3P}$ given in \cite{cdf} by factor 2). 
This value is twice as small as one derived from the
triple-Regge analyses \cite{kklp} at medium large energies.
This is supposed to be due to absorptive corrections which
grossly diminish the survival probability of large rapidity gaps
at high energies \cite{glm}.
One can see energy dependence of $G_{3P}$ even in the energy
range of the CDF experiment \cite{cdf}. 

Assuming the additive quark model (AQM) to be valid
one can write,
\beq
G^{AQM}_{3P}(qN\to XN)\approx 
{1\over3}\,G_{3P}(NN\to XN)\approx
0.5\,\frac{mb}{GeV^2}
\label{3.22}
\eeq
(see below about interference effects).
To compare with this estimate we 
calculate the triple-Pomeron coupling (\ref{3.21})
using the distribution function in the form (\ref{3.12})
and the dipole cross section (\ref{2.24}),
\beq
G^{AQM}_{3P}(qN\to XN) = \frac{27\,\alpha_s}{4}\,
\left(\frac{\sigma_0}{8\pi}\right)^2\,
{\rm ln}\left(\frac{t_1t_2}{t_3^2}\right)\ ,
\label{3.23}
\eeq
where $t_1=b^2(0)/2$, $t_2=t_1+1/\rho_0^2$
and $t_3=2t_1t_2/(t_1+t_2)$. The parameters $\sigma_0$ and
$\rho_0$ are defined in (\ref{2.24}).
We use here a fixed value of $\alpha_s=0.6$ which is an appropriate
approximation for a soft process.

Comparison of Eq.~(\ref{3.23}) with the value (\ref{3.22})
leads to a rough evaluation of the parameter $b_0$ of our potential
(we are not sensitive to $b_1$  since keep $\alpha$ small),
\beq
b^{AQM}(0)\approx 570\,MeV
\label{3.24}
\eeq

Thus, a typical quark-gluon separation is $\sim 1/b(0)\approx 0.4\,fm$
what is roughly the radius of a `constituent' quark.
Note that a substantial modification of (\ref{3.22}) by interference 
of radiation amplitudes for different quarks
is possible.

\subsubsection{Diffractive excitation of nucleons,
\boldmath$N\,N\to X\,N$, beyond the AQM. }

\begin{figure}[tbh]
\includegraphics{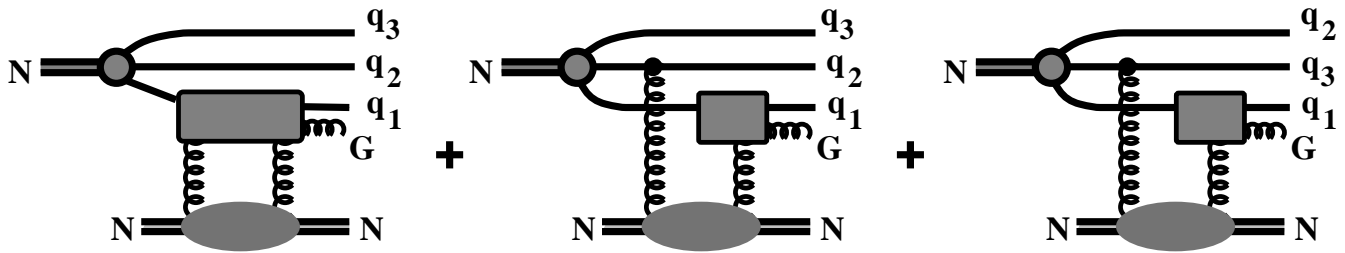}
\begin{center}
\vspace{4cm}
\parbox{13cm}
{\caption[Delta]
{\sl Contributions from projectile valence quarks
to the amplitude of diffractive gluon emission in $NN$
collisions. Six additional graphs resulting from 
the permutation
$\{1\rightleftharpoons2\}$ and $\{1\rightleftharpoons3\}$
have not been plotted.}
\label{fig4}}
\end{center}
\end{figure}

The amplitude of diffractive gluon radiation $N\,N\to 3q\,G\,N$ 
can be represented 
as a superposition of radiation by different 
quarks as shown in Fig.~\ref{fig4}.
In this process the colorless $3q$ system 
($|3q\ra_1$) converts into a color-octet final state 
($|3q\ra_8$). There are two independent octet $|3q\ra$ states 
which differ from each other by their symmetry under a permutation 
of the color indices of the quarks. Correspondingly, the amplitude 
for the process $NN\to |3q\ra_8GN$ is a superposition of two amplitudes
(see below).

The contribution to the amplitude of the first graph in Fig.~\ref{fig4}
reads,
\beq
F^I(NN\to 3qGN)=
\frac{if_{abc}}{\sqrt{3}}\,
\la(3q)_8|\tau_b^{(1)}\,\tau_c^{(2)}
|(3q)_1\ra\, \widetilde\sigma(\vec\rho_1)\,
\Phi(\{\vec r,\alpha\},\vec\rho_1,\alpha_G)
\label{3.28}
\eeq

The second and third graphs in Fig.~\ref{fig4} give correspondingly,
\beq
F^{II}(NN\to 3qGN) =
\frac{if_{abc}}{\sqrt{3}}\,
\la(3q)_8|\tau_b^{(1)}\,\tau_c^{(2)}
|(3q)_1\ra\, 
\biggl[\widetilde\sigma(\rho_2)
-\widetilde\sigma(\vec r_1-\vec r_2)\biggr]\,
\Phi(\{\vec r,\alpha\},\vec\rho_1,\alpha_G)\ ;
\label{3.29}
\eeq
and
\beq
F^{III}(NN\to 3qGN) =
\frac{if_{abc}}{\sqrt{3}}\,
\la(3q)_8|\tau_b^{(1)}\,\tau_c^{(2)}
|(3q)_1\ra\,
\biggl[\widetilde\sigma(\rho_3)
-\widetilde\sigma(\vec r_1-\vec r_3)\biggr]\,
\Phi(\{\vec r,\alpha\},\vec\rho_1,\alpha_G)\ .
\label{3.30}
\eeq
Here $\{\vec r,\alpha\}=(\vec r_1,\vec r_2,\vec r_3;
\alpha_1\alpha_2,\alpha_3)$; $\vec r_i$ are the 
positions of the quarks in the impact parameter plane;
$\vec\rho_i=\vec\rho -
\vec r_i$, where $\vec\rho$ is the position of the gluon;
\beq
\Phi(\{\vec r,\alpha\},\vec\rho_i,\alpha_G/\alpha_i)=
\Psi_{N\to 3q}(\{\vec r,\alpha\})\,
\Psi_{Gq}(\vec\rho_i,\alpha_G)\ ;
\label{3.31}
\eeq
$f_{abc}$ is the structure constant of the color group,
where ``$a$''  is the color index of the radiated  gluon,
and we sum over ``$b$'' and ``$c$''. The Gell-Mann matrices 
$\lambda_c^i=2\,\tau_c^i$ act on
the color index of $i$-th quark.

Using the relation, 
\beq
f_{abc}\,\tau^{(1)}_{b}\,\tau^{(1)}_{c}\,|3q\ra_1=
f_{abc}\,(\tau^{(2)}_{b}\,\tau^{(1)}_{c}+
\tau^{(3)}_{b}\,\tau^{(1)}_{c})\,|3q\ra_1
\label{3.32}
\eeq
one can present the sum of the amplitudes $F^I,\ F^{II}$ and $F^{III}$
in the form,
\beqn
&&F^{(1)}(NN\to 3qGN)=F^{I}+F^{II}+F^{III}\nonumber\\
&=&\frac{if_{abc}}{\sqrt{3}}\,
\Phi(\vec r_i,\alpha_i,\vec\rho-\vec r_1,\alpha_G)\,
\la(3q)_8|\tau_b^{(2)}\,\tau_c^{(1)}\,\Sigma_{12}+
\tau_b^{(3)}\,\tau_c^{(1)}\,\Sigma_{13}
|(3q)_1\ra\ ,
\label{3.33}
\eeqn
where $\Sigma_{ij}=\widetilde\sigma(\vec\rho-\vec r_i)+
\widetilde\sigma(\vec\rho-\vec r_j)-
\widetilde\sigma(\vec r_i-\vec r_j)$.
The index ``1'' in $F^{(1)}(NN\to 3qGN)$ indicates that
the gluon is radiated by the quark $q_1$ in accordance
with Fig.~\ref{fig4}.

The amplitudes $F^{(2)}$ and $F^{(3)}$ is obviously 
related to $F^{(1)}$ by replacement $1\to 2,3$.
Note that the color structure $f_{abc}\,\tau^{(2)}\,\tau^{(3)}$
which is not present in (\ref{3.33}) is not independent
due to the relation,
\beq
f_{abc}\left(\tau^{(1)}\,\tau^{(2)}+\tau^{(2)}\,\tau^{(3)}+
\tau^{(3)}\,\tau^{(1)}\right)\,|3q\ra_1 = 0\ .
\label{3.34}
\eeq
Thus, we are left with only two independent color structures, as was
mentioned above.

The full amplitude for diffractive gluon radiation squared
$|F(NN\to 3qGN)|^2=|F^{(1)}+F^{(2)}
+F^{(3)}|^2$,  summed over all color states of the $3qG$ system,
reads,
\beqn
\sum\limits_f\,\Biggl|F(NN\to 3qGN)\Biggr|^2 &=&
{1\over3}\,\left|\Psi_{N\to 3q}(\{\vec r,\alpha\})\right|^2\,
\left\{\sum\limits_{i=1}^3 \left|\Psi_{qG}(\vec\rho_i,\alpha_G)\right|^2\,
A^{(i)}(\{\vec r\},\vec\rho)\right.\nonumber\\
&-& \left.{\rm Re}\,\sum\limits_{i\not=k}\Psi_{qG}(\rho_i,\alpha_G)\,
\Psi_{qG}(\vec\rho_k,\alpha_G)\,B^{(i,k)}(\{\vec r\},\vec\rho)
\right\}\ ,
\label{3.35}
\eeqn
where
\beqn
A^{(1)}(\{\vec r\},\vec\rho)&=&
\Sigma_{12}^2 + \Sigma_{13}^2
+ \Sigma_{12}\Sigma_{13}\ ,
\label{3.36}\\
B^{(1,2)}(\{\vec r\},\vec\rho)&=&
2\,\Sigma_{12}^2+\Sigma_{12}(\Sigma_{13}+\Sigma_{23})-
\Sigma_{13}\Sigma_{23}\ .
\label{3.37}
\eeqn
The expressions for $A^{(2)},\ A^{(3)}$ and $B^{(1,3)},\ B^{(2,3)}$
are obtained by simply changing the indices.

The effective triple-Pomeron coupling 
results from integrating (\ref{3.35}) over phase space,
\beqn
G_{3P}(NN\to NX)&=&\frac{1}{16\pi}
\int d^2r_1\,d^2r\,_2d^2r_3\,d^2\rho\, d\alpha_1\,d\alpha_2\,d\alpha_3\,
\delta(\vec r_1+\vec r_2+\vec r_3)\nonumber\\
&\times&\delta(1-\alpha_1-\alpha_2-\alpha_3)\,
\sum\limits_f\,\Biggl|F(NN\to 3qGN)\Biggr|^2\ .
\label{3.38}
\eeqn

To evaluate $G_{3P}(NN\to NX)$ we use (\ref{2.13}) for
$\Psi_{qG}(\vec\rho,\alpha)$ and a Gaussian parameterization
for the valence quark distribution in the nucleon,
\beq
\left|\Psi_{N\to 3q}(\{\vec r,\alpha\})\right|^2
\propto {\rm exp}\left[- 
\left(\sum\limits_{i=1}^2 \vec r_i\,^2\right)
\biggl\slash
\la r^2_{ch}\ra_p\right]\ ,
\label{3.39}
\eeq
where $\la r^2_{ch}\ra_p\approx 0.79\pm 0.03\,fm^2$ is the mean square radius 
of the proton \cite{garching}. 
At this point one has to introduce some specific model for the 
$\alpha_i$ distributions. Quite some proposals can be found in the 
literature, and a quantitative analysis will require careful numerical studies. 
For a first qualitative discussion we make the simple ansatz
for the quark momentum distribution in the nucleon, 
$F_q^N(\alpha_1,\alpha_2,\alpha_3)\propto \prod\limits_i 
\delta(\alpha_i-1/3)$ which allows to continue our 
calculations analytically.
The details of the integration of (\ref{3.38}) can be found in Appendix~C.
$G_{3P}$ is a function of the parameter $b(0)$. As a trial 
value we choose Eq.~(\ref{3.24}), 
$b(0)=570\,MeV$, (estimated using the result of 
the additive quark model $G_{3P}(qN\to XN)\approx 0.5\,mb/GeV^2$)
we arrive at $G_{3P}(NN\to XN)\approx 2.4\,mb/GeV^2$. This value
is substantially higher than the experimental value
$G_{3P}(NN\to XN)=1.5\, mb/GeV^2$. This is an obvious manifestation 
of simplifying approximations (the quark additivity) we have done.
In order to fit the experimental value of $G_{3P}(NN\to NX)$ 
after the contributions of the second and third graphs in Fig.~\ref{fig4}
are included we should use in (\ref{3.38})
\beq
b(0)= 650\, MeV\ .
\label{3.40}
\eeq
With this value (\ref{3.23}) gives,
\beq
G_{3P}(qN\to XN)\approx
0.3\,\frac{mb}{GeV^2}\approx {1\over5}\,G_{3P}(NN\to XN)\ ,
\label{3.41}
\eeq
which shows a substantial deviation from the AQM.

\subsubsection{Diffractive gluon radiation by 
a (virtual) photon and mesons.
Breakdown of Regge factorization}

\begin{figure}[tbh]
\includegraphics{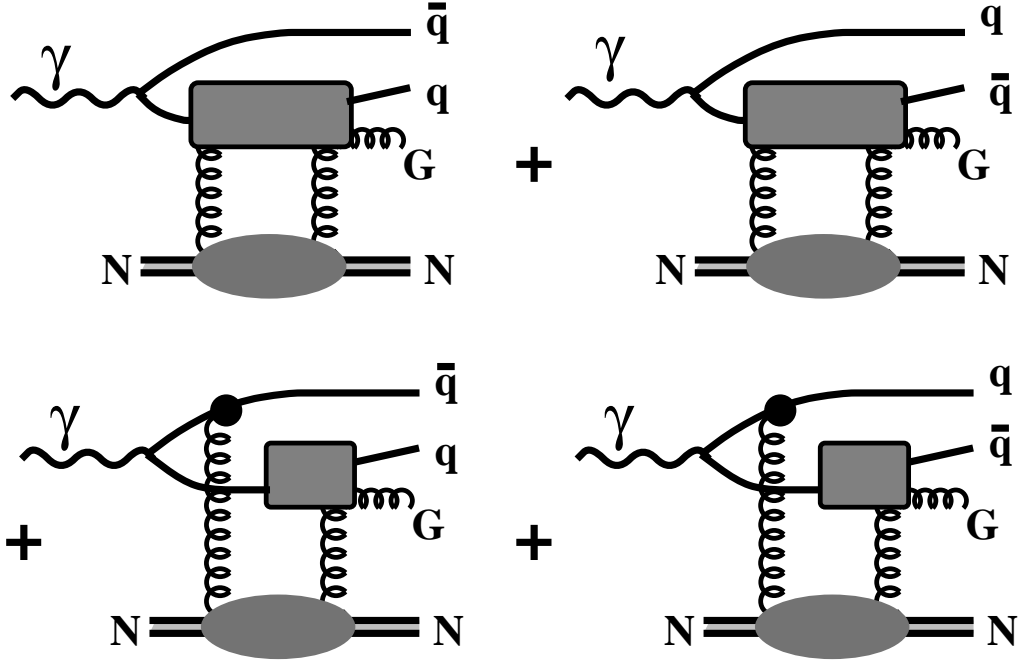}
\begin{center}
\vspace{9cm}
\parbox{13cm}
{\caption[Delta]
{\sl Diagrams for the diffractive radiation of a gluon
in photon-nucleon interaction, $\gamma^* N\to\bar qqGN$.}
\label{fig3}}
\end{center}
\end{figure}

One can use a similar technique to calculate the cross section 
for diffractive gluon radiation by a photon and mesons.
The diffraction amplitude 
$\gamma(M)\,N\to \bar qqG\,N$ is described by
the four diagrams depicted in Fig.~\ref{fig3}.
The first two diagrams correspond to the AQM.
In this approximation the forward ($q_T=0$) 
amplitude $\gamma\,N\to \bar qqG\,N$ reads,
\beqn
&&F^{AQM}(\gamma \,N\to \bar qqG\,N)=
\Psi_{\bar qq}(\alpha,\vec\rho_1-\vec\rho_2)\,
\biggl[F(q\,N\to qG\,N)- F(\bar q\,N\to \bar qG\,N)
\biggr]\nonumber\\
&=&
\Psi_{\bar qq}(\alpha,\vec\rho_1-\vec\rho_2)\,
\left[\Psi_{qG}\left(\frac{\alpha_G}{\alpha},\vec\rho_1\right)\,
\widetilde\sigma(\rho_1) -
\Psi_{\bar qG}\left(\frac{\alpha_G}{1-\alpha},\vec\rho_2\right)\,
\widetilde\sigma(\rho_2)\right]\ ,
\label{3.25}
\eeqn
where $\vec\rho_i=\vec\rho-\vec r_i$ (i=1,2).
$\vec\rho,\ \vec r_{1,2}$ are the radius-vectors of the gluon,
quark and antiquark respectively. The limit $\alpha_G\to 0$ is assumed.

After addition of the last two graphs in Fig.~\ref{fig3}
the amplitude takes the form
(we do not write out its trivial color structure),
\beqn
F(\gamma \,N\to \bar qqG\,N)&=&
\Psi_{\bar qq}(\alpha,\vec\rho_1-\vec\rho_2)\,
\left[\Psi_{qG}\left(\frac{\alpha_G}{\alpha},\vec\rho_1\right) -
\Psi_{\bar qG}\left(\frac{\alpha_G}{1-\alpha},\vec\rho_2\right)\,
\right]\nonumber\\
&\times& \biggl[\widetilde\sigma(\rho_1)+
\widetilde\sigma(\rho_2)-
\widetilde\sigma(\vec\rho_1-\vec\rho_2)\biggr]\ .
\label{3.26}
\eeqn
The detailed calculation of the diagrams depicted in Fig.~\ref{fig3}
is presented in Appendix~B.2. A much simpler and more
intuitive derivation of Eq.~(\ref{3.26}) is suggested in
Appendix~A.3.

If one neglects the nonperturbative effects in Eq.~(\ref{3.26})
($b(\alpha)=0$) this expression coincides with Eq.~(3.4) 
in \cite{bartels}, but is quite different from the
cross section of diffractive gluon radiation
derived in \cite{nz1} (Eq.~(60)) (see footnote$^1$). 
A crucial step in
\cite{nz1} is the transition from 
Fock states which are the eigenstates of interaction, to
the physical state basis. Such a rotation of the $S$-matrix 
leads to a renormalization of the probability
{\it amplitudes}  for the Fock states
(see Appendix~A.1), rather than just
the probabilities as it was assumed in \cite{nz1}. 

The amplitude (\ref{3.26}) is normalized as,
\beq
\frac{d\sigma}{d(ln\alpha_G)\,dq_T^2}\biggr|_{q_T=0}=
\frac{1}{16\pi}\int d^2\rho_1\,d^2\rho_2\,d\alpha\,
\biggl|F(\gamma N\to\bar qqGN)\biggr|^2
\label{3.27}
\eeq
Direct comparison of of the cross section for diffractive 
gluon radiation by a photon calculated with this expression
with data is complicated by contribution of diffraction to
$\bar qq$ states and nondiffractive (Reggeon) mechanisms.
This is why one should first perform a detailed triple-Regge
analysis of data and then to compare (\ref{3.27}) with
the effective triple-Pomeron coupling. Good data for photon 
diffraction are available at lab. energy $E_{\gamma}=100\,GeV$
\cite{100gev}. At this energy, however, there is no true triple-Regge
region which demands $s/M^2\gg 1$ and $M^2\gg 1\,GeV^2$.
Therefore the results of the triple-Regge analysis in \cite{100gev}
cannot be trusted. It is much more appropriate to use available data from
HERA, particularly those in \cite{h1-3p} at $\sqrt{s}=200\,GeV$
where a triple-Regge analysis taking into account four graphs
was performed. The result for the effective triple-Pomeron coupling
\beq
G^{\gamma p}_{3P}(0)=
(8.19\pm 1.6\pm1.34\pm2.22)\,\mu b/GeV^2
\label{3.41a}
\eeq
should be compared with our prediction 
$G^{\gamma p}_{3P}(0)=9\,\mu b/GeV^2$. To estimate the
mean energy for the dipole cross section $\bar s/M^2\ GeV^2$ 
we used the mid value $M^2=250\,GeV^2$ of the interval of $M^2$
measured in \cite{h1-3p} which corresponds to $x_P=0.0064$.
Thus, high-energy data for gluon radiation in diffractive dissociation
of protons and photons give the value (\ref{3.40}) for the parameter of
nonperturbative quark-gluon interaction.

Note that the relative role of ``additive'' (\# 1,2 in Fig.~\ref{fig3}
and \# 1 in Fig.~\ref{fig4}) and ``non-additive'' (\# 3,4 in Fig.~\ref{fig3}
and \# 2,3 in Fig.~\ref{fig4}) graphs depends on the relation
between the three characteristic sizes $R_h=\sqrt{\la r^2_{ij}\ra}$,
$\rho_0$ and $1/b(0)$. In the limit $R_h\gg\rho_0,\ 1/b(0)$ the contribution 
of the ``non-additive'' graphs vanishes and the additive quark model
becomes a good approximation. However, at realistic values of
$R_h\sim 1\,fm$ the ``additive'' and ``non-additive'' contributions
are of the same order and the latter becomes dominant for small $R_h$.
Particularly, this explains why the factorization relation,
\beq
A_{3P}(h\,N\to X\,N)=
\frac{G_{3P}(h\,N\to X\,N)}
{\sigma_{tot}(hN)} = Const\ ,
\label{3.42}
\eeq
{\it i.e.} independent of $h$,
is substantially broken. We expect,
\beqn
A_{3P}(N\,N\to X\,N)&=&0.025\,GeV^{-2}\ ,\nonumber\\
A_{3P}(\pi\,N\to X\,N)&=&0.031\,GeV^{-2}\ ,\nonumber\\
A_{3P}(K\,N\to X\,N)&=&0.042\,GeV^{-2}\ ,\nonumber\\
A_{3P}(\gamma\,N\to X\,N)&=&0.052\,GeV^{-2}\ .
\label{3.43}
\eeqn
We see that our predictions for the triple-Pomeron vertex
as defined from diffractive dissociation
of nucleons and photons are different by almost factor of three.
On top of that, the absorptive corrections which
are known to be larger for diffraction than for elastic scattering 
also contribute to the breaking of Regge factorization.
A manifestation of these correction shows up as deviation 
between the data and the Regge based expectations 
for the energy dependence 
of the diffractive cross section \cite{dino,schlein}.

\subsection{Gluon shadowing in nuclei} 

It is known since long time \cite{kancheli} that the parton
distribution in nuclei is shadowed at small $x$ due to
parton fusion. In QCD this effect corresponds to the nonlinear term
in the evolution equation responsible for gluon recombination
\cite{glr,mq}. 
This phenomenon is very important as soon as one 
calculates the cross section of a hard reaction
(gluon radiation with high $k_T$, prompt photons,
Drell-Yan reaction, heavy flavor production, etc.)
assuming factorization.
Nuclear shadowing of sea quarks is well measured in 
DIS, but for gluons it is poorly known. One desperately
needs to know it to provide predictions for the high-energy
nuclear colliders, RHIC and LHC.

The interpretation of nuclear shadowing 
depends on the choice of the reference frame. In the infinite
momentum frame of the nucleus it looks like parton fusion.
Indeed, the longitudinal spread of the valence quarks in the bound 
nucleons, as well as the internucleon distances, are subject to
Lorentz contraction. Therefore the nucleons are spatially 
well separated. However, the longitudinal spread of partons
at small $x$ contracts much less because they have an $x$ times 
smaller Lorentz factor. Therefore, such partons can overlap
and fuse even if they originate from different nucleons \cite{kancheli}.
Fusion of two gluons into a $\bar qq$ pair leads to
shadowing of sea quarks. If two gluons fuse to a single
gluon it results in shadowing of gluons.

The same phenomenon looks quite differently in the rest frame 
of the nucleus, as shadowing of long-living hadronic
fluctuations of the virtual photon. This resembles the ordinary 
nuclear shadowing for the total cross sections of hadron-nucleus 
interaction. Indeed, the total virtual photoabsorption 
cross section is proportional to the structure function 
$F_2(x,Q^2)$. However, one can calculate in this way only
shadowing of quarks. To predict shadowing of gluons it was suggested
in \cite{frankfurt} to replace the photon by a hypothetical
particle probing gluons. Assuming for the $GG$ fluctuation
of this particle the same distribution function as for $\bar qq$
one may conclude that the
effective absorption cross section providing shadowing is
$9/4$ times larger than for a $\bar qq$ fluctuation of a photon.
Such a simple result cannot be true because of the strong
gluon-gluon interaction which makes their distribution
function quite different (``squeezes'' it). Besides,
the spin structure of the $GG$ distribution function is
different too.

\subsubsection{Nuclear shadowing for longitudinal photons}

Longitudinally polarized photons are known to be a good probe
for the gluon structure function. Indeed, the aligned jet model
\cite{bks} cannot be applied in this case since the
distribution function for longitudinal photons (\ref{1.1}), 
(\ref{1.3}) suppresses the asymmetric $\bar qq$ fluctuations with 
$\alpha \to 0,\ 1$. Therefore, the transverse separation of the 
$\bar qq$ pair is small $\sim 1/Q^2$ and nuclear shadowing
can be only due to shadowing of gluons.
One can also see that from the expression for the cross section
of a small size dipole \cite{fs,barone},
\beq
\sigma_{\bar qq}^{A,N}(r_T,x) \approx
\frac{\pi^2}{3}\,\alpha_s(Q^2)\,
G_N(x,Q^2)\ ,
\label{3.1.1}
\eeq
where $G_N(x,Q^2)=x\,g(x,Q^2)$ is the gluon density and
$Q^2\sim 1/r_T^2$. Thus, we expect nearly the same nuclear shadowing
at large $Q^2$ for the longitudinal photoabsorption cross section and
for the gluon distribution,
\beq
\frac{\sigma^L_A(x,Q^2)}{\sigma^L_N(x,Q^2)}\approx
\frac{G_A(x,Q^2)}{G_N(x,Q^2)}
\label{3.1.2}
\eeq
The estimate for nuclear shadowing for longitudinally polarized 
photons follows.

Nuclear shadowing for photons corresponds to the inelastic
nuclear shadowing as it was introduced for hadrons by
Gribov 30 years ago \cite{gribov}. Therefore, the term 
$\Delta\sigma(\gamma^*A)=\sigma_{tot}(\gamma^*A)-A\,\sigma(\gamma^*N)$
representing
shadowing in the total photoabsorption cross section is proportional 
to the diffractive dissociation cross section $\gamma^*\,N \to X\,N$
\cite{gribov,kk}, considered above.
In the lowest order in the intensity of $XN$ interaction the shadowing 
correction reads,
\beqn
&& \Delta\sigma(\gamma^*A)=
8\pi\,{\rm Re}\int d^2b
\int dM_X^2\,\left.\frac{d^2\sigma(\gamma^*N\to XN)}
{dM_X^2\,dq_T^2}\right|_{q_T=0}\nonumber\\
&\times& \int\limits_{-\infty}^{\infty} dz_1
\int\limits_{-\infty}^{\infty} dz_2\,
\Theta(z_2-z_1)\,\rho_A(b,z_1)\,\rho_A(b,z_2)\,
{\rm exp}\Bigl[-i\,q_L\,(z_2-z_1)\Bigr]\ ,
\label{3.1.3}
\eeqn
where
\beq
q_L=\frac{Q^2+M_X^2}{2\,\nu}\ .
\label{3.1.4}
\eeq
Here $\nu$ is the photon energy; $z_1$ and $z_2$ are the
longitudinal coordinates of the nucleons $N_1$ and $N_2$,
respectively, participating in the diffractive transition
$\gamma^*\,N_1 \to X\,N_1$ and back $X\,N_2\to\gamma^*\,N_2$.

The longitudinal momentum transfer (\ref{3.1.4}) controls
the lifetime (coherence time $t_c$)
of the hadronic fluctuation of the photon, $t_c=1/q_L$.
It is known only if the mass matrix is diagonal, {\it i.e.}
the fluctuations have definite masses. However, in this case
the interaction cross section of the fluctuation has no
definite value. Then one faces a problem of calculation of
nuclear attenuation for the intermediate state $X$ via interaction 
with the nuclear medium. 

This problem can be settled using the Green function formalism
developed above in Section 2.1 \cite{krt,z}.
One should switch to the quark-gluon representation for the 
produced state 
$X=|\bar qq\ra,\ |\bar qqG\ra,\ , |\bar qq2G\ra,\ ...$. 
As one can see below an exact solution is not an easy problem even for 
the two lowest Fock states. For higher states containing two or
more gluons it may be solved in the double--leading--log
approximation which neglects the size of the previous Fock state
and treats a multi-gluon fluctuation as a color  octet--octet dipole.
This is actually what we do in what follows, except the Fock state with
only one gluon leads to a $1/M^2$ mass distribution for diffraction,
while inclusion of multi--gluon components makes it slightly steeper.
This is not a big effect, besides, the nuclear formfactor substantially
cuts off the reachable mass interval (see below). Therefore, we restrict 
the following consideration by the first two Fock states.

For the lowest
state $|\bar qq\ra$ one can write,
\beqn
&& 8\pi\,{\rm Re}\int dM_X^2\, 
\left.\frac{d^2\sigma(\gamma^*N\to XN)}
{dM_X^2\,dq_T^2}\right|_{q_T=0}\,
{\rm exp}\Bigl[-i\,q_L\,(z_2-z_1)\Bigr]\nonumber\\
&=& {1\over2}\,{\rm Re}\int d^2k_T\int\limits_{0}^{1}
d\alpha\,\Bigl|F_{\gamma^*\to\bar qq}(\vec k_T,\alpha)
\Bigr|^2\,{\rm exp}\left[-\,i\,
\frac{\epsilon^2+k_T^2}{2\,\nu\,\alpha(1-\alpha)}\,
(z_2-z_1)\right]\nonumber\\
&\equiv& {1\over2}\,{\rm Re}\int d^2r_1\,d^2r_2
\int\limits_{0}^{1}d\alpha\ 
F^{\dagger}_{\gamma^*\to\bar qq}(\vec r_2,\alpha)\,
G^0_{\bar qq}(\vec r_2,z_2;\vec r_1,z_1)\,
F_{\gamma^*\to\bar qq}(\vec r_1,\alpha)\ ,
\label{3.1.5}
\eeqn
where $\epsilon$ was defined in (\ref{1.2}).

The amplitudes of diffraction $\gamma^*\,N\to X\,N$
in the transverse momentum and coordinate representations
are related by Fourier transform,
\beq
F_{\gamma^*\to\bar qq}(\vec k_T,\alpha)
=\frac{1}{2\pi}\int d^2r\,F_{\gamma^*\to\bar qq}(\vec r_1,\alpha)\,
e^{i\,\vec k_T\cdot\vec r}\ .
\label{3.1.6}
\eeq
This amplitude in the coordinate representation has a factorized form,
\beq
F_{\gamma^*\to\bar qq}(\vec r_1,\alpha)=
\Psi_{\bar qq}(\vec r,\alpha)\,\sigma_{\bar qq}(r)\ .
\label{3.1.7}
\eeq

$G^0_{\bar qq}(\vec r_2,z_2;\vec r_1,z_1)$ in (\ref{3.1.5})
is the Green function of a free propagation of the
$\bar qq$ pair between points $z_1$ and $z_2$.
It is a solution of Eq.~(\ref{2.1}) without
interaction.
\beq
G^0_{\bar qq}(\vec r_2,z_2;\vec r_1,z_1)=
\frac{1}{(2\pi)^2}\int d^2k_T\,
{\rm exp}\left[- i\,\vec k_T\cdot(\vec r_2-\vec r_1) -
\frac{i\,k_T^2\,(z_2-z_1)}{2\,\nu\,
\alpha(1-\alpha)}\right]\ ;
\label{3.1.8}
\eeq
The boundary condition for the Green function is,
\beq
G^0_{\bar qq}(\vec r_2,z_2;\vec r_1,z_1)
\Bigr|_{z_2=z_1}= \delta(\vec r_2-\vec r_1)\ .
\label{3.1.9}
\eeq

In Eq.~(\ref{3.1.8} the phase shift on the distance 
$z_2-z_1$ is controlled by the transverse momentum
squared as one could expect from Eqs.~(\ref{3.1.3})-(\ref{3.1.4})
where it depends on the $M_X^2$. However, Eq.~(\ref{3.1.5}
is written now in the coordinate representation and contains
no uncertainty with the absorption cross section, as different from
(\ref{3.1.3}). In order to include the effects of absorption
of the intermediate state $X$ into (\ref{3.1.5}) one should replace
the free Green function $G^0_{\bar qq}(\vec r_2,z_2;\vec r_1,z_1)$
by the solution of the Schr\"odinger equation (\ref{2.1})
with imaginary potential (\ref{2.3}). 
This was done in paper \cite{krt} and the results have 
demonstrated a substantial deviation of nuclear shadowing
from usually used approximations for transverse photons.
One should also include
the real part of the potential which 
takes into account the nonperturbative interaction between
$q$ and $\bar q$ as it is described in Section 2.1. 
This is important only for nuclear shadowing of 
transverse photons and and low $Q^2$ longitudinal photons
and is beyond the scopes of present paper.
Therefore, we skip further discussion of nuclear shadowing for
the $|\bar qq\ra$ pair and switch to the next
Fock component $|\bar qqG\ra$.

For the intermediate state (\ref{3.2}) $X=\bar qqG$ 
Eq.~(\ref{3.1.5}) is modified as,
\beqn
&& 8\pi\,\int dM_X^2\,
\left.\frac{d^2\sigma(\gamma^*N\to XN)}
{dM_X^2\,dq_T^2}\right|_{q_T=0}\,
{\rm cos}[q_L\,(z_2-z_1)]\nonumber\\
&\Rightarrow&\ 
{1\over2}\,{\rm Re}\int d^2x_2\,d^2y_2\,d^2x_1\,d^2y_1
\int d\alpha_q\,d{\rm ln}(\alpha_G)\nonumber\\ 
&\times& F^{\dagger}_{\gamma^*\to\bar qqG}
(\vec x_2,\vec y_2,\alpha_q,\alpha_G)\ 
G_{\bar qqG}(\vec x_2,\vec y_2,z_2;\vec x_1,\vec y_1,z_1)\ 
F_{\gamma^*\to\bar qqG}(\vec x_1,\vec y_1,\alpha_q,\alpha_G)\ ,
\label{3.1.10}
\eeqn
where $\alpha_q$ and $\alpha_G$ are the fractions of the photon 
light cone momentum carried by the quark and gluon respectively.
The amplitude of diffraction $\gamma^*\,N\to X\,N$ depends
on the $q$-$\bar q$ transverse separation $\vec x$ and
the distance $\vec y$ from the gluon to the center of
gravity of the $\bar qq$ pair (we switch to these variables from
the previously used $\vec\rho_{1,2}$ for the sake of convenience,
it simplifies the expression for kinetic energy).

The Schr\"odinger equation for the Green function
$G_{\bar qqG}$ describing propagation of the $\bar qqG$ system through
a medium including interaction with the environment as well as
between the constituent has the form,
\beqn
&& i\,\frac{d}{dz_2}\;
G_{\bar qqG}(\vec x_2,\vec y_2,z_2;\vec x_1,\vec y_1,z_1)
\nonumber\\
&=& \left\{\frac{Q^2}{2\nu} - 
\frac{\alpha_q+\alpha_{\bar q}}
{2\,\nu\,\alpha_q\,\alpha_{\bar q}}\,
\Delta(\vec x_2) -
\frac{1}{2\,\nu\,\alpha_G(1-\alpha_G)}\,
\Delta(\vec y_2) +
V(\vec x_2,\vec y_2,z_2,\alpha_q,\alpha_G)\right\}
\nonumber\\ &\times&
G_{\bar qqG}(\vec x_2,\vec y_2,z_2;\vec x_1,\vec y_1,z_1)\ ,
\label{3.1.11}
\eeqn
with the boundary condition,
\beq
G_{\bar qqG}(\vec x_2,\vec y_2,z_2;\vec x_1,\vec y_1,z_1)
\Bigr|_{z_2=z_1}=
\delta(\vec x_2-\vec x_1)\,
\delta(\vec y_2-\vec y_1)\ .
\label{3.1.12}
\eeq

The imaginary part of the potential 
$V(\vec x_2,\vec y_2,z_2,\alpha_q,\alpha_G)$
in (\ref{3.1.11}) is proportional to the 
interaction cross section for the $\bar qqG$ system with
a nucleon,
\beqn
&& 2\,{\rm Im}V(\vec x_2,\vec y_2,z_2,\alpha_q,\alpha_G)
\nonumber\\ &=&
\left\{{1\over8}\,\sigma_{\bar qq}(\vec x) -
{9\over8}\,\left[\sigma_{\bar qq}
\left(\vec y-\frac{\alpha_{\bar q}}{1-\alpha_G}\,
\vec x\right) + \sigma_{\bar qq}
\left(\vec y-\frac{\alpha_{\bar q}}{1-\alpha_G}\,
\vec x\right)\right]\right\}\,\rho_A(b,z)\ .
\label{3.1.13}
\eeqn
The real part of this potential responsible for the nonperturbative
interaction between the quarks and gluon is discussed below.

If the potential ${\rm Im}V(\vec x_2,\vec y_2,z_2,\alpha_q,\alpha_G)$
is a bi-linear function of $\vec x$ and $\vec y$ then Eq.~(\ref{3.1.11}
can be solved analytically. Nevertheless, the general case of 
nuclear shadowing for a three-parton system is quite 
complicated and we should simplify the problem.

Let us consider nuclear shadowing for longitudinally
polarized photons with high $Q^2$. The latter means
that one can neglect the eikonal attenuation for the
$\bar qq$ Fock component of the longitudinal
photon, {\it i.e.} 
\beq
Q^2\gg 4\,C\,\la T_A\ra \approx 1\,GeV^2\ ,
\label{1.1.14}
\eeq
where $C$ is the factor in Eq.~(\ref{2.3a}) and
$\la T_A\ra$ is the mean nuclear thickness function.

As different from the case of transversely polarized photons
which distribution function (\ref{1.1}) - (\ref{1.2})
contains $\bar qq$ pairs with large separation
($\alpha \to 0,\ 1$) even at large $Q^2$, in longitudinally
polarized photons small size ($\sim 1/Q$) $\bar qq$ pairs always 
dominate \cite{ks,bks}. This property suggest a few
simplifications for the following calculations.
\begin{enumerate}

\item
One can neglect at large $Q^2$ the nonperturbative
$\bar qq$ interaction and use the perturbative photon
wave function (\ref{1.1}) - (\ref{1.3}).

\item
One can simplify the expression for the diffractive 
amplitude $\gamma^*\,N \to \bar q\,q\,G\,N$ introduced in
(\ref{3.1.10}) relying on smallness of the typical
$\bar qq$ separation $|\vec x|\sim 1/Q$ in comparison with
the distance between the $\bar qq$ and the gluon
$|\vec y \sim 1/b_0\approx 0.3\,fm$.

\item
One can also simplify the equation (\ref{3.1.11}) for the
Green function $G_{\bar qqG}$ fixing $\vec x=0$ in 
the expression (\ref{3.1.12}) for the nonperturbative potential
${\rm Im}V(\vec x_2,\vec y_2,z_2,\alpha_q,\alpha_G)$.
This leads in (\ref{3.1.11}) to a factorized dependence on 
variables $\vec x$ and $\vec y$.
\end{enumerate}

As a result of these approximations and $\alpha_G\to 0$
we arrive at,
\beq
F_{\gamma^*\to\bar qqG}(\vec x,\vec y,
\alpha_q,\alpha_G) =
-\Psi^L_{\bar qq}(\vec x,\alpha_q)\,
\vec x\cdot\vec\nabla\,
\Psi_{qG}(\vec y)\,
\sigma_{\bar GG}(\vec y)\ ,
\label{3.1.15}
\eeq
where
\beq
\Psi_{qG}(\vec y)=\lim\limits_{\alpha_G\to 0} 
\Psi_{qG}(\alpha_G,\vec y)\ ,
\label{3.1.16}
\eeq
and 
\beq
\sigma_{GG}(r,s)={9\over4}\,\sigma_{\bar qq}(r,s)\ .
\label{3.1.17}
\eeq
As soon as we neglect the size of the color-octet $\bar qq$ pair,
it interacts a gluon, this is why one can replace
$\sigma_{\bar qqG}$ by the dipole cross section $\sigma_{GG}$.
The latter is larger than $\sigma_{\bar qq}$ by the Casimir 
factor $9/4$.

In this case the tree-body Green function factorizes to a product of 
two-body ones,
\beq
G_{\bar qqG}(\vec x_2,\vec y_2,z_2;\vec x_1,\vec y_1,z_1)
\Rightarrow 
G_{\bar qq}(\vec x_2,z_2;\vec x_1,z_1)\;
G_{GG}(\vec y_2,z_2;\vec y_1,z_1)\ ,
\label{3.1.18}
\eeq
where $G_{\bar qq}(\vec x_2,z_2;\vec x_1,z_1)$
is the ``free'' Green function of the $\bar qq$ pair,
and $G_{GG}(\vec y_2,z_2;\vec y_1,z_1)$ describes propagation
of the $GG$ dipole which constituents interact with each other, 
as well as with the nuclear medium.
\beq
i\,\frac{d}{d\,z_2}\;
G_{GG}(\vec y_2,z_2;\vec y_1,z_1) = 
\left[ - \frac{\Delta(\vec y_2)}
{2\,\nu\,\alpha_G(1-\alpha_G)} +
V(\vec y_2,z_2)\right]\,
G_{GG}(\vec y_2,z_2;\vec y_1,z_1)\ ,
\label{3.1.19}
\eeq
where
\beq
2\,{\rm Im}\,V(\vec y,z) =
-\sigma_{GG}(\vec y)\,\rho_A(b,z)\ .
\label{3.1.20}
\eeq

On analogy to (\ref{2.5}) we assume the real part of the 
potential has a form
\beq
{\rm Re}\,V(\vec y,z) =
\frac{\tilde b^4\,y^2}
{2\,\nu\,\alpha_G(1-\alpha_G)}\ ,
\label{3.1.21}
\eeq
where $\tilde b \sim b_0$.

To simplify the estimate we 
assume that $\sigma_{GG}(r,s)\approx C_{GG}(s)\,r^2$,
where $C_{GG}(s) =\\ d\,\sigma_{GG}(r,s)/d\,r^2_{r=0}$.

The solution of Eq.~(\ref{3.1.19}) has a form,
\beqn
&& G_{GG}(\vec y_2,z_2;\vec y_1,z_1) =
\frac{A}{2\pi\,{\rm \sinh}(\Omega\,\Delta z)}
\nonumber\\ &\times& 
{\rm exp}\left\{-\frac{A}{2}\,
\left[(y_1^2+y_2^2)\,{\rm coth}(\Omega\,\Delta z) -
\frac{2\vec y_1\cdot\vec y_2}{{\rm sinh}(\Omega\,\Delta z)}
\right]\right\}\ ,
\label{3.1.22}
\eeqn
where
\beqn
A&=&\sqrt{{\tilde b}^4-i\,\alpha_G(1-\alpha_G)\,\nu\,C_{GG}\,
\rho_A}\nonumber\\
\Omega &=& \frac{i\,A}
{\alpha_G(1-\alpha_G)\,\nu}\nonumber\\
\Delta z&=&z_2-z_1\ .
\label{3.1.23}
\eeqn

The quark-gluon wave function in (\ref{3.1.15}) has a form similar to
(\ref{2.21}),
\beq
\Psi_{qG}(\vec y)=
\frac{2}{\pi}\,\sqrt{\frac{\alpha_s}{3}}\,
\frac{\vec e\cdot\vec y}{y^2}\,
{\rm exp}\left[-{b^2\over2}\,y^2\right]\ .
\label{3.1.24}
\eeq

Now we have all the components of the amplitude (\ref{3.1.15})
which we need to calculate the nuclear shadowing correction
(\ref{3.1.10}). Integration in $\vec x_{1,2}$ and $\vec y_{1,2}$
can be performed analytically. 
\beqn
&& 8\pi\,\int dM_X^2\,
\left.\frac{d^2\sigma(\gamma^*N\to XN)}
{dM_X^2\,dq_T^2}\right|_{q_T=0}\,
{\rm cos}(q_L\,\Delta z)\nonumber\\
&=& 
{\rm Re}\int d\alpha_q\,d{\rm ln}(\alpha_G)\,
\frac{16\,\alpha_{em}\,(\sum\limits_{F}Z_q^2)\,
\alpha_s(Q^2)\,C_{GG}^2}
{3\,\pi^2\,Q^2\,\tilde b^2}
\left[(1-2\,\zeta-\zeta^2)\,e^{-\zeta} +
\zeta^2\,(3+\zeta)\,{\rm E_1}(\zeta)\right]
\nonumber\\
&\times&\, 
\left[\frac{t}{w}+\frac{{\rm sinh}(\Omega\,\Delta z)}
{t}\,{\rm ln}\left(1-\frac{t^2}{u^2}\right)
+ \frac{2\,t^3}{u\,w^2} + 
\frac{t\,{\rm sinh}(\Omega\,\Delta z)}
{w^2} + \frac{4\,t^3}{w^3}\right]\ ,
\label{3.1.24a}
\eeqn
where
\beqn
\zeta &=& i\,x\,m_N\,\Delta z\ ;\nonumber\\
t &=& \frac{A}{\tilde b^2}\ ;\nonumber\\
u &=& t\,{\rm cosh}(\Omega\,\Delta z) +
{\rm sinh}(\Omega\,\Delta z)\ ;\nonumber\\
w &=& (1+t^2)\,{\rm sinh}(\Omega\,\Delta z) +
2\,t\,{\rm cosh}(\Omega\,\Delta z)\ .
\label{3.1.25}
\eeqn

\begin{figure}[tbh]
\includegraphics{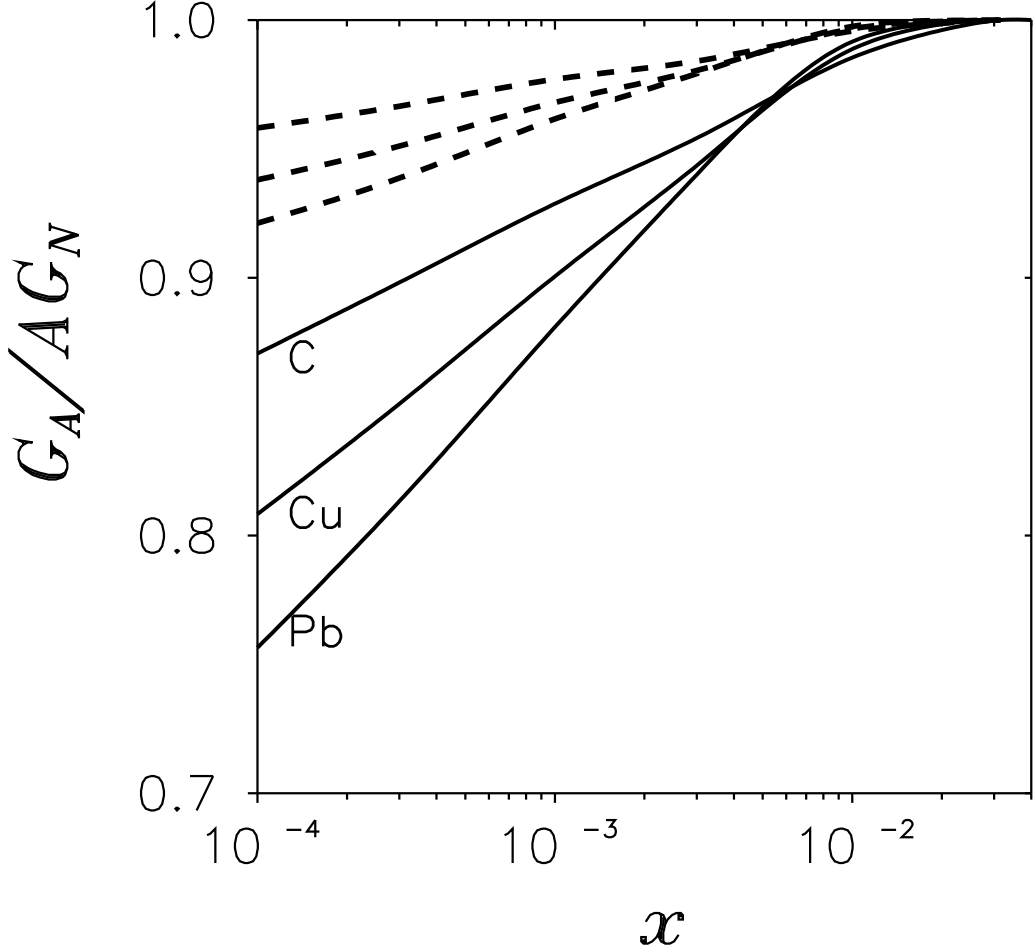}
\begin{center}
\vspace{13cm}
\parbox{13cm}
{\caption[Delta]
{\sl Ratio of the gluon distribution functions
in nuclei (carbon, copper and lead)
and nucleons at small Bjorken $x$
and $Q^2 = 4\,GeV^2$ (solid curves) and $40\,GeV^2$
(dashed curves).}
\label{glue}}
\end{center}
\end{figure}

The rest integration in (\ref{3.1.24a}) can be performed 
numerically. We calculated the ratio $R^G_{A/N}=
G_A(x,Q^2)/a\,G_N(x,Q^2)$
for the gluon distribution functions for small values
of Bjorken $10^{-4} < x < 10^{-1}$ and high
$Q^2 = 10\,GeV^2$. We found $R^G_{A/N}$ almost independent of $Q^2$
at higher $Q^2$. The results are depicted in Fig.~\ref{glue}.

One can see that in contrast to the quark distribution 
the onset of nuclear shadowing for gluons starts at quite
small $x\sim 10^{-2}$. This is because the photon fluctuations
containing gluons are heavier than $\bar qq$ fluctuations.
Correspondingly, the lifetime of such fluctuations is
shorter (or $q_L$ is smaller) and they need a smaller $x$
to expose coherent effects like nuclear shadowing. 
One can expect an antishadowing effect at medium $x\sim 0.1$
like in $F_2(x,Q^2)$ which should push the crossing point
$G_A(x,Q^2)/G_N(x,Q^2)=1$ down to smaller $x$. Discussion of the 
dynamics of antishadowing (swelling of bound nucleons, etc.)
goes beyond the scopes of this paper. 

A similar approach to the problem of gluon shadowing is developed
in \cite{fs1} which relates shadowing to the diffractive
radiation of gluons. 
Note that a delayed onset of gluon shadowing (at $x < 0.02$)
is also expected in \cite{fs1}. However, this is a result of an
{\it ad hoc} parameterization for antishadowing, rather than calculations.
The phase shift factor ${\rm cos}(q_L\Delta z)$ which controls 
the onset of shadowing in (\ref{3.1.10}), (\ref{3.1.24a}) 
is neglected in \cite{fs1} assuming that $x$ is sufficiently small.
However, nuclear shadowing for gluons does not saturate even at 
very small $x$ because of the $1/M^2$ form of the mass dependence
of diffractive radiation of gluons (triple-Pomeron diffraction).
The smaller the $x=Q^2/2m_N\nu$ is, 
the higher masses are allowed by the nuclear 
form factor ($q_L=(Q^2+M^2)/2\nu < 1/R_A$) 
to contribute to the shadowing.

Our results also show that $R^G_{A/N}$  steeply
decreases down to small $x$ and seems to have a tendency to
become negative. It would not be surprising for heavy nuclei
if our shadowing correction corresponded to double scattering
term only. However, the expression (\ref{3.1.10}) includes
all the higher order rescattering terms. The source of the trouble 
is the obvious breaking down of the unitarity limit 
$\sigma_{diff} < \sigma_{tot}$. This problem is well known and easily
fixed by introducing the unitarity or absorptive corrections
which substantially slow down the growth of the diffractive cross section.
Available data for diffraction $pp\to pX$ clearly demonstrate
the effect of unitarity corrections \cite{dino,schlein}. One may expect
that at very high energies the relative fraction of diffraction
decreases. We restrict ourselves with this word of caution
in the present paper and postpone a further study of unitarity 
effects for a separate publication, as well as the effects
of higher Fock components containing more than one gluon.
Those corrections also become more important at small $x$.

Note that quite a strong nuclear suppression for gluons 
$G^A(x,Q^2)/G^N(x,Q^2) <\\ F^A_2(x,Q^2)/F^N_2(x,Q^2)$
was predicted in \cite{frankfurt} basing 
on the fact that the cross section
of a color octet-octet dipole contains the factor $9/4$ compared to
$\sigma_{\bar qq}$. However, it is argued above in section~3.1 
and confirmed by the following calculations that the observed smallness
of the diffractive cross section of gluon radiation
shows that that the strong nonperturbative interaction of gluons
substantially reduces the size of fluctuations including the gluon.
The situation is much more complicated and cannot be reduced
to a simple factor $9/4$.

A perspective method for calculation of
nuclear shadowing for gluons was suggested in
the recent publication \cite{fs2}. Experimental data for diffractive
charm production can be used to estimate 
the effect. This seems to be more reliable 
than pure theoretical calculations performed above.
Indeed, the transverse separation of a heavy flavored $\bar QQ$
pair is small even at low $Q^2$, and may be assumed to be much smaller
that the mean distance between the $\bar QQ$ and the gluon. 
Unfortunately, the available data
obtained at HERA have quite poor accuracy. The results
from H1 \cite{h1} and ZEUS \cite{zeus} experiments are different
by almost factor of two. Besides, the theoretical analysis
\cite{penn1,penn2} 
which is needed to reconstruct the diffractive cross section
of charm production from production of $D^*$ in a limited
phase space, introduces substantial uncertainty.
According to \cite{penn2} the realistic solutions for
the diffractive charm production differ by a factor of five.
In this circumstances we suppose our calculation for nuclear 
shadowing of gluons seems to be more reliable. 

Note that we expect much weaker nuclear shadowing for gluons 
than it was predicted in \cite{eqw,ku,fs1}. For instance at $x=10^{-3}$
and $Q^2=4\,GeV^2$ 
we expect $G_A/A\,G_N \approx 0.9$, while a much stronger suppression 
$G_A/A\,G_N \approx 0.6$ \cite{eqw}, even $G_A/A\,G_N \approx 0.3$ 
\cite{ku,fs1} was predicted for $A\approx 200$ at $Q^2=4\,GeV^2$. 

It is instructive to compare the gluon shadowing at high $Q^2$
with what one expects for hadronic
reactions at much smaller virtualities.
One should expect more shadowing at smaller $Q^2$, however,
the soft gluon shadowing evaluated in the next section turns out
to be much weaker than one predicted in \cite{eqw,ku,fs1} at high $Q^2$.

At the same time, quite a different approach to the problem of gluon shadowing
based on the nonlinear GLR evolution
equation \cite{glr} used in \cite{agl} led to the results
pretty close to ours.

\subsubsection{Nuclear shadowing for soft gluons}

\noi
{\sl (i) Hadronic diffraction and gluon shadowing}\\ 
The hadron--nucleus total cross section is known
to be subject to usual Glauber (eikonal) \cite{glauber}
shadowing and Gribov's inelastic corrections \cite{gribov}.
Those corrections are controlled by
the cross section of diffractive dissociation
of the projectile hadron $h\,N\to X\,N$ which contains particularly
the triple-Pomeron contribution. The latter as was shown above is
related to gluon shadowing in nuclei. Namely, absorption of the
incoming hadron can be treated as a result of interaction with
the gluon cloud (in the infinite momentum frame
of the nucleus) of bound nucleons at small $x$.
A substantial part of this absorption is reproduced by 
the eikonal approximation which assumes the gluon density to be
proportional to the number of bound nucleons. However,
evolution of the gluon density including gluon fusion
(see \cite{kancheli} and \cite{glr,mq} for high $Q^2$) 
results in reduction of the gluon
density compared to one used in the eikonal 
approximation. Such a reduction 
makes nuclear matter more transparent for protons
\cite{kn}.

That part of nuclear shadowing which comes from
diffractive excitation of the valence quark component of the 
projectile  hadron corresponds in terms of the triple-Regge phenomenology
to the $PPR$ term in the diffractive cross section.
In eigenstate representation for the interaction Hamiltonian
the same effect comes from the dependence of the elastic amplitude on
positions of the valence quarks in the impact parameter plane \cite{zkl}.
On top of that, the projectile hadron can dissociate via gluon radiation
which corresponds to the triple-Pomeron term in diffraction. It can also
be interpreted in the infinite momentum frame of the nucleus as
a reduction of the density of gluons which interact with the hadron.
This relation gives a hint how to approach the problem of gluon
shadowing at small $x$ for soft gluons.

Let us model this situation in eigenstate 
representation with two Fock states for the projectile hadron,
\beq
|h\ra=(1-w)\,|h\ra_v + w\,|h\ra_G\ ,
\label{3.1.25a}
\eeq
where $|h\ra_v$ and $|h\ra_G$ are the components without (only valence
quarks) and with
gluons which can be resolved at the soft scale. 
We assume them to be eigenstates of interaction with 
eigenvalues $\sigma_v$ and $\sigma_G$ respectively.
The relative 
weights are controlled by the parameter $w$. 
The hadron-nucleon and hadron-nucleus total cross sections
can be represented as \cite{kl,zkl},
\beq
\sigma^{hN}_{tot}=\sigma_v + w\,\Delta\sigma\ ,
\label{3.1.25b}
\eeq
where $\Delta\sigma=\sigma_G - \sigma_v$, and
\beq
\sigma^{hA}_{tot}= 2\int d^2b\,\left\{
\left[1-\,{\rm exp}\left(-{1\over2}\,\sigma_v\,T(b)
\right)\right]
+ w\,\left[{\rm exp}\left(-\,{1\over2}\,\sigma_v\,T(b)\right)-
{\rm exp}\left(-\,{1\over2}\,\sigma_G\,T(b)\right)\right]\right\}
\label{3.1.25c}
\eeq
This cross section is smaller than one given by the eikonal Glauber
approximation \cite{glauber}, and the difference is known as Gribov's
inelastic corrections \cite{gribov}. The Glauber's cross section 
can be corrected by replacing 
the nuclear thickness function by a reduced one, $T(b) \Rightarrow 
\widetilde T(b) < T(b)$, which is related to the reduced gluon 
density in the nucleus,
\beq
\frac{G_A(x,b)}{G_N(x,b)}=
\frac{\widetilde T(b)}{T(b)}\ .
\label{3.1.25d}
\eeq

Thus, nuclear shadowing for soft gluons can be evaluated comparing the total
cross section (\ref{3.1.25c}) with the modified Glauber approximation,
\beq
\sigma^{hA}_{tot}= 2\int d^2b\,
\left[1-\,{\rm exp}\left(-{1\over2}\,
\sigma^{hN}_{tot}\,\widetilde T(b)\right)\right]\ .
\label{3.1.25e}
\eeq
Expanding both expressions in small parameters $\Delta\sigma\,T$
and $\sigma_v\,\Delta T$, where $\Delta T(b)=T(b)-\widetilde T(b)$,
(they are indeed small even for heavy nuclei) we get,
\beq
\Delta T(b)=\frac{w\,(\Delta\sigma)^2}
{4\,\sigma^{hN}_{tot}}\,T^2(b)
\left[1-{1\over6}\,\Delta\sigma\,T(b)
+ O\biggl((\Delta\sigma T)^2\biggr)\right]\ .
\label{3.1.25f}
\eeq
We left here only the leading terms and omitted for the sake of
simplicity the terms containing higher powers of $w$.

According to relation (\ref{b.6a}) 
$w\,(\Delta\sigma)^2/16\pi$ is the forward cross section
of diffractive gluon radiation which corresponds to the triple-Pomeron
part of the diffraction cross section $h\,N \to X\,N$. Therefore, the
correction (\ref{3.1.25f}) to the nuclear thickness function can
be expressed in terms of the effective cross section,

\beq
\sigma_{eff}= \frac{w\,(\Delta\sigma)^2}
{\sigma^{hN}_{tot}} = 
16\,\pi\,A_{3P}(hN\to XN)\,
{\rm ln}\left(\frac{M^2_{max}}{M^2_{min}}\right)\ ,
\label{3.1.26}
\eeq
where $M^2_{max}\approx 2\,\sqrt{3}\,s/(m_N\,R_A)$
is the upper cut off for the diffractive mass spectrum
imposed by the the nuclear formfactor.
The bottom cut off depends on $M^2$-dependence for
the triple-Pomeron diffraction at small masses which is
poorly known. At high energies under consideration
this uncertainty related to the choice of $M^2_{min}$
is quite small. We fix $M_{min}=2\,GeV$. 

Within an approximate Regge factorization scheme 
$A_{3P}(hN \to XN)$ defined in (\ref{3.42}) is an
universal constant (see, however, (\ref{3.43})).
Therefore, the driving term in (\ref{3.1.25f}) 
and gluon shadowing are independent
of our choice for hadron $h$, a result which could be expected. 

Data on diffractive reaction $p\,p\to p\,X$
fix the triple-Pomeron coupling ({\it e.g.} see in
\cite{kklp,dino,schlein}) with much better certainty
than for other reactions (including data for diffractive DIS).
The value of 
$A_{3P}$ varies from $0.075\,GeV^{-2}$ at medium high energies
to $0.025\,GeV^{-2}$ at Tevatron energy (see (\ref{3.43})).
Correspondingly, the effective cross section for $A\approx 200$
ranges as $\sigma_{eff} \approx 3.5 - 5.5\,mb$. This is an order
of magnitude smaller than the value used in \cite{fs1} at high $Q^2$. 
It is very improbable that $\sigma_{eff}$ can grow (so much!) with
$Q^2$.

It is silently assumed in Eq.~(\ref{3.1.25c})
that the energy is sufficiently high to freeze the fluctuations,
{\it i.e.} there is no mixing between the Fock components during
propagation through the nucleus. If, however, the energy is not high, 
or the effective mass of the excitation is too large, one should take care 
of interferences and represent (\ref{3.1.25d}), (\ref{3.1.25f}) 
in the form (compare to \cite{kk,murthy})
\beqn
&&\frac{G_A(x)}{A\,G_N(x)} = 1\,-\,
8\pi\,A_{3P}(pp\to pX)\,
{\rm Re}\int d^2b
\int\limits_{M_{min}^2}^{\infty}  
\frac{dM_X^2}{M_X^2}
\int\limits_{-\infty}^{\infty} dz_1
\int\limits_{-\infty}^{\infty} dz_2\,
\Theta(z_2-z_1)\nonumber\\ &\times& 
\rho_A(b,z_1)\,\rho_A(b,z_2)\,
{\rm exp}\Bigl[-i\,q_L\,(z_2-z_1)\Bigr]\,
{\rm exp}\Bigl[-\,{1\over2}\,\sigma_{abs}
\int\limits_{z_1}^{z_2} dz\,\rho_A(b,z)\Bigr]\ ,
\label{3.1.27}
\eeqn
where $\sigma_{abs} = \Delta\sigma$, and 
we exponentiated the expression in square brackets in the
r.h.s. of (\ref{3.1.25d}).

The important difference between (\ref{3.1.27}) and the usual
expression \cite{kk,murthy} for inelastic corrections 
is absence of absorption for the initial ($z < z_1$) and final
($z > z_2$) protons in (\ref{3.1.27}). This is a natural
result, since
proton absorption (mostly of eikonal type) has no relevance
to gluon shadowing. 

Absorption $\sigma_{abs}=\Delta\sigma$ in 
intermediate state ($z_1 > z > z_2$) 
is much smaller than $\sigma^{NN}_{tot}$ and is related to
the amplitude of diffractive gluon radiation (see (\ref{b.6a})).
One can estimate $\sigma_{abs}$ assuming Regge factorization.
In this case $\sigma_{eff}$ is universal and can be applied
even to a quark, {\it i.e.} $h=q$. This makes sense in our model
due to short range nature of the nonperturbative gluon interactions.

demanding Eq.~(\ref{3.1.27})
to reproduce correctly the ``frozen'' limit of $q_L\to 0$.
This needs $\sigma_{abs}=\sigma_{eff}$, as was actually guessed
in \cite{fs1}. 

However, the discussion following Eq.~(\ref{2.33}) shows that
after it is averaged over the quark-gluon separation the
absorptive cross section gains an extra factor, 
$\sigma_{abs}=1.5\,\sigma_{eff}$.

We performed numerical estimates for $A=200,\ 64$ and $12$ 
assuming a constant
nuclear density $\rho_A(r)=\rho_0\,\Theta(R_A-r)$ with 
$\rho_0=0.157\,fm^{-3}$ and $R_A=1.15\,A^{1/3}\ fm$.
In this case integration in (\ref{3.1.27}) can be performed
analytically and the result reads,
\beqn
\frac{G_A(x)}{A\,G_N(x)} &=& 1\,-\,
\frac{1}{3\,\upsilon^3\,{\rm ln}(M^2_{max}/M^2_{min})}\,
\left\{\left[3-{3\over2}\,\upsilon^2 + \upsilon^3 -
3\,(1+\upsilon)\,e^{-\upsilon}\right]
\right.\nonumber\\ &\times &
\left[{\rm ln}\left(\frac{s}{m_N\,R_A\,(M_{min}^2-m_N^2)}\right)
- \gamma\right] +
\left[3-{3\over2}\,\upsilon^2+\upsilon^3\right]\,\Bigl[\gamma+
{\rm ln}\upsilon - Ei(-\upsilon)\Bigr]
\nonumber\\
&+& \left.\left[{3\over2}\,\upsilon^2 - {11\over2}+\left(
{11\over2} + {5\over2}\,\upsilon - \upsilon^2\right)\,
e^{-\upsilon}\right]\right\}\ ,
\label{3.1.28}
\eeqn
where $\upsilon = {3\over2}\,\sigma_{eff}\,\rho_0\,R_A$,
$\gamma=0.5772$ is the Euler constant, and $Ei(z)$
is the integral exponential function. The value of $x$ 
can be evaluated as $x=4\,\la k^2\ra/s$, where $\la k^2\ra
\sim 1/b_0^2$ is the mean transverse momentum squared in the
quark -- gluon system.

\begin{figure}[tbh]
\includegraphics{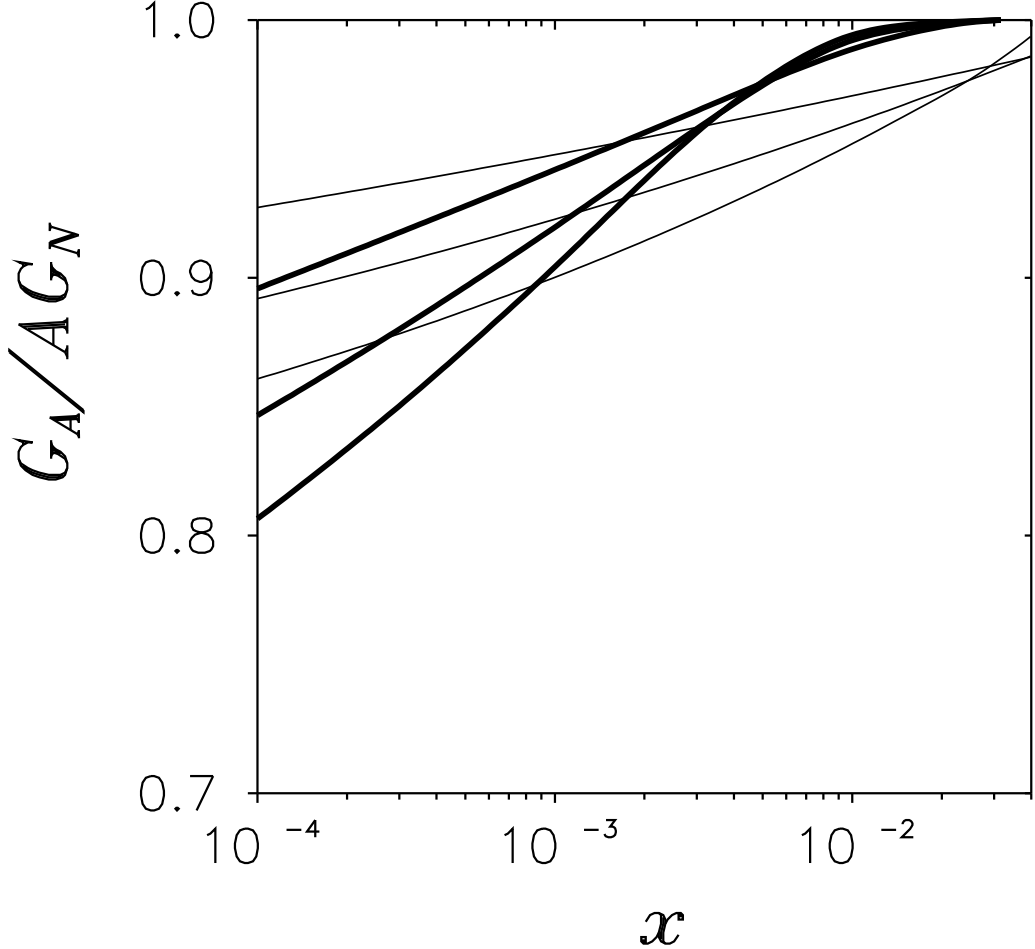}
\begin{center}
\vspace{13cm}
\parbox{13cm}
{\caption[Delta]
{\sl The same as in Fig.~\ref{glue}, 
but for soft gluons. The thin curves are obtained
with (\ref{3.1.28}) using data for the triple-Pomeron
contribution to diffraction $pp\to pX$.
The thick curves are predicted using the Green function method.}
\label{soft}}
\end{center}
\end{figure}

The results of numerical calculations with Eq.~\ref{3.1.28}
for gluon shadowing are depicted in Fig.~\ref{soft}
by thin solid curves for lead, copper and carbon
(from bottom to top) as function of $x$.
Shadowing for soft gluons turns out to be much weaker than predicted
in \cite{fs1,fs2} for high $Q^2$. This contradicts the natural expectation
that the softer gluons are, the stronger shadowing should be.

\medskip

\noi{\sl (ii) The Green function formalism}\\
One can also use the Green function formalism to calculate nuclear
shadowing for soft gluon radiation. It provides a better treatment
of multiple interactions and phase shifts 
in intermediate state. In contrast to the
above approach which uses a constant average value for $\sigma_{eff}$,
in the Green function formalism the absorption cross section as well as
the phase shift are functions of longitudinal coordinate. This is also
a parameter-free description, all the unknowns have already been 
fixed by comparison with data.

As usual, we treat shadowing for soft gluons as a contribution
of the gluonic Fock component to shadowing of the projectile-nucleus
total cross section. One can use as a soft projectile a real photon,
a meson, even a single quark. Indeed, the mean quark-gluon separation
$1/b_0\approx 0.3\,fm$ is much smaller that the quark-antiquark separation
in a light meson or a $\bar qq$ fluctuation of a photon. For this reason
one can neglect in (\ref{3.5b}) the interference 
between the amplitudes of gluon radiation
by the $q$ and $\bar q$. Since the gluon contribution to the cross section
corresponds to the difference between total cross sections for
$|\bar qqG\ra$ and $|\bar qq\ra$ components, the quark spectator cancels
out and the radiation cross section is controlled by the quark-gluon wave
function and color octet ($GG$) dipole cross section.

Thus, the contribution to the total hadron-nucleus cross section which comes 
from gluon radiation has the form,
\beq
\sigma^{hA}_G = \int\limits_x^1
\frac{d\,\alpha_G}{\alpha_G}\int d^2b\,
P(\alpha_G,\vec b)\ ,
\label{3.1.29}
\eeq
where
\beqn
P(\alpha_G,\vec b) &=& 
\int\limits_{-\infty}^{\infty} 
dz\,\rho_A(b,z)\int d^2r\,
\Bigl|\Psi_{qG}(\vec r,\alpha_G)\Bigr|^2\,
\sigma_{GG}(r,s)
\nonumber\\ & -\, &
{1\over2}\,{\rm Re}\int\limits_{-\infty}^{\infty}
dz_1\,dz_2\,\Theta(z_2-z_1)\,
\rho_A(b,z_1)\,\rho_A(b,z_2)\int d^2r_1\,d^2r_2\,
\label{3.1.30}\\
&\times&
\Psi^*_{qG}(\vec r_2,\alpha_G)
\sigma_{GG}(r_2,s)\,
G_{GG}(\vec r_2,z_2;\vec r_1,z_1)\,
\sigma_{GG}(r_1,s)\,
\Psi_{qG}(\vec r_1,\alpha_G)\nonumber\ .
\eeqn
Here the energy and Bjorken $x$ are related as
$s=2m_N\nu=4b_0^2/x$. The explicit solution for the
Green function $G_{GG}(\vec r_2,z_2;\vec r_1,z_1)$
in the case of $\sigma_{GG}(r,s)=C_{GG}(s)\,r^2$ and a
constant nuclear density is given by Eq.~(\ref{3.22}).
Note that the $r^2$ approximation for the dipole cross section
is justified by the small value of 
$\la r^2\ra = 1/b_0^2\approx 0.1\,fm^2$.

Integrations in (\ref{3.1.30}) can be performed analytically,
\beq
P(\alpha_G,\vec b) =
\frac{4\,\alpha_G}{3\,\pi}\,
{\rm Re}\,ln(W)\ ,
\label{3.1.31}
\eeq
where
\beq
W = ch(\Omega\,\overline{\Delta z}) +
\frac{A^2+b_0^2}{2\,A\,b_0^2}\,
sh(\Omega\,\overline{\Delta z})\ ,
\label{3.1.32}
\eeq
\beq
\overline{\Delta z}=2\,\sqrt{R_A^2-b^2}\ .
\label{3.1.33}
\eeq
We use here the same notations as in Eqs.~(\ref{3.1.22}) - (\ref{3.1.23}).

The results of calculations are depicted in Fig.~\ref{soft} by
thick curves for lead, copper and carbon (from bottom to top).
They demonstrate about the same magnitude of shadowing as
was calculated above using hadronic basis. However, the onset of
shadowing is delayed down to $x<0.01$. We believe that this result 
is trustable since the Green function approach treats phase 
shifts and attenuation in nuclear matter more consistently.

Comparing predicted shadowing for soft gluons in Fig.~\ref{soft}
and one at $Q^2=4\,GeV$ in Fig.~\ref{glue} we arrive at a surprising
conclusion that shadowing is independent of scale.
A small difference is within the accuracy of calculations.
This is a nontrivial result since calculations were done using
very different approximations. Shadowing of hard gluons was estimated
assuming that the $\bar qq$ pair is squeezed 
to a size $\sim 1/Q$ much smaller 
than the transverse separation between the gluon and the 
$\bar qq$. On the contrary,
radiation of soft gluons is dominated by  configurations
with a distant $q$ and $\bar q$ surrounded by small gluon clouds.
The fact that shadowing appears to be the same is a result
of existence of the semihard scale $b_0^2$ (which should be compared
with $Q_{eff}^2 < Q^2/4$). At larger virtualities shadowing
decreases as one can see from comparison of $Q^2=4\,GeV^2$ with
$16\,GeV^2$ in Fig.~\ref{glue}.

\subsection{Nonperturbative effects 
in the transverse momentum distribution of 
gluon bremsstrahlung}

As soon as the strength of the nonperturbative quark-gluon 
interaction is fixed, we are in a position to calculate
the cross section of gluon bremsstrahlung for a high
energy quark interacting with a nucleon or a nuclear
target and to compare the results with the perturbative
QCD calculations \cite{kst}. 

\subsubsection{Nucleon target}

The transverse momentum distribution of soft gluons ($\alpha_G\ll 1$)
reads \cite{kst},
\beqn
\frac{d\,\sigma}{d({\rm ln}\alpha_G)\,d^2k_T}&=&
\frac{1}{2\,(2\pi)^2}\,\int d^2r_1\,d^2r_2\,
\overline{\Psi^{\dagger}_{qG}(\vec r_1,\alpha_G)\,
\Psi_{qG}(\vec r_2,\alpha_G)}\,{\rm exp}\Bigl[i\,\vec k_T
(\vec r_1-\vec r_2)\Bigr]\nonumber\\ &\times&
\Bigl[\sigma_{GG}(r_1)+
\sigma_{GG}(r_2)-\sigma_{GG}(\vec r_1-\vec r_2)\Bigr]\ .
\label{3.44}
\eeqn
Here the overline means that we sum over all possible polarizations
of the radiated gluon and recoil quark and average over the 
polarization of the initial quark. In our model for the quark-gluon 
distribution function including nonperturbative effects we get,
\beq
\overline{\Psi^{\dagger}_{qG}(\vec r_1,\alpha_G)\,
\Psi_{qG}(\vec r_2,\alpha_G)} =
\frac{4\,\alpha_s}{3\,\pi^2\,r_1^2\,r_2^2}\,
\vec r_1\cdot \vec r_2\,
{\rm exp}\left[-\,\frac{b_0^2}{2}\,(r_1^2+r_2^2)\right]\ .
\label{3.45}
\eeq

The cross section $\sigma_{GG}(r)$ in (\ref{3.44}) has the form
(\ref{3.1.17}.

We performed calculations for the transverse momentum distribution
of gluons for two parameterizations of the dipole cross section,\\
(I) one which is given by (\ref{2.24}) which is constant at 
$\rho^2\gg\rho_0^2$. For the sake of convenience we change the notation 
here, $s^2=2/\rho_0^2=0.125\,GeV^{-2}$;\\
 (II) the dipole approximation
(\ref{2.3a}) with $C=\sigma_0\,s^2/2$. Only this parameterization 
is used for nuclear targets because it allows to perform integrations 
analytically (of course one can do numerical calculation for
any shape of the cross section).

Correspondingly, we obtain for the differential radiation cross
section,
\beq
\frac{d\,\sigma^N_I}{d({\rm ln}\alpha_G)\,d^2k_T} =
\frac{3\,\alpha_s\,\sigma_0}{\pi^2}\,
F(k_T^2,b_0^2,s^2)\ ,
\label{3.48}
\eeq
where
\beqn
&&F(k_T^2,b_0^2,s^2)=
\frac{1}{2\,k_T^2}\,\Omega_1\,(\Omega_1-2\,\Omega_2)
\nonumber\\ &+&
\frac{1}{4\,s^2}\,\left[{\rm Ei}\left(\frac{k_T^2}{2\,s^2}\right)
- 2\,{\rm Ei}\left(\frac{k_T^2}{2\,s^2}\,x_1\right)
+{\rm Ei}\left(\frac{k_T^2}{2\,s^2}\,x_2\right)\right]
\label{3.49}
\eeqn
\beqn
\Omega_1&=&1 - {\rm exp}\left(-\frac{k_T^2}{2\,b_0^2}\right)\ ;
\nonumber\\
\Omega_2&=&1 - {\rm exp}\left[-\frac{k_T^2}{2\,(b_0^2+s^2)}\right]\ ;
\nonumber\\
x_1&=&\frac{b_0^2}{b_0^2+s^2}\ ;\ \ \ \ x_1=\frac{b_0^2}{b_0^2+2\,s^2}\ ;
\nonumber
\label{3.50}
\eeqn
and ${\rm Ei}(z)$ is the exponential integral function.

In the case of parameterization {\rm II} it is convenient to
represent the dipole cross section in the form,
\beq
\sigma_{\bar qq}(r)=\sigma_0\,s^2\,\frac{d}{d\,s^2}\,\left[ 1 - 
{\rm exp}\left(-{1\over2}\,s^2\,r^2\right)\right]_{s^2=0}\ .
\label{3.51}
\eeq
Then the differential cross section reads,
\beq
\frac{d\,\sigma^N_{II}}{d({\rm ln}\alpha_G)\,d^2k_T} =
\frac{3\,\alpha_s\,\sigma_0\,s^2}{\pi^2}\,
F_1(k_T^2,b_0^2,s^2)\ ,
\label{3.52}
\eeq
where
\beqn
F_1(k_T^2,b_0^2,s^2)&=&\frac{d}{d\,s^2}\,
F(k_T^2,b_0^2,s^2)\biggr|_{s^2=0}\nonumber\\
&=& \Lambda_1^2-\Lambda_1\,\Lambda_2+{1\over2}\,
\Lambda_2^2\ ;
\label{3.53}
\eeqn
\beqn
\Lambda_1&=& \frac{1}{k_T^2}\,\Omega_1 = \frac{1}{k_T^2}\,
\left[1 - {\rm exp}\left(-\frac{k_T^2}{2\,b_0^2}
\right)\right]\ ;\nonumber\\ 
\Lambda_2 &=& \frac{1}{k_T^2}\,
{\rm exp}\left(-\frac{k_T^2}{2\,b_0^2}\right)\ .
\nonumber
\label{3.54}
\eeqn

\begin{figure}[tbh]
\includegraphics{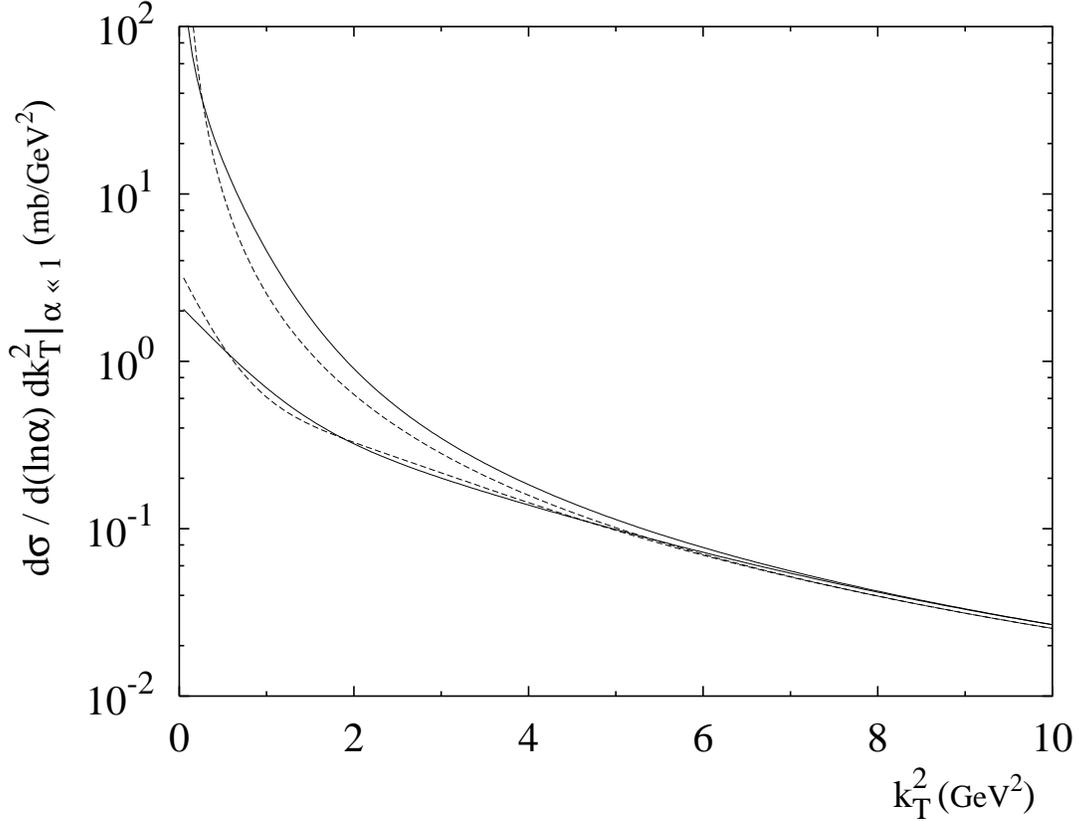}
\begin{center}
\vspace{12cm}
\parbox{13cm}
{\caption[Delta]
{\sl Transverse momentum distribution for gluon
bremsstrahlung by a quark scattering on a nucleon target.
The solid and dashed curves correspond to
parameterizations {\rm I}
and  {\rm II} for the dipole cross section,
respectively. The upper curves show the results of the 
perturbative QCD predictions \cite{kst}, the bottom curves
correspond to the full calculation including the
nonperturbative interaction of the radiated gluon.}
\label{tav1}}
\end{center}
\end{figure}

The results of calculations for variants {\rm I}
and  {\rm II} are depicted in Fig.~\ref{tav1} by
solid and dashed curves respectively. 
The two upper curves
correspond to perturbative calculations, while the two bottom
ones include the nonperturbative effects.
The strong interaction between
gluon and quark leads to a substantial decrease in the 
mean transverse size of the quark-gluon fluctuation.
Therefore, the mean transverse momentum of the radiated
gluons increases. The 
nonperturbative interaction has especially strong effect at small
transverse momentum $k_T$, where the radiation
cross section turns out to be suppressed by almost two 
orders of magnitude compared to the perturbative QCD expectations.

Note that intensive gluon radiation originating from multiple
nucleon interactions in relativistic heavy ion collisions is
found \cite{hk-ab,hhk} to be an 
important alternative source for suppression of charmonium
production rate and is able to explain the corresponding data from
the NA50 experiment at CERN SPS.
The found strong suppression of gluon bremsstrahlung
by the nonperturbative interaction relevant only to
small $\alpha\ll1$. However, it may substantially reduce the influence of
prompt gluons on charmonium production if is important at large
$\alpha$ as well. This is to be checked.

\subsubsection{Nuclear targets}

In the case of nuclear targets Eq.~(\ref{3.44}) holds, but 
$\sigma_{GG}(r)$ has the form, 
\beq
\sigma^A_{GG}(r)= 2\,\int d^2B\,\left\{
1-{\rm exp}\left[-\,
{9\over8}\,\sigma_{\bar qq}(r)\,T(B)\right]\right\}\ ,
\label{3.47}
\eeq

\begin{figure}[tbh]
\includegraphics{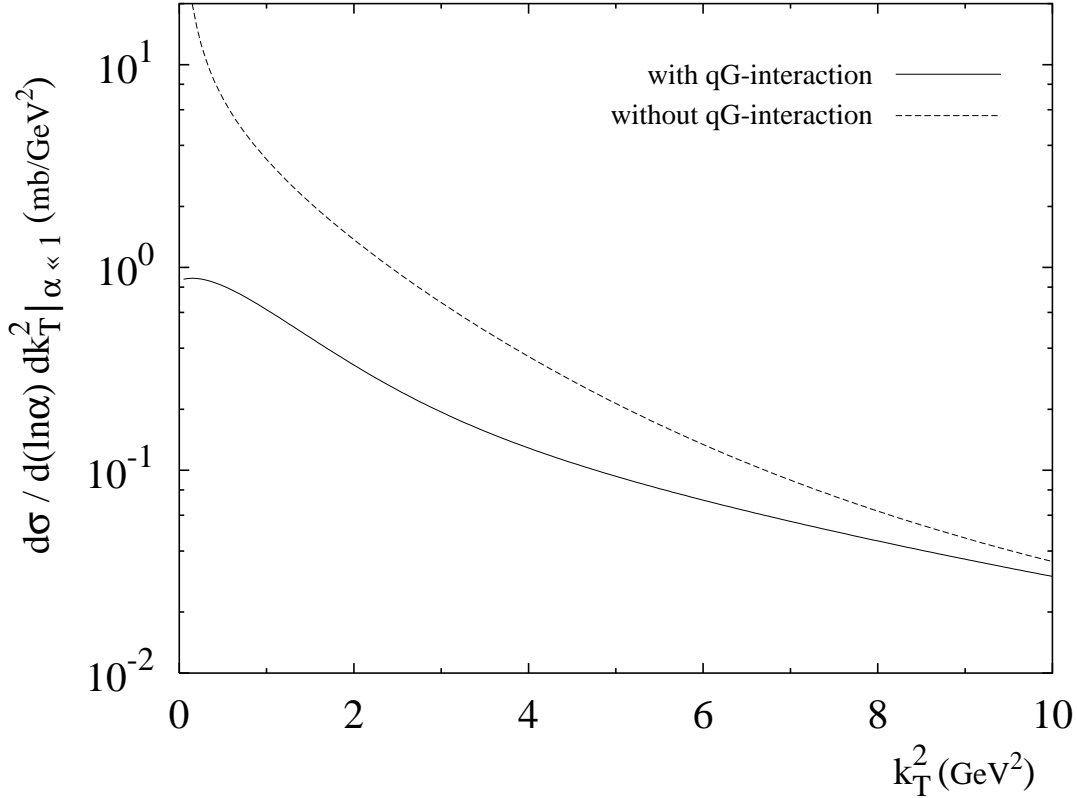}
\begin{center}
\vspace{12cm}
\parbox{13cm}
{\caption[Delta]
{\sl The differential cross section 
per bound nucleon of soft gluon
bremsstrahlung in quark-copper collisions.
The solid and dashed curves correspond to
calculations with and without the nonperturbative 
effects respectively.}
\label{tav2}}
\end{center}
\end{figure}

\begin{figure}[tbh]
\includegraphics{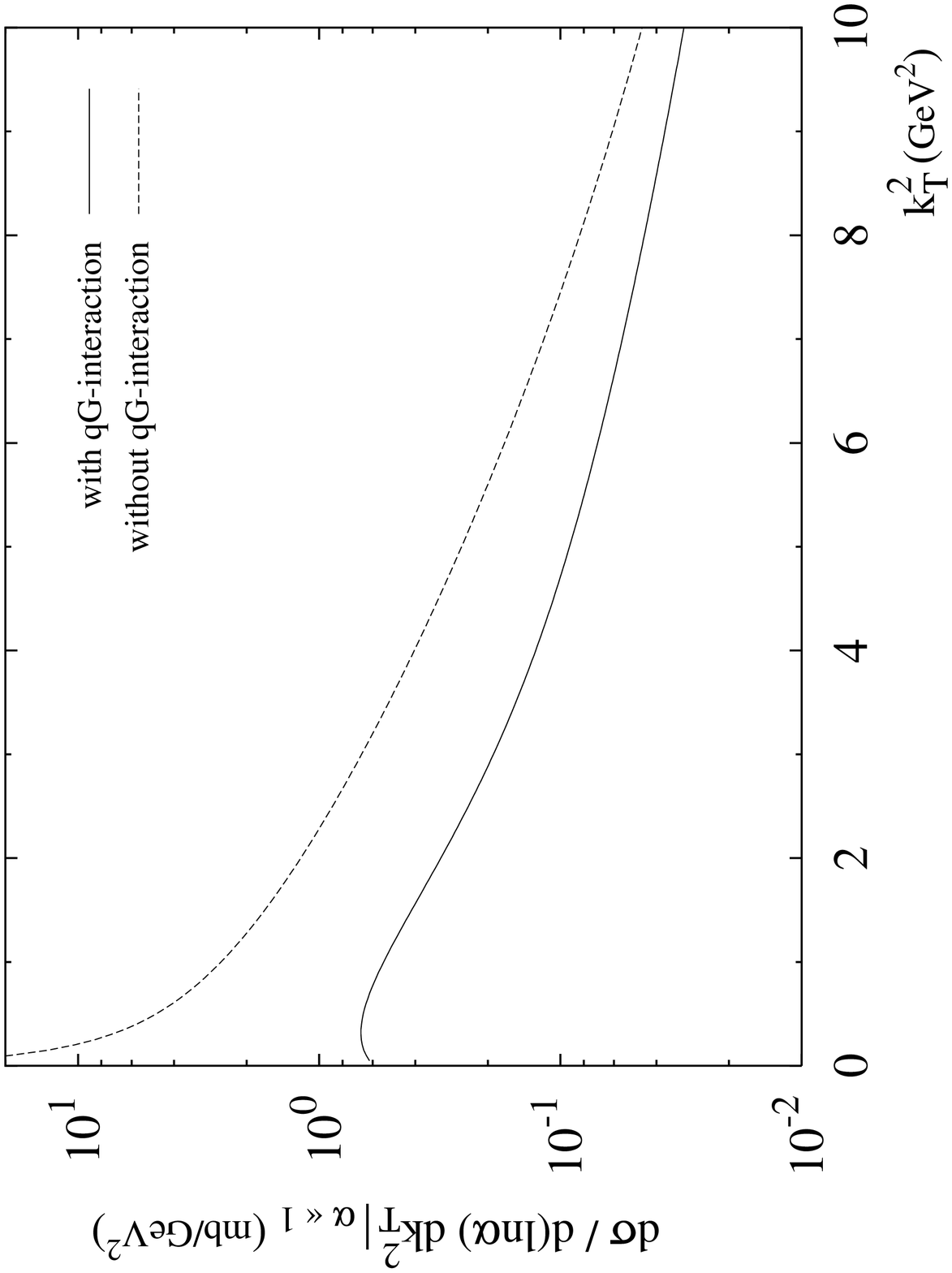}
\begin{center}
\vspace{12cm}
\parbox{13cm}
{\caption[Delta]
{\sl The same as in Fig.~\ref{tav2},
but for a lead target.}
\label{tav4}}
\end{center}
\end{figure}

Our calculations for gluon radiation in the interaction 
of a quark with a nuclear target are performed 
only in the parameterization {\rm II}
for the sake of simplicity. For heavy nuclei this approximation
can be quite good due to a strong color filtering
effect which diminishes the contribution from large size dipoles.
The transverse momentum distribution has the form,
\beq
\frac{d\,\sigma^A_{II}}{d({\rm ln}\alpha_G)\,d^2k_T} =
\frac{8\,\alpha_s}{3\,\pi^2}\int d^2B\,
F(k_T^2,b_0^2,S^2(B))\ ,
\label{3.55}
\eeq
where
\beq
S^2(B)= {9\over8}\,\sigma_0\,s^2\,T(B)\ .
\label{3.56}
\eeq
For numerical calculations we use the approximation
of constant nuclear density, $\rho_A(r)=3A/(4\pi R_A^3)\,
\Theta(R_A-r)$. The results for the radiation cross section
per bound nucleon with (solid curve) 
and without (dashed) the nonperturbative effects
are compared in Figs.~\ref{tav2} and \ref{tav4} for copper and lead
targets respectively. 
Obviously the nonperturbative interaction generates very large 
nuclear effects.

\begin{figure}[tbh]
\includegraphics{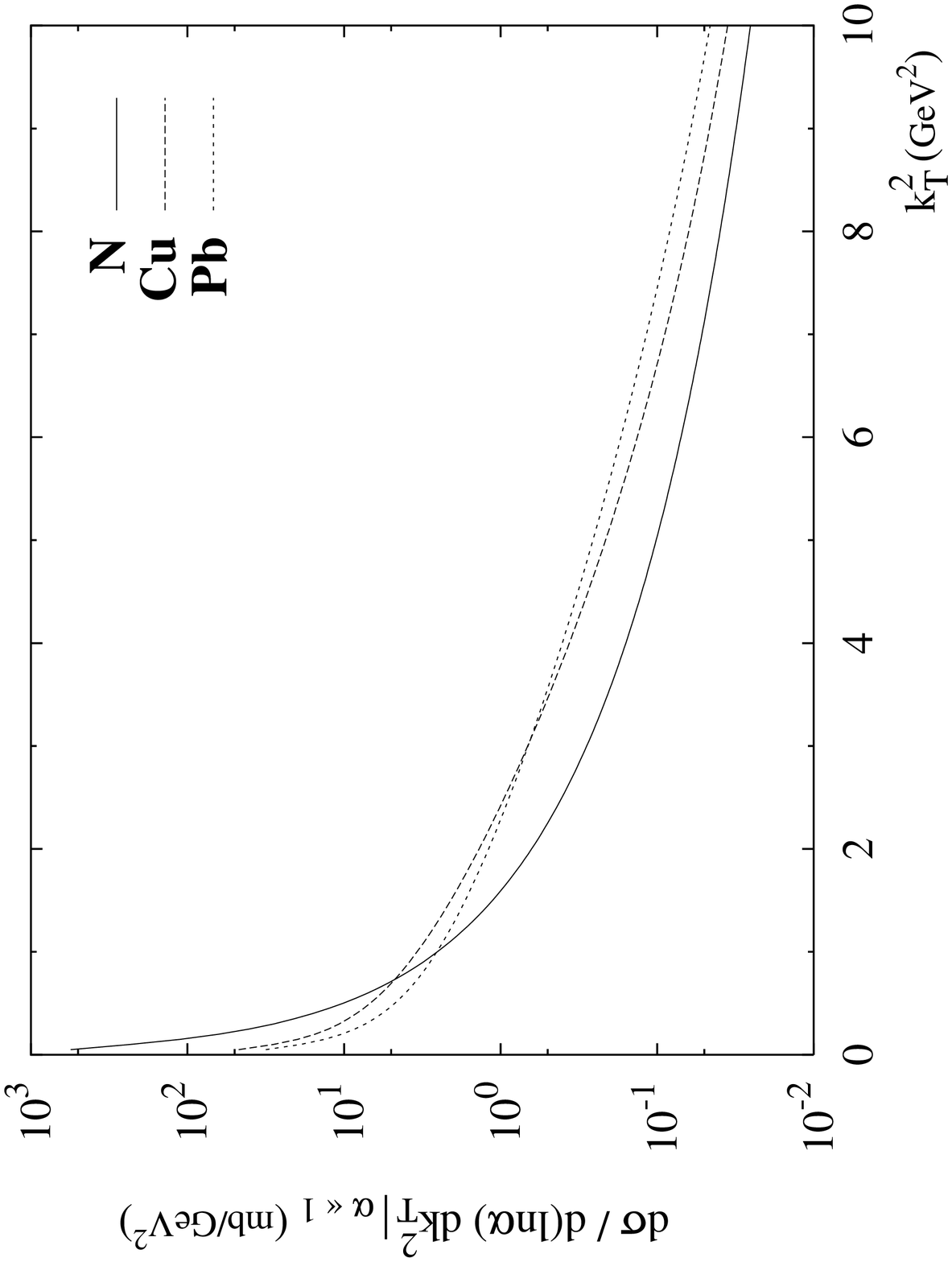}
\begin{center}
\vspace{11cm}
\parbox{13cm}
{\caption[Delta]
{\sl Comparison of the cross sections of gluon
radiation per nucleon in the perturbative 
QCD limit for collisions of a quark
with a nucleon (solid curve), copper (dashed curve)
and lead (dotted curve) versus the transverse momentum 
squared of the gluon.}
\label{tav0}}
\end{center}
\end{figure}

\begin{figure}[tbh]
\includegraphics{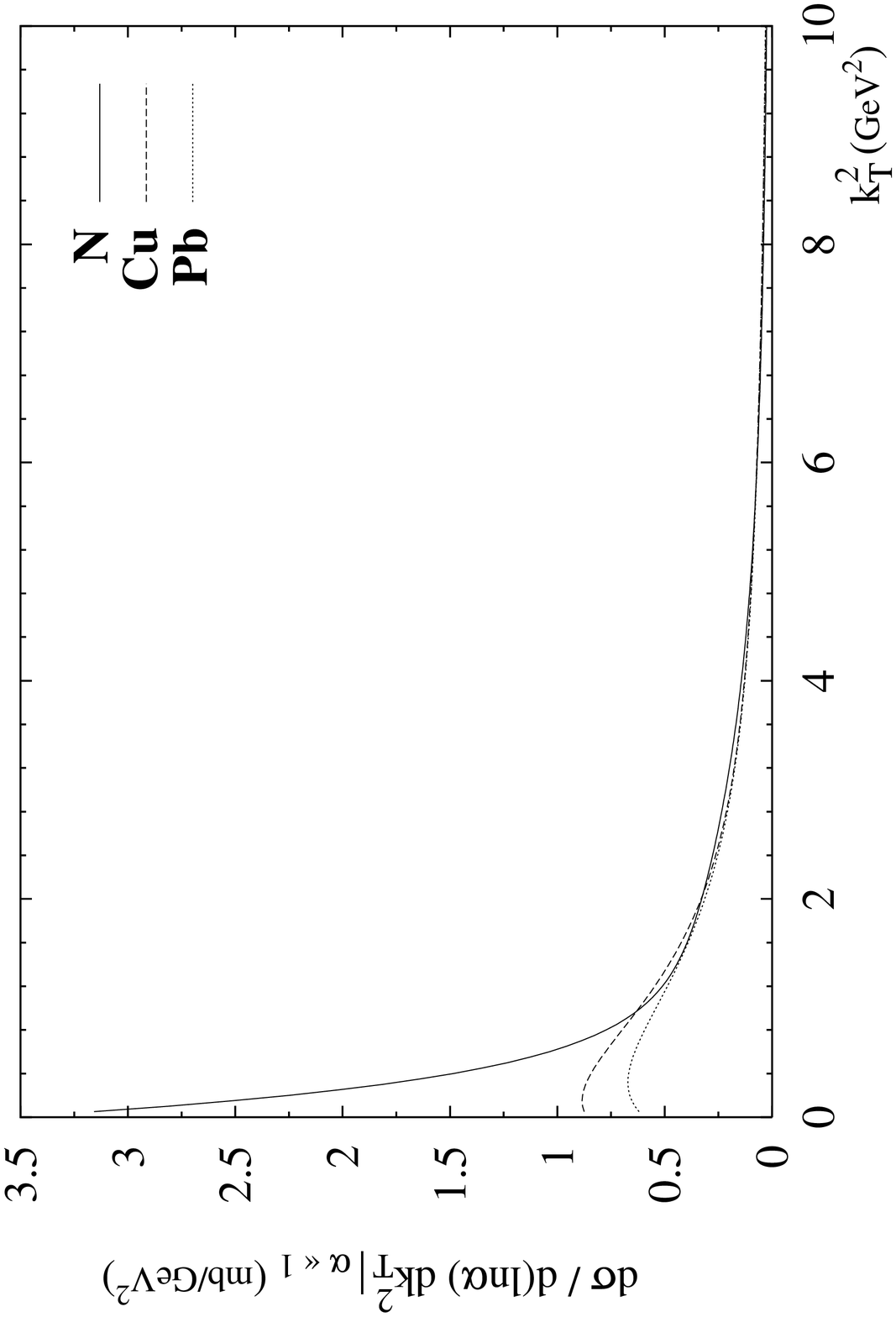}
\begin{center}
\vspace{9.5cm}
\parbox{13cm}
{\caption[Delta]
{\sl The same as in Fig.~\ref{tav0}, 
but the nonperturbative interaction of gluons is included.}
\label{tavB}}
\end{center}
\end{figure}

The nuclear effects are emphasized by a direct comparison 
in Figs.~\ref{tav0} and \ref{tavB} for different
targets, a nucleon, copper and lead, 
including and excluding the nonperturbative interaction
respectively. 
We see that the difference between a free and a bound nucleon
at small $k_T$ is substantially reduced by the nonperturbative interaction.
Indeed, the interaction squeezes  the quark-gluon fluctuation
and reduces the nuclear effects. Besides, the region of antishadowing is
pushed to larger values of $k_T$.

This manifestation of the nonperturbative interaction implies that
gluon saturation which is an ultimate form of shadowing
should happen with a smaller gluon density compared to
the expectations \cite{mv,m99} based on perturbative calculations.
On the other hand, the saturation region spreads up to higher values of
$k_T$.

\section{Summary and outlook}

We explicitly introduced a nonperturbative interaction 
between partons into the evolution equation for
the Green function of a system of quarks and gluons.
The shape of the $\bar qq$ potential is chosen to reproduce
the light cone wave function of mesons. The magnitude
of the potential is adjusted to reproduce data for
photoapsorptive cross sections on nucleons and nuclei
and data on diffractive dissociation of photons into $\bar qq$ pairs.

Based on theoretical arguments and experimental facts
we expect a much stronger interaction for a
quark-gluon pair than for a quark-antiquark pair. Indeed, data
on diffractive dissociation of hadrons and photons into
high mass states show that the cross section is amazingly small,
what is usually phrased as evidence that the triple-Pomeron
coupling is small. We have performed calculations for diffractive
gluon radiation (responsible for the production of high mass excitations)
including the nonperturbative effects, and fixed the strength
of the quark-gluon potential. We found a very simple and intuitive
way to get the same results as direct calculations of
Feynman diagrams. Both approaches lead to the same
diffractive cross section which in the limit of
perturbative QCD coincides with the result of a recent calculation
\cite{bartels} for the process $\gamma^*N\to \bar qqGN$.
We conclude that the previous analogous calculations \cite{nz1}
are incorrect.

We adjusted the quark-gluon potential to data for the diffractive
reaction $pp\to Xp$ which have the best accuracy and cover the 
largest range of energies and masses. We predicted the single 
diffractive cross sections for pions, kaons and photons and
find a substantial violation of Regge factorization.

We calculated nuclear shadowing for longitudinally
polarized photons which are known to serve as 
a sensitive probe for the gluon distribution, 
using the Green function technique
developed in \cite{krt} describing the evolution of
a $\bar qqG$ system propagating through nuclear matter.
The evolution equation includes the phase shift which 
depends on the effective mass of the fluctuation, 
nuclear attenuation which depends on the transverse separation and 
energy, and the distribution over transverse separation and
longitudinal momenta of the partons which is essentially affected
by the nonperturbative interaction of the gluon. The latter
substantially reduces the effect of nuclear shadowing of 
gluons. We have found an $x$ dependence for gluons which is 
quite different
from that for quarks. These differences are far beyond
the simple Casimir factor $9/4$.

Nuclear shadowing for soft gluons is essentially controlled by
the nonperturbative effects.
It turns out to be rather weak similar to what is found at
$Q^2\sim 4\,GeV^2$. Such a scale invariance at low and medium
high virtualities is a consequence of the strong nonperturbative
interaction of gluons which introduces a semihard scale
$\sim 4\,b_0^2=1.7\,GeV^2$.

The nonperturbative interaction changes dramatically the
transverse momentum distribution of gluon bremsstrahlung
by a high energy quark interacting with a nucleon or
a nucleus. The gluon radiation cross section at small $k_T$ 
turns out to be suppressed by nearly two orders of magnitude
compared to the expectations from perturbative QCD 
\cite{mueller,kst}. Although these results concern the gluons
radiated with $\alpha_G\to 0$, it might also suppress 
gluon bremsstrahlung at larger $\alpha_G$ which is predicted
\cite{hk-ab} to contribute to the break up of charmonia produced in
relativistic heavy ion collisions.

This effect is especially strong for nuclear targets where
the nonperturbative interaction of radiated gluons
creates a forward minimum in 
the transverse momentum distribution.
This suppression is
an additional contribution to nuclear shadowing calculated
perturbatively in \cite{mueller,kst} which also leads to
a suppression of small transverse momenta.
The results of our calculations presented in Figs.~\ref{tav2},
\ref{tav4} include both phenomena.

Nuclear shadowing for small transverse momenta of the radiated
gluons is the same effect as the saturation of parton densities
at small $x$ in nuclei as seen in the infinite
momentum frame of the nucleus. This phenomenon is expected 
to be extremely important for the problem of quark-gluon 
plasma formation in relativistic heavy ion collisions.
On the one hand, a growth of the mean transverse momentum of 
radiated gluons increases the produced transverse energy, on the other
hand, it leads to a higher probability for such gluons
to escape the interaction region without collisions,
{\it i.e.} the gluon gas may not reach equilibrium
\cite{al-bnl}.

{\bf Acknowledgements}: We are grateful to Yuri Ivanov and J\"org
Raufeisen for their constant  assistance in numerical 
calculations and to J\"org H\"ufner, Mikkel Johnson, Andrei Leonidov 
and Hans-J\"urgen Pirner for
useful discussions. A substantial part of this work was done when A.V.T.
was employed by  the Institut f\"ur
Theoretische Physik der Universit\"at, Heidelberg and was supported by
the Gesellschaft f\"ur
Schwerionenforschung, GSI, grant HD H\"UF T.

\def\appendix{\par
 \setcounter{section}{0}
 \setcounter{subsection}{0}
 \def\thesection{Appendix \Alph{section}}
 \def\thesubsection{\Alph{section}.\arabic{subsection}}
 \def\theequation{\Alph{section}.\arabic{equation}}
 \setcounter{equation}{0}}

\appendix

\section{Diffraction} 
\setcounter{equation}{0}

\subsection{General consideration}

In this section we present a general analysis of
diffraction based on the eigenstate decomposition.

The off-diagonal diffractive scattering is a direct consequence of the fact
that the interacting particles (hadrons, photon)
are not eigenstates of the 
interaction Hamiltonian \cite{pom,gw}. They can be decomposed
in a complete set of such eigenstates $|k\ra$ \cite{kl,mp},
\beq
|h\ra=\sum\limits_{k}\,C^h_k\,|k\ra\ ,
\label{b.1}
\eeq
where $C^h_k$ are the amplitudes for the decomposition which obey
the orthogonality conditions,
\beqn
\sum\limits_{k}\,\left(C^{h'}_k\right)^{\dagger}\,C^h_k
&=&\delta_{h\,h'}\ ;
\nonumber\\
\sum\limits_{h}\,\left(C^{h}_l\right)^{\dagger}\,C^h_k
&=&\delta_{lk}\ .
\label{b.2}
\eeqn

We denote by $f_k=i\,\sigma_k/2$ the eigenvalues of the
elastic amplitude operator $\hat f$. We assume that the amplitude 
is integrated over impact parameter, {\it i.e.} that
the forward scattering elastic amplitude is normalized as
$|f_k|^2=4\,\pi\,d\sigma_k/dt|_{t=0}$.  
We can then express the hadronic amplitudes, the elastic
$f_{el}(hh)$ and off diagonal diffractive $f_{dd}(hh')$ 
amplitudes as,
\beq
f_{el}(hh)=2i\,\sum\limits_k\,\left|C^h_k\right|^2\,\sigma_k
\equiv 2i\,\la\sigma\ra\ ;
\label{b.3}
\eeq
\beq
f_{dd}(hh')=
2i\,\sum\limits_k\, (C^{h'}_k)^{\dagger}\,C^h_k\,\sigma_k\ .
\label{b.4}
\eeq
Note that if all the eigen amplitudes are equal the diffractive
amplitude (\ref{b.4}) vanishes due to the orthogonality relation,
(\ref{b.2}). The physical reason is obvious. If all the $f_k$ are 
identical 
the interaction does not affect the coherence between the different 
eigen components $|k\ra$ of the projectile hadron $|h\ra$. Therefore,
off diagonal transitions are possible only due to differences 
between the $f_k$'s. For instance, in the two channel case,
\beq
f_{dd}(hh')=2i\,(C^h_2)^{\dagger}\,C^h_1\,
(\sigma_1-\sigma_2)\ .
\label{b.5}
\eeq

If one sums over all final states in the diffractive cross section
one can use the completeness condition (\ref{b.2}).
Excluding the elastic channels one gets \cite{kl,mp,zkl},
\beq
16\pi\,\frac{d\sigma^h_{dd}}{dt}\biggr|_{t=0}=
\sum\limits_i \left|C^h_i\right|^2\sigma^2_i
- \biggl(\sum\limits_i \left|C^h_i\right|^2\sigma_i\biggr)^2
\equiv \la\sigma_i^2\ra - \la\sigma\ra^2\ ,
\label{b.6}
\eeq
This formula is valid only for the total (forward) diffractive
cross section and cannot be used for exclusive channels.

\subsection{Diffractive excitation of a quark, \boldmath$q\to qG$}

In this case we can restrict ourselves to the first two
Fock components of the quark, a bare quark $|q\ra$ and
$|qG\ra$. Therefore, we can use Eq.~(\ref{b.5}). 
Thus, we arrive at the following
expression for the forward amplitude of diffractive 
dissociation into a $qG$ pair with transverse separation $\vec\rho$,
\beq
f_{dd}(q\to qG)\biggr|_{q_T=0}=
i\,\Psi_{qG}(\alpha,\vec\rho)\,
\biggl[\sigma_{qG}(\rho)-\sigma_q\biggr]\ .
\label{b.5a}
\eeq
Both cross sections, $\sigma_{qG}$ and $\sigma_{q}$
are infra-red divergent, but this divergence is obviously the same
and cancels in (\ref{b.5a}).

To regulate the divergence
we can introduce a small gluon mass $m_G$, which will not
enter the final result, and impose that for separations $r\gg 1/m_G$
the dipole cross section is given by the additive quark limit,
$\sigma_{\bar qq}(r\gg 1/m_G)= 2\,\sigma_q$.
To find the convergent part of $\sigma_{qG}(\rho)-\sigma_q$
we can make use of Eq.~(\ref{3.5}).
Let us choose in (\ref{3.5}) $r_1\ll 1/m_G$
and $r_2\gg 1/m_G$. Then the {\it l.h.s.} of (\ref{3.5}) 
saturates at $\sigma_q+\sigma_{qG}(r_1)$. Here $\sigma_{qG}(r_1)$
is different from $\sigma_{\bar q}$ due to the color dipole moment of
the $qG$ system, {\it i.e.} due to $r_1\not= 0$. Then (\ref{3.5})
is modified to,
\beq
\sigma_q+\sigma_{qG}(r_1)=\frac{9}{8}\,
\biggl\{\sigma_{\bar qq}(r_1)+2\,\sigma_q\biggr\}\,
-\,{2\over8}\,\sigma_q\ .
\label{b.5b}
\eeq
From this relation we obtain the combination of cross sections
at the {\it r.h.s.} of Eq.~(\ref{b.5a}) which takes the form,
\beq
f_{dd}(q\to qG)\biggr|_{q_T=0}=
i\,\Psi_{qG}(\alpha,\vec\rho)\,
{9\over8}\,\sigma_{\bar qq}(\rho)\ .
\label{b.5c}
\eeq
Thus, we derived Eq.~(\ref{3.17}) in a simple and intuitive way.
A more formal derivation based on direct calculation of Feynman
diagrams is presented in Appendix~B.1.

\subsection{Diffractive gluon radiation by a \boldmath$\bar qq$
pair}

The diffractive amplitude of gluon radiation by
a $\bar qq$ pair, $\bar q\,q\to\bar q\,q\,G$,
can be easily derived in this approach.
We restrict ourselves to two Fock components
$|\bar qq\ra$ and $|\bar qqG\ra$. Then the
distribution amplitudes $C^h_k$ get the meaning
of distribution functions for these Fock
states, namely $\Psi_{\bar qq}(\vec r_1-\vec r_2,\alpha)$ and
$\Psi_{G\bar qq}(\vec\rho_1,\vec\rho_2,\alpha,\alpha_G)$,
where the transverse coordinates are defined in (\ref{3.25}).
Summation over $k$ in (\ref{b.1}) - (\ref{b.4}) now means
integration over the transverse separations and summation
over the Fock components. According to (\ref{b.4}) - (\ref{b.5})
the diffractive amplitude $f_{dd}(\bar qq\to\bar qqG)$
reads,
\beq
f_{dd}(\bar qq\to\bar qqG)=
2\int d^2\rho_1\,d^2\rho_2\,
\Psi_{\bar qqG}(\vec\rho_1,\vec\rho_2,\alpha,\alpha_G)\,
\left[\sigma_{G\bar qq}(\vec\rho_1,\vec\rho_2) - 
\sigma_{\bar qq}(\vec\rho_1 - \vec\rho_2)\right]\ .
\label{b.6a}
\eeq
Here we make use of the obvious relation $C^{\bar qq}(\vec r)
= \delta(\vec r)$.
The total cross sections for the two Fock components $\bar qq$
and $G\bar qq$ are introduced in (\ref{2.3}) and (\ref{3.5}).

The distribution amplitude for the $G\bar qq$ fluctuation 
in the limit of $\alpha_G\to 0$ is easily 
guessed. Indeed, in this limit the impact parameters of the $q$ 
and $\bar q$ are not affected by gluon radiation. Therefore, the
$\Psi_{\bar qqG}$ 
should be a product of the $\bar qq$ distribution function
in the projectile hadron (photon) times the sum of the gluon distribution
amplitudes corresponding to radiation of the gluon by $q$ or $\bar q$,
\beq
\Psi_{\bar qqG}(\vec\rho_1,\vec\rho_2,\alpha,\alpha_G)=
\Psi_{\bar qq}(\vec\rho_1-\vec\rho_2,\alpha)\,
\left[\Psi_{qG}\left(\vec\rho_1,\frac{\alpha_G}{\alpha}\right) -
\Psi_{\bar qG}\left(\vec\rho_2,\frac{\alpha_G}{1-\alpha}\right)\right]\ ,
\label{b.7}
\eeq
where $\Psi_{\bar qq}$ and $\Psi_{qG}$ are defined in
(\ref{2.12}), (\ref{2.13}) and in (\ref{3.12}) respectively.
Thus, we have arrived at Eq.~(\ref{3.26}). A more formal derivation
based on the calculation of Feynman graphs is presented in the
next Appendix. 

After integration over $(\vec\rho_1 + \vec\rho_2)$ in (\ref{b.6a})
the amplitude of diffractive gluon radiation  turns out to be
proportional to the difference $\Delta\sigma(\vec\rho_1-\vec\rho_2)$
between the cross sections of the colorless systems $G\bar qq$ and $\bar qq$.
This is a straightforward consequence of the general property
of off-diagonal diffractive amplitudes given in (\ref{b.5}).

These conclusions are also valid for diffractive gluon radiation
by a photon $\gamma\ N \to \bar q\,q\,G\ N$. At first glance
presence of a third channel, the photon, may change the situation
and gluon radiation amplitude may not be proportional to 
$\Delta\sigma$. This is not true, however,  since
the relative weights of the $\bar qq$
and $\bar qqG$ components of the photon are the same as above
as soon as they are generated perturbatively. 

In the limit of purely perturbative interactions the same result as 
our Eq.~(\ref{3.26}) was obtained recently in \cite{bartels} (Eq.~(3.4)).
However, the  cross section for diffractive gluon radiation
derived earlier in \cite{nz1} (Eq.~(60)) is not proportional to
$(\Delta\sigma)^2$, but contains a linear term.
We think that this is a consequence of improper application of relation 
(\ref{b.6})  to an exclusive channel.

\subsection{Diffractive electromagnetic radiation}

The forward amplitude for photon (real or virtual)
radiation by a quark is similar to that for gluon
radiation (\ref{b.5a}), except that the photon does not
interact strongly and one has to replace $\sigma_{qG}$
by $\sigma_q$,
\beq
f_{dd}(q\to q\gamma^*)\biggr|_{q_T=0}=
i\,\Psi_{q\gamma^*}(\alpha,\vec\rho)\,
\biggl[\sigma_{q}-\sigma_q\biggr]
\,=\,0\ .
\label{b.8}
\eeq
Thus, in order to radiate a photon the quark has to
get a kick from the target, no radiation happens if 
the momentum transfer to the target is zero.

This conclusion is different from the expectation for
diffractive Drell-Yan pair production of \cite{k}.
The latter was based on the conventional formula (\ref{b.6a})
which cannot be used for an exclusive channel (as well as
for gluon radiation). Therefore, the diffractive Drell-Yan
cross section should be much smaller than estimated in 
\cite{k}.

Nevertheless, a hadron as a whole can radiate diffractively a photon
without momentum transfer as two of its quarks can participate
in diffractive scattering, each of them may getting a momentum transfer, 
while the total momentum transfer is zero.

\section{Diffraction: Feynman diagrams}
\setcounter{equation}{0}

\subsection{\boldmath$q\,N\to q\,G\,N$}

For the example of diffractive excitation of a quark,
\beq
q\,N\to q\,G\,N\ ,
\label{c.1}
\eeq
we demonstrate in the following the techniques and approximations
we use for the calculation of more complicated diffractive processes.

We use the following notations for the kinematics of (\ref{c.1}):
$\vec k_T$ and $\vec p_T$ are the transverse momenta of the 
final gluon and quark respectively; $\alpha$ is the fraction 
of the initial light--cone momentum carried by the gluon;
$\vec q_T=\vec k_T+\vec p_T$ is the total transverse  
momentum of the final quark and gluon, and
$\vec\kappa_T=(1-\alpha)\vec k_T-\alpha\vec p_T$
appears further on, when the transverse separations
$\vec r_G=\vec b+(1-\alpha)\vec\rho$ and
$\vec r_q=\vec b+\alpha\vec\rho$ 
are inserted: $\vec k_T\cdot\vec r_G+\vec p_T\cdot\vec r_q=
(\vec k_T + \vec p_T)\cdot\vec b +
\Bigl((1-\alpha)\vec k_T+\alpha\vec p_T\Bigr)\cdot\vec\rho$.

We normalize the amplitude of (\ref{c.1}) according to
\beqn
\frac{d\sigma(qN\to qGN)}{d{\rm ln}\,\alpha\,
d^2\kappa_T\,d^2q_T} &=&
{1\over3}\,\sum\limits_{\mu,nu,s}
\left|A^{(\mu,\nu)}_s\left(\vec q_T,\vec\kappa_T,\alpha\right)
\right|^2\nonumber\\
&=& {1\over 3}\,\sum\limits_{s}{\rm Tr}\,
\left[A^{\dagger}_s\left(\vec q_T,\vec\kappa_T,\alpha\right)\,
A_s\left(\vec q_T,\vec\kappa_T,\alpha\right)\right]\ ,
\label{c.2}
\eeqn
where
\beq
A^{(\mu,\nu)}_s=\left(q^{\mu}\right)^{\dagger}\,
\hat A_s\,q^{\nu}\ ,
\label{c.3}
\eeq
and $q^{\nu(\mu)}$ are the color spinors of the quark in the
initial and final states; $s$ is the color index of the radiated gluon.

We assume that at high energies one can neglect the ratio of the 
real to imaginary parts of the amplitude for reaction (\ref{c.1}).
Then one can apply the generalized optical theorem 
({\sl Cutkosky rules} \cite{cut}),
\beq
\hat A(a\to b)={i\over2}\,\sum\limits_c \hat A^{\dagger}(b\to c)\,
\hat A(a\to c)\ .
\label{c.4}
\eeq
here $\sum\limits_c$ includes not only a sum over intermediate 
channels, but also an integration over the intermediate particle
momenta.

To simplify the problem we switch to the impact parameter representation,
\beq
\hat A(\vec b,\vec\rho) = 
\frac{1}{(2\,\pi)^2}\,
\int d^2q\,d^2\kappa_T\,
\hat A(\vec q_T,\vec\kappa_T)\,
{\rm exp}\,\left(-i\,\vec q_T\vec b -
i\,\vec\kappa_T\vec\rho\right)\ .
\label{c.5}
\eeq
Since the initial impact parameters are preserved during 
the interaction we sum only over intermediate channels in 
this representation.

We use the Born approximation, {\it i.e.} the lowest order
in $\alpha_s$, for the sake of clarity, and generalization is 
straightforward.
In this case and for $a=\{qN\},\ b=\{qGN\}$ only two
intermediate states are possible in (\ref{c.4}):
$c_1=\{q\,N_8^*\}$ and $c_2=\{q\,G\,N_8^*\}$,
where $N_8^* \equiv |3q\ra_8$ is the octet color state of the
$3q$ system produced when the nucleon absorbs the exchanged gluon.

One should sum in (\ref{c.4}) over all excitations $f$ of the $N_8^*$,
\beqn
\hat A_s(qN\to qGN)&=&\sum\limits_f\,
\left[ \hat A^{\dagger}_s(qGN\to qN^*_8)\,
\hat A(qN\to qN^*_8)\right.\nonumber\\ &+& \left.
\sum\limits_{s'}\,\hat A^{\dagger}_{s's}(qGN\to qGN^*_8)\,
\hat A_{s'}(qN\to qGN^*_8)\right]\ .
\label{c.6}
\eeqn
Here $s'$ is the color index of the gluon in the intermediate state.

We skip the simple but lengthy details of calculation of 
the amplitudes on the {\it r.h.s} of (\ref{c.6})
and present only the results.
\beq
\hat A(qN\to qN^*_8)=
\tau_{r}\,
\left\la f\left|\hat\gamma_{r}(\vec b_1)\right|i\right\ra\ ;
\label{c.7}
\eeq
\beqn
\hat A_s'(qN\to qGN^*_8)&=&
\left[\tau_{s'}\,\tau_{r}\,
\left\la f\left|\hat\gamma_{r}(\vec b_1)\right|i\right\ra
- \tau_{r}\,\tau_{s'}\,
\left\la f\left|\hat\gamma_{r}(\vec b_2)\right|i\right\ra
\right.\nonumber\\
&-& \left.i\,f_{s'rp}\,\tau_p\,
\left\la f\left|\hat\gamma_{r}(\vec b_3)\right|i\right\ra\right]\,
\frac{\sqrt{3}}{2}\,
\Psi_{qG}(\alpha,\vec\rho)\ ;
\label{c.8}
\eeqn
\beqn
\hat A_s(qGN\to qN^*_8)&=&
\left[\tau_{r}\,\tau_{s}\,
\left\la f\left|\hat\gamma_{r}(\vec b_1)\right|i\right\ra
- \tau_{s}\,\tau_{r}\,
\left\la f\left|\hat\gamma_{r}(\vec b_2)\right|i\right\ra
\right.\nonumber\\
&-& \left.i\,f_{rsp}\,\tau_p\,
\left\la f\left|\hat\gamma_{r}(\vec b_3)\right|i\right\ra\right]\,
\frac{\sqrt{3}}{2}\,
\Psi_{qG}(\alpha,\vec\rho)\ ;
\label{c.9}
\eeqn
\beq
\hat A_{ss'}(qGN\to qGN^*_8)=
\delta_{ss'}\,\tau_r\,
\left\la f\left|\hat\gamma_{r}(\vec b_2)\right|i\right\ra
+i\,f_{ss'r}\,
\left\la f\left|\hat\gamma_{r}(\vec b_3)\right|i\right\ra\ .
\label{c.10}
\eeq
Here $\vec b_1=\vec b$; $\vec b_2=\vec b-\alpha\,\vec\rho$ are
the impact parameters of the projectile and ejectile quarks 
in reaction (\ref{c.1}), respectively;
$\vec b_3=\vec b + (1-\alpha)\,\vec\rho$ is the impact 
parameter of the radiated gluon; $\vec\rho$ 
is the transverse separation inside the $qG$ system,
and $\vec b$ is the distance from its center of gravity
to the nucleon target; $\Psi_{qG}(\alpha,\vec\rho)$ is the
distribution function for the $qG$ pair;
$\lambda_r=2\,\tau_r$ are the Gell-Mann matrices; 
$f_{rsp}$ is the structure constant for the $SU(3)$ group.
The matrices $\hat\gamma_r(\vec b_k),\ (k=1,2,3)$
are the operators in coordinate and color space for the 
target quarks,
\beq
\hat\gamma_r(\vec b_k)=
\sum\limits_{j=1}^3\, \tau^{(j)}_r\,
\chi(\vec b_k-\vec s_i)
\label{c.11}
\eeq
\beq
\chi(\vec\beta)={1\over\pi}\int d^2q\,
\frac{\alpha_s(q)\,{\rm exp}(i\,\vec q_T\vec\beta)}
{q^2+\Lambda^2}\ ,
\label{c.12}
\eeq
where $\vec s_i$ is the transverse distance between the
$j$-th valence quark of the target nucleon and its 
center of gravity; the matrices $\tau^{(j)}_r$
act on the color indices of this quark.
The matrix elements $\la f|\hat\gamma_r(\vec b_k)|i\ra$
between the initial $i=N$ and final $f=N_8^*$ states
are expressed through the wave functions of these states.
The effective infra-red cut off $\Lambda$ in
(\ref{c.12}) does not affect our results, which are
infra-red stable due to color screening effects.

Substitution of (\ref{c.7})--(\ref{c.10}) into (\ref{c.6}) results in,
\beqn
\hat A_s(qN\to qGN) &=& {i\over2}\,
\left\{\tau_s\,\tau_{r}\,\tau_{r'}\,
|\Phi_{rr'}(\vec b_1,\vec b_1) -
\tau_r\,\tau_{s}\,\tau_{r'}\,
|\Phi_{rr'}(\vec b_2,\vec b_1)+
i\,f_{rsp}\,\tau_{p}\,\tau_{r'}\,
|\Phi_{rr'}(\vec b_3,\vec b_1)
\right.\nonumber\\
&+& \tau_r\,\tau_{s}\,\tau_{r'}\,
|\Phi_{rr'}(\vec b_1,\vec b_2)-
\tau_s\,\tau_{r}\,\tau_{r'}\,
|\Phi_{rr'}(\vec b_2,\vec b_2)-
i\,f_{rsp}\,\tau_{p}\,\tau_{r'}\,
|\Phi_{rr'}(\vec b_2,\vec b_3)
\nonumber\\ 
 &-& \left.
i\,f_{ss'r}\,\left[\tau_{s'}\,\tau_{r'}\,
|\Phi_{rr'}(\vec b_1,\vec b_3) -
\tau_{r'}\,\tau_{s'}\,
|\Phi_{rr'}(\vec b_2,\vec b_3) -
i\,f_{s'r'p}\,\tau_p\,\Phi_{rr'}(\vec b_3,\vec b_3)
\right]\right\}\nonumber\\
&\times&\,\frac{\sqrt{3}}{2}\,
\Psi_{qG}(\alpha,\vec\rho)\ ,
\label{c.13}
\eeqn
where
\beq
\Phi_{rr'}(\vec b_k,\vec b_l)=
\sum\limits_{f}\,
\left\la i\left|\hat\gamma_r(\vec b_k)
\right|f\right\ra\,
\left\la f\left|\hat\gamma_{r'}(\vec b_l)
\right|i\right\ra\ .
\label{c.14}
\eeq

We sum in (\ref{c.13}) over all excitations of the two color
octet states of the $3q$ system.
To have a complete set of states we have to include also 
color singlet and decuplet $|3q\ra$ states. 
As these states cannot be produced via single 
gluon exchange, they do not contribute and 
we can simply extend the summation in
(\ref{c.14}) to the complete set of states and get,
\beq
\Phi_{rr'}(\vec b_k,\vec b_l)=
\left\la i\left|\hat\gamma_r(\vec b_k)
\hat\gamma_{r'}(\vec b_l)
\right|i\right\ra\ .
\label{c.15}
\eeq

In the matrix element (\ref{c.15}) 
we average over color indices of the valence
quarks and their relative coordinates in the target nucleon.
To do so one should use the relation,
\beq
\left\la \tau^{(j)}_r\cdot
\tau^{(j')}_{r'}\right\ra_{|3q\ra_1} =
\left\{ \begin{array}{cc}
{1\over6}\,\delta_{rr'}&\ \ (j=j')\\
-{1\over12}\,\delta_{rr'}&\ \ (j\not= j')
\end{array}\right.
\label{c.16}
\eeq
Then, Eq.~(\ref{c.15}) can be represented as,
\beq
\Phi_{rr'}(\vec b_k,\vec b_l)=
{3\over4}\,\delta_{rr'}\,S(\vec b_k,\vec b_l)\ ,
\label{c.17}
\eeq
where $S(\vec b_k,\vec b_l)$ is a scalar function of two 
vector variables,
\beq
S(\vec b_k,\vec b_l)=
{2\over9}\,\int d\{s\}\,\left[
\sum\limits_{j=1}^3\,
\chi(\vec b_k-\vec s_j)\,\chi(\vec b_l-\vec s_j)-
{1\over2}\, \sum\limits_{j\not=j'}\, 
\chi(\vec b_k-\vec s_j)\,\chi(\vec b_l-\vec s_{j'})
\right]\,
\left|\Phi_{3q}(\{s\})\right|^2\ .
\label{c.18}
\eeq
This function is directly related to the $\bar qq$ dipole
cross section (\ref{2.3a}),
\beq
\sigma_{\bar qq}(\vec\rho_1-\vec\rho_2)=
\int d^2b\,\left[S(\vec b+\vec\rho_1,\vec b+\vec\rho_1)
+ S(\vec b+\vec\rho_2,\vec b+\vec\rho_2)-
2\,S(\vec b+\vec\rho_1,\vec b+\vec\rho_2)\right]\ .
\label{c.19}
\eeq

According to (\ref{c.17}) and (\ref{c.18}) the function 
$\Phi(\vec b_k,\vec b_l)$ is symmetric under the replacement 
$\vec b_k\rightleftharpoons\vec b_l$. Therefore,
the terms proportional to $\Phi(\vec b_1,\vec b_2)$
and $\Phi(\vec b_2,\vec b_1)$ in (\ref{c.13}) cancel,
as well as the terms proportional to $\Phi(\vec b_1,\vec b_3)$
and $\Phi(\vec b_3,\vec b_1)$. At the same time,
the terms proportional to $\Phi(\vec b_2,\vec b_3)$
and $\Phi(\vec b_3,\vec b_2)$ add up.

Making use of the relations,
\beqn
\tau_r\tau_r&=&4/3\nonumber\\
f_{ss'r}\,f_{s'rp}&=&3\delta_{sp}\nonumber\\
-i\,f_{rsp}\,\tau_p\,\tau_r&=&
{3\over2}\,\tau_s\ ,
\label{c.20}
\eeqn
we arrive at the final result for the amplitude of diffractive
dissociation of a quark $(q\,N\to q\,G\,N)$ in impact 
parameter representation,
\beqn
\hat A_s(\vec b,\vec\rho,\alpha)&=&
\frac{i\,3\,\sqrt{3}}{16}\,
\tau_s\,\Psi_{qG}(\alpha,\vec\rho)\,
\left\{{4\over3}\,\left[S(\vec b_1,\vec b_1) -
S(\vec b_2,\vec b_2)\right]\right.
\nonumber\\
&+& \left.3\,\left[S(\vec b_1,\vec b_3)-
S(\vec b_3,\vec b_3)\right]\right\}\ .
\label{c.21}
\eeqn

The diffraction amplitude in momentum representation reads,
\beq
\hat A_s(\vec q_T,\vec\kappa_T,\alpha)=
\frac{1}{(2\pi)^2}\,
\int d^2b\,d^2\rho\,\hat A_s(\vec b,\vec\rho,\alpha)\,
{\rm exp}\,\left(i\,\vec q_T\vec b + 
i\,\vec\kappa_T\vec\rho\right)\ .
\label{c.22}
\eeq

Using (\ref{c.19}) and the above mentioned symmetry of 
$S(\vec\rho_1,\vec\rho_2)$ we obtain a very simple
expression for the forward ($q_T=0$) diffraction amplitude
which is related to the dipole cross section,
\beq
\hat A_s(0,\vec\kappa_T,\alpha)=
-\,\frac{i\,9\,\sqrt{3}}
{32\,(2\pi)^2}\,\tau_s\,
\int d^2\rho\,\Psi_{qG}(\alpha,\vec\rho)\,
\sigma_{\bar qq}(\vec\rho)\,
e^{i\,\vec\kappa_T\vec\rho}\ .
\label{c.23}
\eeq
Eventually, the forward diffractive dissociation cross section
of a quark reads,
\beqn
\left.\frac{d\sigma}{d({\rm ln}\,\alpha)\,
d^2q_T}\right|_{q_T=0} &=&
{1\over3}\,\int d^2\kappa_T\,
\sum\limits_s\, {\rm Tr}\,
\hat A^{\dagger}_s(0,\vec\kappa_T,\alpha)\,
\hat A_s(0,\vec\kappa_T,\alpha)\nonumber\\
&=& \frac{1}{(4\pi)^2}\,
\int d^2\rho\,
\left|\Psi_{qG}(\alpha,\vec\rho)\,
{9\over8}\,\sigma_{\bar qq}(\vec\rho)\right|^2\ .
\label{c.24}
\eeqn
We should emphasize that all above calculations are done
for an arbitrary $\alpha$.

\subsection{Diffractive gluon radiation by a \boldmath$\bar qq$ pair}
\setcounter{equation}{0}

Gluon radiation is an important contribution to the diffractive dissociation
of a (virtual) photon,
\beq
\gamma^*\,N\to \bar q\,q\,G\,N\ .
\label{d.1}
\eeq
In analogy to the previous section we make use of
the generalized unitarity relation,
\beqn
\hat A_s(\gamma^*N\to \bar qqGN) &=&
{i\over 2}\,\sum\limits_f\,
\left[\hat A_s(\bar qqGN\to\bar qqN^*_8)\,
\hat A(\gamma^*N\to\bar qqN^*_8)\right.
\nonumber\\ &+& \left.
\sum\limits_{s'}\,\hat A_{ss'}(\bar qqGN\to\bar qqGN^*_8)\,
\hat A_{s'}(\gamma^*N\to\bar qqGN^*_8)\right]\ ,
\label{d.2}
\eeqn
where the amplitudes are defined as follows,
\beq
\hat A(\gamma^*N\to\bar qqN^*_8) = 
\left[\tau_r\,\left\la f \left|
\hat\gamma_r(\vec b_1)\right|i\right\ra +
\bar\tau_r\,\left\la f \left|
\hat\gamma_r(\vec b_2)\right|i\right\ra
\right]\,\Psi_{\bar qq}(\vec\rho_1-\vec\rho_2,\alpha)\,
\left|\bar qq\right\ra_1\ ;
\label{d.3}
\eeq
\beqn
&&\hat A_{s'}(\gamma^*N\to \bar qqN^*_8) =
\frac{i\sqrt{3}}{2}\, f_{s'rp}\,
\left\{ \tau_p\,\left[
\left\la f \left|
\hat\gamma_r(\vec b_1)\right|i\right\ra -
\left\la f \left|
\hat\gamma_r(\vec b_3)\right|i\right\ra
\right]\,\Psi_{qG}(\vec\rho_1)\right.
\nonumber\\ &+& \left.
\bar\tau_p\,\left[ \left\la f \left|
\hat\gamma_r(\vec b_2)\right|i\right\ra
- \left\la f \left|
\hat\gamma_r(\vec b_3)\right|i\right\ra\right]\,
\Psi_{qG}(\vec\rho_2)\right\}\,
\Psi_{\bar qq}(\vec\rho_1-\vec\rho_2,\alpha)\,
\left|\bar qq\right\ra_1\ ;
\label{d.4}
\eeqn
\beqn
\hat A_s(\bar qqGN\to\bar qqN^*_8) &=&
-\,\frac{i\,\sqrt{3}}{2}\,
f_{srp}\,\left\{
\tau_p\,\left[\left\la f \left|
\hat\gamma_r(\vec b_1)\right|i\right\ra-
\left\la f \left|
\hat\gamma_r(\vec b_3)\right|i\right\ra
\right]\,\Psi_{qG}(\vec\rho_1)
\right.\nonumber\\ &+& \left.
\bar\tau_p\,\left[\left\la f \left|
\hat\gamma_r(\vec b_2)\right|i\right\ra -
\left\la f \left|
\hat\gamma_r(\vec b_3)\right|i\right\ra
\right]\,\Psi_{qG}(\vec\rho_2)\right\}\ ;
\label{d.5}
\eeqn
\beqn
\hat A_{ss'}(\bar qqGN\to\bar qqGN^*_8) &=&
\tau_r\,\left\la f \left|
\hat\gamma_r(\vec b_1)\right|i\right\ra\,
\delta_{ss'} +\bar\tau_r\,
\left\la f \left|
\hat\gamma_r(\vec b_2)\right|i\right\ra\,
\delta_{ss'} \nonumber\\
&+& i\,f_{ss'r}\,\left\la f \left|
\hat\gamma_r(\vec b_3)\right|i\right\ra\ .
\label{d.6}
\eeqn
Here $\vec b_1=\vec b+\vec r_1$, $\vec b_2=\vec b+\vec r_2$,
$\vec b_3=\vec b+\vec\rho$ are the impact parameters 
of the quark, antiquark and gluon respectively;
$\vec b$ is the photon impact parameter;
$\vec\rho_{1,2}=\vec\rho-\vec r_{1,2}$;
$\Psi_{\bar qq}$ and $|\bar qq\ra$ are spatial and color
parts of the $\bar qq$-component of the photon wave function,
respectively.
The matrices $\tau_r=\lambda_r/2$ and $\bar\tau_r=\lambda^*_r/2$ 
act on the color indices of quark and antiquark respectively.
The indices $s,\ s'$ mark the color states of the gluons in intermediate
and final states.

Note that the condition of color neutrality of the singlet
state $|\bar qq\ra$ leads to the relation,
\beq
\left(\tau_r+\bar\tau_r\right)\,
\left|\bar qq\right\ra = 0
\label{d.7}
\eeq

Substitution of (\ref{d.2}) - (\ref{d.6}) into (\ref{d.1})
leads to the following expression for the amplitude
of diffractive dissociation of the photon,
\beqn
&&
\hat A(\gamma^*N\to\bar qqGN) =
\frac{i\,3\,\sqrt{3}}{16}\,
\left\{\left(
i\,f_{srp}\,
\left[\tau_p\,\tau_r +
\tau_r\,\tau_p\right]\,
\left[s(\vec b_1,\vec b_1) - s(\vec b_3,\vec b_1)\right]
\right.\right.\nonumber\\ &+& \left.
i\,f_{srp}\,\left[\tau_p\,\bar\tau_r+\bar\tau_r\,\tau_p\right]\,
\left[s(\vec b_1,\vec b_2)-s(\vec b_3,\vec b_2)\right] +
f_{ss'r}\,f_{s'rp}\,\tau_p\,
\left[s(\vec b_3,\vec b_1)-s(\vec b_3,\vec b_3)\right]\right)\,
\Psi_{qG}(\vec\rho_1)\nonumber\\ &+&
\left( i\,f_{srp}\,\left[\bar\tau_p\,\bar\tau_r +
\bar\tau_r\,\bar\tau_p\right]\,
\left[ s(\vec b_2,\vec b_2)-s(\vec b_2,\vec b_3)\right] +
i\,f_{srp}\,\left[\bar\tau_p\,\tau_r +
\tau_r\,\bar\tau_p\right]\,
\left[ s(\vec b_2,\vec b_1)-s(\vec b_3,\vec b_1)\right] 
\right.\nonumber\\ &+& \left.\left.
f_{ss'r}\,f_{s'rp}\,\bar\tau_p\,
\left[ s(\vec b_3,\vec b_2)-s(\vec b_3,\vec b_3)\right]
\right)\,\Psi_{qG}(\vec\rho_2)\right\}\,
\left|\bar qq\right\ra_1\,\Psi_{\bar qq}(\vec\rho_1-
\vec\rho_2,\alpha)\ ,
\label{d.8}
\eeqn
where we made use of the completeness condition, $\sum_f\,
|f\ra\la f|=1$ (see Appendix B.1).

In order to simplify Eq.~(\ref{d.8}) we apply a 
few relations as follows.
Since $f_{srp}=-f_{spr}$ we find
\beq
f_{srp}\,\left[\tau_p\,\tau_r +
\tau_r\,\tau_p\right] =
f_{srp}\,\left[\bar\tau_p\,\bar\tau_r +
\bar\tau_r\,\bar\tau_p\right] = 0\ .
\label{d.9}
\eeq
Then, relying on the condition (\ref{d.7}) we find,
\beqn
\left(\tau_p\,\bar\tau_r + 
\bar\tau_r\,\tau_p\right)\,\left|\bar qq\right\ra_1 &=&
2\,\tau_p\,\bar\tau_r\,\left|\bar qq\right\ra_1
\nonumber\\ &=&\,-\,2\,\tau_p\,\tau_r\,\left|
\bar qq\right\ra_1\ ;
\label{d.10}
\eeqn
\beqn
\left(\bar\tau_p\,\tau_r + 
\tau_r\,\bar\tau_p\right)\,\left|\bar qq\right\ra_1 &=&
2\,\bar\tau_p\,\tau_r\,\left|\bar qq\right\ra_1
\nonumber\\ &=&\,-\,2\,\bar\tau_p\,\bar\tau_r\,\left|
\bar qq\right\ra_1\ .
\label{d.11}
\eeqn
We also use the relations
\beqn
i\,f_{srp}\,\tau_p\,\tau_r &=& 
{3\over2}\,\tau_s\ ;\nonumber\\
i\,f_{srp}\,\bar\tau_p\,\bar\tau_r &=& 
{3\over2}\,\bar\tau_s\ ;\nonumber\\
f_{ss'r}\,f_{s'rp} &=& 
3\,\delta_{sp}\ ,
\label{d.12}
\eeqn
and the symmetry condition, $s(\vec b_k,\vec b_l)=
s(\vec b_l,\vec b_k)$, and eventually arrive at a
modified form of Eq.~(\ref{d.8})
\beqn
\hat A_s(\gamma^*N\to\bar qqGN) &=&
\frac{9\,\sqrt{3}}{16}\,
\left[\tau_s\,\Psi_{qG}(\vec\rho_1) +
\bar\tau_s\,\Psi_{qG}(\vec\rho_2)\right]\,
\left|\bar qq\right\ra_1\,
\Psi_{\bar qq}(\vec\rho_1-\vec\rho_2,\alpha)\,
\nonumber\\ &\times&\,
\left[s(\vec b_2,\vec b_3)+s(\vec b_1,\vec b_3)-
s(\vec b_1,\vec b_2)-s(\vec b_3,\vec b_3)\right]\ .
\label{d.13}
\eeqn
The last factor in square brackets can be represented as,
\beqn
P(\vec b_1,\vec b_2;\vec b_3) &\equiv &
s(\vec b_2,\vec b_3)+s(\vec b_1,\vec b_3)-
s(\vec b_1,\vec b_2)-s(\vec b_3,\vec b_3)\nonumber\\
&\equiv& {1\over2}\,\left\{
\left[s(\vec b_1,\vec b_1)+
s(\vec b_2,\vec b_2) - 2\,s(\vec b_1,\vec b_2)\right] -
\left[s(\vec b_2,\vec b_2)+s(\vec b_3,\vec b_3)-
2\,s(\vec b_2,\vec b_3)\right]\right.\nonumber\\
&-& \left.\left[s(\vec b_1,\vec b_1) + s(\vec b_3,\vec b_3)
-2\,s(\vec b_1,\vec b_3)\right]\right\}\ .
\label{d.14}
\eeqn
Then, the forward diffraction amplitude ($q_T=0$) 
in impact parameter representation has the form,
\beqn
&& \frac{1}{2\,\pi}\,\int d^2b\,
\hat A_s(\vec b,\vec\rho_1\vec\rho_2)\nonumber\\
&=&\, -\,\frac{i}{4\,\pi}\,
\Sigma(\vec\rho_1,\vec\rho_2)\,
\left[\frac{\sqrt{3}}{2}\,\tau_s\,\Psi_{qG}(\vec\rho_1)+
\frac{\sqrt{3}}{2}\,\bar\tau_s\,\Psi_{qG}(\vec\rho_2)
\right]\,\left|\bar qq\right\ra_1\,
\Psi_{\bar qq}(\vec\rho_1-\vec\rho_2,\alpha)\ ,
\label{d.15}
\eeqn
where $\Sigma(\vec\rho_1,\vec\rho_2)$ is introduced 
in (\ref{3.33}).

From (\ref{d.15}) one easily gets the forward 
diffractive cross section,
\beqn
\left.\frac{d\sigma(\gamma^*N\to\bar qqGN)}
{d({\rm ln}\,\alpha_G)\,dq_T}\right|_{
\begin {array}{c}q_T=0\\ 
\alpha_G\to 0 \end{array}} &=&
\frac{1}{(4\,\pi)^2}
\int d^2\rho_1\,d^2\rho_2\,d\alpha\,
\Biggl|\Psi_{\bar qq}(\vec\rho_1-\vec\rho_2,
\alpha)\,\nonumber\\ &\times&
\biggl[\Psi_{qG}(\vec\rho_1)-
\Psi_{qG}(\vec\rho_2)\biggr]\,
\Sigma(\vec\rho_1,\vec\rho_2)\Biggr|^2\ .
\label{d.16}
\eeqn

\subsection{Diffractive photon radiation, \boldmath$q\,N\to\gamma\,q\,N$} 
\setcounter{equation}{0}

Diffractive electromagnetic radiation is calculated
in analogy to what was done in Appendix B.1 for gluon radiation.
Since the photon does not interact with the gluonic field of
the target the structure of all the amplitudes
in the relation,
\beqn
\hat A(qN\to q\gamma N)&=&
{i\over2}\,\sum\limits_f\left[
\hat A^{\dagger}(q\gamma N\to qN^*_8)\,
\hat A(qN\to qN^*_8)\right.\nonumber\\ 
&+& \left.
\hat A^{\dagger}(q\gamma N\to q\gamma N^*_8)\,
\hat A(qN\to q\gamma N^*_8) \right]\ ,
\label{e.1}
\eeqn
turns out to me much simpler.
\beq
\hat A(qN\to qN^*_8)=
\tau_r\,
\left\la f\left|\hat\gamma_r(\vec b_1)
\right| i\right\ra\ ;
\label{e.2}
\eeq
\beq
\hat A(q\gamma N\to q\gamma N^*_8)=
\tau_r\,
\left\la f\left|\hat\gamma_r(\vec b_2)
\right| i\right\ra\ ;
\label{e.3}
\eeq
\beqn
\hat A(qN\to q\gamma N^*_8)&=&
\hat A(q\gamma N\to q\gamma N^*_8)
\nonumber\\&=&
\tau_r\,\left[
\left\la f\left|\hat\gamma_r(\vec b_1)
\right| i\right\ra -
\left\la f\left|\hat\gamma_r(\vec b_2)
\right| i\right\ra 
\right]\,\Psi_{q\gamma}(\vec\rho,\alpha)
\label{e.4}
\eeqn
Here $\vec b_1=\vec b$, $\vec b_2=\vec b-\alpha\,\vec\rho$
are the impact parameters of the quark before and after radiation of
the photon; $\vec\rho$ is the transverse separation between
the quark and photon in the final state; and $\alpha$ is the
fraction of the quark light cone momentum carried away by the photon.
$\Psi_{q\gamma}(\vec\rho,\alpha)$ is the distribution function for
the $q\gamma$ fluctuation of the quark.
The initial, $|i\ra$, and final, $|f\ra$, states of the
target, as well as the operators $\hat\gamma(\vec b_k)$ ($k=1,2$)
are the same as in Appendix B.1.

After substitution of (\ref{e.2}) -- (\ref{e.4})
into (\ref{e.1}) we get,
\beq
\hat A(qN\to q\gamma N) =
{i\over2}\,\left\{
\tau_r\,\tau_{r'}\,\left[
\Phi_{rr'}(\vec b_1,\vec b_1) -
\Phi_{rr'}(\vec b_1,\vec b_2) +
\Phi_{rr'}(\vec b_2,\vec b_1) -
\Phi_{rr'}(\vec b_2,\vec b_2)\right]
\right\}\ .
\label{e.5}
\eeq
Here the functions $\Phi_{rr'}(\vec b_k,\vec b_l)$
are defined in Appendix B.1 Then, the amplitude 
in impact parameter representation reads,
\beq
\hat A(\vec b,\vec\rho)=
{i\over 2}\,\left[s(\vec b_1,\vec b_1) -
s(\vec b_2,\vec b_2)\right] =
{i\over 2}\,\left[s(\vec b,\vec b) -
s(\vec b-\alpha\,\vec\rho,\vec b-\alpha\,\vec\rho)\right]\ .
\label{e.6}
\eeq
After Fourier transform to the momentum
representation we get for the forward diffractive
amplitude of photon radiation,
\beq
\left.A(\vec q_T,\vec\kappa_T)\right|_{q_T=0} =
\left.\frac{1}{(2\pi)^2}\,
\int d^2b\,d^2\rho\,
{\rm exp}\,\left(i\,\vec q_T\,\vec b +
i\,\vec\kappa_T\,\vec\rho\right)\,
\hat A(\vec q_T,\vec\kappa_T)\right|_{q_T=0}
\,=\,0\ .
\label{e.7}
\eeq
Thus, the direct calculation of Feynman diagrams confirms
our previous conclusion (Appendix A.4) that a quark does not 
diffractively emit electromagnetic radiation if the momentum
transfer with the target is zero (as different from
the statement in \cite{k}). A hadron, however, can radiate
in forward scattering.

\section{The triple-Pomeron coupling}

\setcounter{equation}{0}

In the limit of vanishing quark and gluon masses the quark-gluon 
wave function (\ref{3.12}) retains only the second term $\propto
\vec\Phi_1$ which has the form (\ref{2.21}).
Bi-linear combinations of this wave function averaged over final 
polarizations can be represented as follows,
\beqn
\left|\Phi_{1}(\vec\rho_i,\alpha)\right|^2&=&
\frac{1}{(2\pi)^2}\,
\int\limits_{b(\alpha)^2}^{\infty}dt\,e^{-t\rho_i^2}
\ ,\label{a.1}\\
\vec\Phi_{1}(\vec\rho_i,\alpha)\cdot\vec\Psi_1(\vec\rho_k)&=&
\frac{\vec\rho_i\cdot\vec\rho_k}{(2\pi)^2}
\int\limits_{b^2(\alpha)/2}^{\infty}du
\int\limits_{b^2(\alpha)/2}^{\infty}dt\,
e^{-t\rho_i^2-u\rho_i^2}\ .
\label{a.2}
\eeqn

This together with (\ref{2.24}) and (\ref{3.39}) allows to 
integrate analytically 
over the coordinates of the quarks and the gluon
in (\ref{3.38}).
Finally integrating over $t$ and $u$ we arrive at,
\beq
G_{3P}(NN\to XN) = 
\frac{\alpha_s}{(4\pi)^2}\left({9\over8}\,\sigma_0\right)^2\,
\biggl[F_1(x,z)-F_2(x,z)\biggr]\ ,
\label{a.3}
\eeq
where $x=b^2(0)\rho_0^2$, $z=z_N=2\la r^2\ra_p/\rho_0^2$, and
\beqn
F_1(x,z)&=&{\rm ln}\left[\frac{(x+1)^2}{x(x+2)}\right] +
2s_1\,{\rm ln}\left[\frac{(x+1)(x+s_1)}{x(1+x+s_1)}\right]
\nonumber\\
&+&{2\over3}\,s_2\left[2\,{\rm ln}\left(\frac{s_1x+s_2}{xs_1}\right)
- {\rm ln}\left(\frac{x+2s_2}{x}\right)\right]\nonumber\\
&+&{1\over3}\,s_3s_4\left[2\,{\rm ln}\left(\frac{s_1x+s_3s_4}{xs_1}\right)
- {\rm ln}\left(\frac{x+2s_4}{x}\right)\right]\ .
\label{a.4}
\eeqn
Here 
\beqn
&&s_1=\frac{1}{1+z},\ \ \ s_2=\frac{1}{1+2z},\ \ \
s_3=\frac{2}{2+z},\nonumber\\
&&s_4=\frac{2}{2+3z},\ \ \ s_5=\frac{4}{4+3z}\ ;
\label{a.5}
\eeqn
\beq
z\,F_2(x,z)=\sum\limits_{i=1}^{14}\frac{g(i)}{\beta(i)}\,
{\rm ln}\left[\frac{\delta(i)\,\gamma(i)}
{\delta(i)\,\gamma(i)-\beta^2(i)}\right]\ ,
\label{a.6}
\eeq
where
\beq
\begin{array}{cccccc}
i=1 && g={2/3}, & \beta={1/z}+2, & 
\gamma=\delta={x/2}+\beta\ ;&\\
i=2 &&  g=2, & \beta={1/z}+1, & 
\gamma=\delta={x/2}+\beta\ ; & \\
i=3 &&  g=-{10/3}, & \beta={1/z}+1, & 
\delta={x/2}+\beta, & 
\gamma=\delta+1\ ;\\
i=4 &&  g=1, & \beta={1/z}, & 
\gamma=\delta={x/2}+\beta\ ; & \\
i=5 &&  g=-4, & \beta={1/z}, &
\delta={x/2}+\beta, &
\gamma=\delta+1\ ;\\
i=6 &&  g={5/3}, & \beta={1/z}+2, &
\gamma=\delta={x/2}+\beta+1\ ; &\\
i=7 &&  g=2, & \beta={1/z}, &
\delta={x/2}+\beta, &
\gamma=\delta+2\ ;\\
i=8 &&  g={s_4/3}, & \beta={1/z}+{1/2}, &
\gamma=\delta={x/2}+\beta\ ; &\\
i=9 &&  g=-2s_5/3, & \beta={1/z}+s_5/4, &
\delta={x/2}+\beta, &
\gamma=\delta+1\ ;\\
i=10 &&  g=2s_5/3, & \beta={1/z}+1+s_5/4, &
\gamma=\delta={x/2}+\beta\ ; &\\
i=11 &&  g=-{s_4/3}, & \beta={1/z}-s_4/2 &
\gamma=\delta={x/2}+\beta +s_4\ ; &\\
i=12 &&  g=-2s_5/3, & \beta={1/z}+1-s_5/4, &
\gamma=\delta={x/2}+\beta+s_5/2\ ; &\\
i=13 &&  g=-2s_5/3, & \beta={1/z}-s_5/4, &
\delta={x/2}+\beta+s_5/2, &
\gamma=\delta+1\ ;\\
i=14 &&  g=2s_4/3, & \beta={1/z}, &
\delta={x/2}+\beta+1/2, &
\gamma={x/2}+\beta+s_4/2\ .
\end{array}
\label{a.7}
\eeq

The effective triple-Pomeron coupling $G_{3P}(MN\to XN)$ for 
diffractive dissociation of a meson $M$ can be calculated
in a similar way assuming a Gaussian shape of the
quark wave function of the meson,
\beq
\left|\Psi_{M\to\bar qq}(\vec r)\right|^2=
\frac{1}{\pi R^2}\,e^{-r^2/R^2}\ ,
\label{a.8}
\eeq
where $R^2=8\,\la r^2_M\ra_{ch}/3$. The triple-Pomeron coupling
is smaller by a factor $2/3$ (different number of valence quarks)
and has a form similar to (\ref{a.3}), 
\beq
G_{3P}(MN\to XN) =
\frac{2\,\alpha_s}{3\,(4\pi)^2}\left({9\over8}\,\sigma_0\right)^2\,
\biggl[F^M_1(x,z_M)-F^M_2(x,z_M)\biggr]\ ,
\label{a.9}
\eeq
but $z_M=R^2/\rho_0^2\not=z_N$
and the functions $F^M_{1,2}$ are different too.
The expression for $F^M_1(x,z_M)$
results from $F_1(x,z)$ via the replacement $s_3\to 1$, $s_4\to s_2$
and $z\to z_M$.

The expression for $F_2(x,z_M)$ follows from
$F_2(x,z)$ after moderate  modifications in (\ref{a.7}):
$g(1)=1$, $g(3)=-4$, $g(6)=2$, all $g(i)=0$ for $i\geq 8$
and $z\to z_M$.

In the case of diffractive dissociation of a photon,
the calculations are more complicated since the spatial
distribution of quarks in the photon is very
different from a Gaussian. Nevertheless, it
can be represented as a superposition of Gaussians,
\beq
\left|\Psi_{\gamma\to\bar qq}(\vec r,\alpha)\right|^2=
\frac{\alpha_{em}\,N_c\,\sum\, Z_q^2}{2\,\pi}
\int\limits_0^{1/a^2(\alpha)}
\frac{dR^2}{R^2}\left(\frac{1}{\pi R^2}\,e^{-r^2/R^2}\right)\ .
\label{a.10}
\eeq
Then, the effective coupling $G_{3P}(\gamma N\to XN)$
takes a form similar to (\ref{a.3}) and (\ref{a.9}),
\beq
G_{3P}(\gamma N\to XN) =
\frac{2\alpha_{em}\alpha_s\, N_c\,\sum\, Z_q^2}
{6\,(2\pi)^3}\left({9\over8}\,\sigma_0\right)^2\,
\biggl[F^{\gamma}_1(x,z_{\gamma})-F^{\gamma}_2(x,z_{\gamma})\biggr]\ ,
\label{a.11}
\eeq
where $z_{\gamma}=[a^2(\alpha)\,\rho_0^2]^{-1}$ and 
\beq
F^{\gamma}_{1,2}(x,z_{\gamma})=
\int\limits_0^{z_{\gamma}}
\frac{dv}{v}F^M_{1,2}(x,v)\ .
\label{a.12}
\eeq

\end{document}